\documentclass[usenatbib]{mnras}

\usepackage{graphicx}	
\usepackage{amsmath,amssymb,amsfonts,mathrsfs} 
\usepackage{bm}		
\usepackage{soul}
\usepackage[flushleft]{threeparttable}
\usepackage{multirow}
\usepackage{xspace}
\usepackage{Custom_Functions}

\defcitealias{puchwein2015}{P15}		
\defcitealias{puchwein2019}{P19}		
\defcitealias{haardt2012}{HM12}		
\defcitealias{khaire2015a}{KS15}		
\defcitealias{khaire2019a}{KS19}		
\defcitealias{onorbe2017}{OH17}		
\defcitealias{faucher2020}{FG20}		
\defcitealias{gaikwad2018}{G18}	

\usepackage[usenames, dvipsnames]{color}
\definecolor{mycolor}{RGB}{0,128,0}
\definecolor{newaddcolor}{RGB}{255,0,255}

\usepackage[normalem]{ulem}

\usepackage{subfiles} 

\title[Thermal state of IGM at $2 \leq z \leq 4$]{A consistent  and robust measurement of the thermal state of the IGM at $2 \leq z \leq 4$ 
from a large sample of  \lya forest spectra: Evidence for late and rapid HeII reionization}

\author[Gaikwad et.al]{Prakash Gaikwad$^{1,2}$\thanks{E-mail: \href{pgaikwad@ast.cam.ac.uk}{pgaikwad@ast.cam.ac.uk}}, 
    Raghunathan Srianand$^{3}$,
    Martin G. Haehnelt$^{1,2}$ and 
    \newauthor{Tirthankar Roy Choudhury$^{4}$}
    \\
    $^{1}$Institute of Astronomy, University of Cambridge, Madingley Road, Cambridge, CB3 0HA, UK \\
    $^{2}$Kavli Institute for Cosmology, University of Cambridge, Madingley Road, Cambridge, CB3 0HA, UK \\
    $^{3}$Inter-University Centre for Astronomy and Astrophysics (IUCAA), Post Bag 4, Pune 411007, India \\
    $^{4}$National Centre for Radio Astrophysics, Tata Institute of Fundamental Research, Pune 411007, India}

\date{}

\pubyear{2019}

\begin{document}
\label{firstpage}
\pagerange{\pageref{firstpage}--\pageref{lastpage}}
\maketitle


\begin{abstract}
We characterise the thermal state of the intergalactic medium (IGM) in ten
redshift bins in the range \zrange{2}{4} with a sample of 103 high
resolution, high \SNR \lya forest spectra using four different  flux
distribution statistics. Our measurements are  calibrated with mock spectra
from a  large suite  of hydrodynamical simulations post-processed  with our
thermal IGM evolution code \citecode , finely sampling  amplitude  and slope of
the expected temperature-density  relation.  The thermal parameters inferred
from our   measurements of the flux power spectrum,  Doppler parameter
distribution, as well as  wavelet and  curvature statistics agree
well within their respective  errors and all clearly show  the  peak in
temperature  and  minimum in slope  of the temperature density relation
expected from  \HeII reionization. Combining our measurements from the different
flux statistics  gives  $T_0=(14750 \pm 1322)$K for the peak temperature at mean
density and a corresponding minimum slope  $\gamma = 1.225 \pm 0.120$. The peak
in the  temperature evolution occurs around $z \approx 3$, in agreement with previous
measurements that had suggested the presence of  such a peak, albeit with a large scatter.  Using \citecode, we also calculate the thermal state of
the IGM  predicted by  five  widely used (spatially homogeneous) UV-background
models. The rather rapid thermal evolution inferred by our measurements is
well reproduced by two  of the models, if  we assume (physically well
motivated) non-equilibrium evolution with  photo-heating rates that  are
reduced by a moderate factor of $\sim 0.7-0.8$.  The other three  models
predict \HeII reionization to be more extended with  a higher temperature peak occurring somewhat earlier than our measurements suggest. 
\end{abstract}
\begin{keywords}
cosmology: large-scale structure of Universe - methods: numerical - galaxies: intergalactic medium - QSOs: absorption lines
\end{keywords}


\section{Introduction}
\label{sec:introduction}

\lya\ absorption seen in the spectra of distant bright QSOs (Quasi Stellar
Objects) allow one to probe the thermal and ionization
history of the Intergalactic Medium (IGM) in addition to constraining
cosmological parameters.  The thermal state of the IGM  is often characterized
by normalization $(T_0)$ and slope $(\gamma)$ of the temperature-density
relation \citep[TDR, $T=T_0 \Delta^{\gamma-1}$;][]{hui1997}, while  the
ionization state of the IGM is characterized by \HI and \HeII photo-ionization
rates \citep[e.g.][]{haardt1996}.  These parameters have been  measured using
high-resolution, high signal-to-noise spectroscopic observations in conjunction
with high-resolution hydrodynamical simulations.

\vspace{8mm}
Accurate measurements of thermal parameters and photo-ionization rates of the
IGM have  been  used to place constraints on (i) the epoch and extent of \HI
and  \HeII reionization
\citep{miralda1994,theuns2002c,worseck2011,puchwein2015,worseck2016,sanderbeck2016,gaikwad2019,worseck2019,sanderbeck2020},
(ii) ionizing ultra-violet background (UVB) models that are important inputs
for cosmological hydrodynamical simulations \citep[][hereafter
\citetalias{haardt2012,khaire2015a,onorbe2017,khaire2019a,puchwein2019,faucher2020}
respectively]{mcdonald2001b,bolton2005,faucher2009,haardt2012,khaire2015a,onorbe2017,khaire2019a,puchwein2019,faucher2020},
(iii) the escape fraction of \HI\ ionizing photons from galaxies and the
relative contribution of galaxies and QSOs to the total \HI ionizing
background \citep{faucher2008,khaire2016,khaire2017}, (iv) and the effect of non-radiative
processes like heating by cosmic rays \citep[][]{Nath1993,samui2005}, blazars
\citep[][]{Chang2012, Puchwein2012} and dark matter annihilation
\citep[][]{Cirelli2009,liu2020}.

He~{\sc ii} reionization is expected to be completed in the redshift range $2\le z\le 4$.
He~{\sc ii} reionization injects energy into the IGM and thereby broadens  the absorption features \citep{schaye2000}, 
decreases the neutral fraction of hydrogen \citep{rauch1997,bolton2007}  and leads to additional pressure  smoothing of the density and flux fields \citep{gnedin1998,peeples2010}. Accurate measurements of thermal parameters in this redshift range can therefore provide important constraints on the models of He~{\sc ii} reionization.

The thermal parameters of the IGM have been  measured  by (i)
decomposing absorption features  into  multi-component Voigt profiles and
identifying the lower envelope of the line width (b-parameter) vs \HI column
density \citep[\NHI; ][]{schaye1999,schaye2000,bolton2014,hiss2018,telikova2019},
(ii) measuring the small scale suppression in the transmitted flux power
spectrum \citep{theuns2000a,mcdonald2003,kim2004,walther2019,boera2019}, (iii)  using
curvature (defined as, $\kappa = F^{\prime\prime} /[1+(F^\prime)^2]^{3/2}$
where $F^\prime$ and $F^{\prime\prime}$ are the first and second derivatives of
the normalised flux with respect to the wavelength) statistics
\citep{becker2011,boera2014,hamsa2015} or (iv) characterising Fourier modes  of
the flux distribution  in the range sensitive to thermal parameters using
wavelet analysis \citep{zaldarriaga2002,theuns2002b,lidz2010,garzilli2012}. 
Even if the same observational data is used, the sensitivity to small changes
in thermal parameters and associated systematic uncertainties are found to be
different for different flux statistics (see \S \ref{sec:results} for a
    detailed discussion of this). Ideally, one would thus like to calculate
    all these different flux statistics simultaneously and consistently for the
    same data set to obtain  joint best fit values of the thermal parameters
    and the associated errors.  In order to obtain robust and consistent
    results  it is also important  to perform such an analysis using a
    consistent set of model parameters.

As simulated data are an integral part of the parameter estimation  from  \lya
forest data, the reliability and accuracy of the extracted parameters depend on
the assumptions involved in the simulations and the  ability to perform a wide
range of simulations that span a sufficiently wide parameter space with good
{\it i.e.} not too coarse sampling. Generating such a set of simulations
accounting for the  relevant physical effects is  challenging.  At any given
epoch the pressure smoothing of the density field depends on the past thermal
history \citep{gnedin1998,peeples2010,kulkarni2015,nasir2016,rorai2017} and one
needs self-consistent high-resolution cosmological hydrodynamical simulations
of the IGM \citep{springel2005,almgren2013,bolton2017}.  Practically one
achieves a range of $T_0$ and $\gamma$ in simulations by varying photo-heating
rates of \HI and \HeII as a function of redshift \citep[see for
example,][]{schaye2000,becker2011}.  Such simulations are computationally expensive which
limits the thermal parameter space that can be probed  at any given redshift.  

As noted before, most simulations used for this purpose are performed assuming
a uniform ionizing background and ionization equilibrium.  While these are
good approximations after \HeII reionization is complete such simulations do
not capture all relevant  physics before and during  \HeII reionization, i.e.,
\zrange{2.7}{4.0} \citep[see][]{puchwein2015,puchwein2019,gaikwad2019}.  For
this one  needs simulations that also  account for the non-equilibrium effects
during \HeII reionization.  Ideally one would also like simulations to
incorporate the spatially inhomogeneous nature of \HeII reionization and the
corresponding fluctuations of  the UV background. However, such
simulations will not only require to include radiative transfer, but also large
box sizes as well as high resolution and thus very high dynamic range to
capture the relevant physical processes accurately.

The main aim of this work is to obtain a consistent measurement of thermal
parameters using the larger  data sets that have become publicly available
recently. Thanks to compilations like \kodiaqdr \citep[][from KECK/HIRES
archival data]{omeara2015,omeara2017} and UVES \squaddr \citep[][from VLT/UVES
archival data]{murphy2019}, large samples of QSO spectra with high resolution
($\sim 6$ \kmps adequate to  resolve the thermally broadened \lya absorption
lines) and high \SNR are now available for analysis. These samples  have
dramatically increased the number of available QSO spectra  that have been
reduced and continuum normalised  using uniform techniques.

Additional impetus  to perform  a consistent measurement of thermal parameters
for these large samples comes from two tools we have developed over the past few
years:  (i) the \citefullform \citep[\citecode,][]{gaikwad2017a} and  (ii) the
\viperfullform \citep[{\sc viper},][]{gaikwad2017b}, an automatic Voigt profile
fitting code that decomposes  \lya\ absorption spectra into Voigt profile
components. {\citecode} not only allows us to construct models with a wide range of
thermal and ionization histories efficiently without running full
hydrodynamical simulations, but also enables us to calculate the non-equilibrium
evolution of the thermal and ionization state of the gas \citep[see][for
details]{gaikwad2019}. \citecode\ has been  shown to reproduce the results of
full hydro-simulations to well within 10 percent
accuracy \citep[][]{gaikwad2018}.  {\sc viper} runs on  parallel architectures
and allows us to perform a Voigt profile  analysis for  large samples  of
observed and simulated spectra consistently   \citep[see][for a clustering
analysis using Voigt profile components]{maitra2019,maitra2020}. Thanks to
these two codes, we are in a position to measure the physical parameters of the
IGM from  \lya\ forest data by simultaneously using the different
statistics mentioned above. 

We present here the measurements of thermal parameters ($T_0$ and
$\gamma$) over the redshift range \zrange{2}{4} in  10 redshift
bins of width $\Delta z = 0.2$ using  103 high resolution,  high
\SNR QSO spectra drawn from the \kodiaqdr sample. %
The paper is organized as follows, in \S \ref{sec:observation} we present the details of the
observational data used in this work and compare our new
measurements of the mean flux as a function of redshift, \HI\ column density
distribution, flux PDF and power-spectrum with measurements  in the literature. We
describe our simulations and explain how we generate simulated spectra for a
wide range of finely spaced thermal parameters in \S \ref{sec:simulation}. We
provide details of four different flux statistics of \lya forest used to measure
thermal parameters in \S \ref{sec:method} and in online supplementary 
appendix~\ref{app:statistics-sensitivity}.
%
In \S \ref{sec:results} we show our  measurements of thermal parameters,
present our error analysis and compare our measurements with measurements in
the literature. In \S \ref{sec:uvb-model-comparison} we show predictions for the
thermal parameter evolution for different UVB models. Finally we summarize in
\S \ref{sec:summary}.  Note that we make all the measurements from
observational data available to the community as online supplementary material.

The default cosmological parameters used here  are
$(\Omega_{\Lambda},\Omega_{m},\Omega_{b},\sigma_8,n_s, h,Y) =
(0.69,0.31,0.0486,0.83,0.96,0.674,0.24)$, consistent with a flat
$\Lambda$CDM cosmology \citep{planck2014}.  All distances are given in comoving
units unless specified.  We have expressed \GHI in units of $10^{-12} \: {\rm
s}^{-1}$ denoted by \GTW.  We often refer to $T_0,\gamma$ as thermal
parameters in this work.

\section{Observations}
\label{sec:observation}
\InputFig{{Redshift_Coverage_delta_z_0.2}.pdf}{75}%
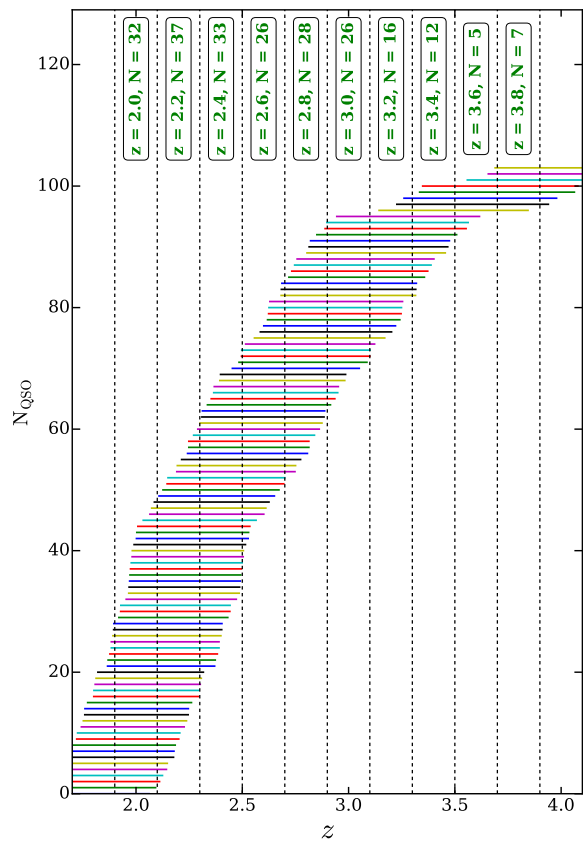
{Each horizontal line shows the \lya redshift range corresponding to the
    wavelength range between \lya and  \lyb emission for individual QSOs in our
    sample.  We exclude QSO proximity regions of $10$ pMpc from the QSO
    towards us.  The spectra are divided in ten redshift bins, each of size
    $\Delta =0.2$ centered on $z=2.0,2.2,\cdots,3.8$ as shown by the vertical
    dashed lines.  The number of lines of sight in a given redshift bin are shown in the
    boxes.
}{\label{fig:redshift-coverage}}

\begin{table}
\centering
\caption{Observed mean \lya\ transmitted flux}
\begin{tabular}{cccc}
\hline  \hline 
$z \: \pm \: dz$  & ${\rm N_{los}}$ & Median SNR & \Fmean $\pm$ \dFmean \\
\hline
$2.0 \: \pm \: 0.1$ & 32 & 14.7 & $0.8690 \: \pm \: 0.0214$ \\
$2.2 \: \pm \: 0.1$ & 37 & 21.2 & $0.8261 \: \pm \: 0.0206$ \\
$2.4 \: \pm \: 0.1$ & 33 & 19.2 & $0.7919 \: \pm \: 0.0210$ \\
$2.6 \: \pm \: 0.1$ & 26 & 19.3 & $0.7665 \: \pm \: 0.0216$ \\
$2.8 \: \pm \: 0.1$ & 28 & 19.7 & $0.7398 \: \pm \: 0.0212$ \\
$3.0 \: \pm \: 0.1$ & 26 & 23.4 & $0.7105 \: \pm \: 0.0213$ \\
$3.2 \: \pm \: 0.1$ & 16 & 22.0 & $0.6731 \: \pm \: 0.0223$ \\
$3.4 \: \pm \: 0.1$ & 12 & 22.4 & $0.5927 \: \pm \: 0.0247$ \\
$3.6 \: \pm \: 0.1$ & 5 & 22.8 & $0.5320 \: \pm \: 0.0280$ \\
$3.8 \: \pm \: 0.1$ & 7 & 22.3 & $0.4695 \: \pm \: 0.0278$ \\
\hline \hline
\end{tabular}
\\
\label{tab:observations}
\end{table}

We use observed spectra from the second data release of the  \kodiaqfullform
(\kodiaqdr) survey
\citep{omeara2015,omeara2017}\footnote{\url{https://www2.keck.hawaii.edu/koa/public/koa.php}}.
The sample consists of 300 QSO spectra  with  emission redshifts  $z\le 5.3$.
All available spectra are continuum normalised and the data product provides
normalised flux and the associated error as a function of wavelength.  Many of
these QSOs were observed more than once with different exposure times. We
co-added all the spectra using the procedure described in online supplementary
appendix \ref{app:spectra-coaddition}.  

In total there are 214 QSO spectra  that cover the \lya forest in the range
$1.9 \leq z \leq 4$.  We manually checked all the coadded spectra  and excluded
spectra if one or more of the following criteria are satisfied: 
 (i) sightline  does not (partially or fully) contain \lya forest in the
 redshift range $1.9 \leq z \leq 4$, 
(ii) sightline contains Damped \lya (DLA) or sub-DLA systems, 
(iii) the sightline contains large spectral gaps (see online supplementary
appendix \ref{app:spectra-coaddition} for details), 
or (iv) the median \SNR per pixel along the sightlines is smaller than $5$. 
After excluding the QSO spectra with above criteria, the resulting sample
consists of 103 QSO spectra.  \figref{fig:redshift-coverage} shows the \lya
redshift range corresponding to the wavelength range between the \lya and
\lyb emission lines of the 103 QSOs in our sample.  The \lya absorption
close to the QSOs is expected to be influenced by the enhanced ionizing
flux  due to  the QSOs
\citep{carswell1982,kulkarni1993,Srianand1996,lidz2007,calverley2011,bolton2012}. We
exclude such biased QSO proximity regions (i.e 10 pMpc from the QSO towards
us) in our subsequent analysis.  This choice of QSO proximity region size corresponds to the quasar rest frame wavelength of $\sim 1166$  \AA~at $z=3$. To avoid possible contamination from intrinsic O~{\sc iv} absorption we consider only regions with rest wavelength greater than 1050~\AA\ at the quasar emission redshift.

In order to measure the evolution of thermal parameters, we divide our sample
into ten redshift bins centered on $z=2.0,2.2,2.4,\cdots,3.6~{\rm and}~ 3.8$ with  a bin
width of $\Delta z = 0.2$.  \figref{fig:redshift-coverage} shows the  number of
QSO spectra contributing to our sample in each of these ten redshift bins.  The
properties of the observed spectra in these redshift bins are summarized in
\tabref{tab:observations}.   The median \SNR of the observed sample is $>10$ in
all  redshift bins.  The observed \lya forest regions are usually contaminated
by metal lines that produce  narrow absorption features and potentially could
bias our measurements of $T_0$ and $\gamma$.  To account for this we manually
identified contaminating metal line absorption using the list of known
intervening metal line systems along these sightlines.  It is not always
possible to identify all the metal lines contaminating the \lya forest.
However, as we fit all the observed spectra with our automated Voigt profile
fitting routine \viper \citep[][]{gaikwad2017b} we can mitigate this.  We treat
all lines  with $b \leq 8$ \kmps as metal lines.  Finally, we replace all
metal lines by continuum and add  noise to the replaced regions.  We have
checked the effect of residual metal line contamination on our  $T_0-\gamma$
measurements and found the effect to be marginal (see online supplementary
appendix \ref{app:error-analysis}). 

There could be additional errors in the observed flux due to continuum fitting
uncertainties. Continuum fitting uncertainties depend not only on the number of
unabsorbed spectral regions used in the fit but also on the observed \SNR per
pixel in these regions and the QSO spectral energy distribution (SED). Since
the true QSO continuum is unknown, an exact quantification of the contribution of
continuum fitting uncertainty to the error in the normalised flux is not
available for the \kodiaqdr sample. The continuum fitting uncertainty is
expected to be of the order of a few percent for moderate \SNR data
\citep{omeara2015}. We thus allow for the possibility of a  systematic error of $\pm5$ per cent in the normalised flux for our final $T_0$ and $\gamma$
measurements to account for continuum uncertainties.

\subsection{Comparison of observed statistics with previous measurements}
\label{subsec:stat-comparison}
\InputFig{FMEAN_Literature_Comparison_With_Errorbars.pdf}{75}%
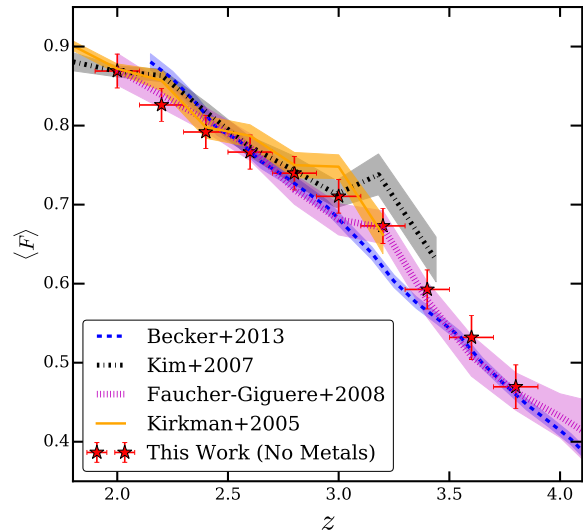
{ The redshift evolution of the observed mean flux from this work (red stars
    with errorbar) is compared with those from
    \citet[][]{kirkman2005,kim2007,faucher2008,becker2013b}. The shaded region
    show $1\sigma$ uncertainty on \Fmean for literature measurements. The \Fmean
    evolution in \citet{becker2013b} is obtained from a large number of moderate
    resolution and low SNR SDSS spectra,  while the \Fmean evolution in
    \citet{kirkman2005,kim2007,faucher2008} were obtained using high resolution
    QSO absorption spectra.  Our $\langle F \rangle$ measurements (based on \kodiaq DR2
    sample) are in agreement with \citet[][maximum deviation $\sim 2
    \sigma$]{kirkman2005,becker2013b}. However, there is a slight increase in
    \Fmean at \zrange{2.9}{3.3} compared to \citet{becker2013b}.  The
    presence or absence of this possible excess in \Fmean has only a marginal
    effect on our measured $T_0-\gamma$ but may be interesting
    in its own right as we will discuss later (see \S \ref{sec:observation}
    for details).
}{\label{fig:mean-flux-comparison-literature}}

\InputFigCombine{FPDF_FPS_CDDF_Literature_Comparison_Shaded_Errorbars.pdf}{175}%
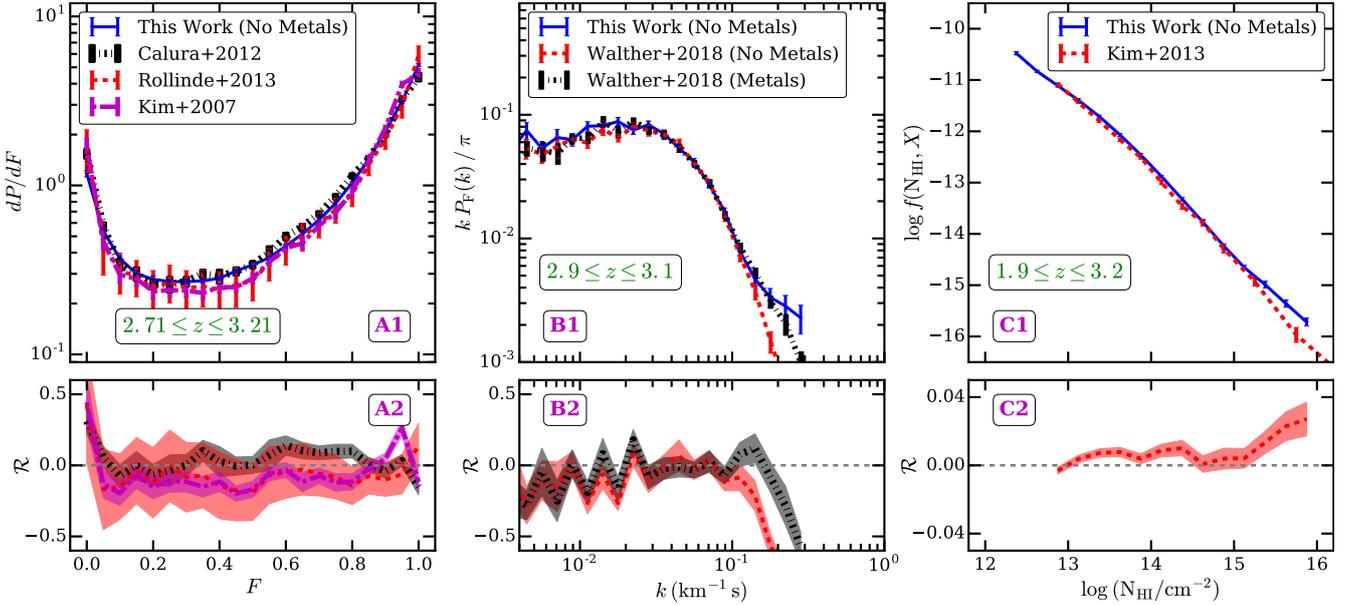
{ Panels A1, B1 and C1 show  a comparison of our measurements  (blue solid
curves) of the flux PDF/power spectrum (FPDF/FPS) and \HI\ column density
distribution (CDDF) with other measurements in the literature. Panels A2, B2
and C2 show the differences  between our measurements and those from the
literature.  The shaded regions show the $1\sigma$ uncertainty of the measurements.
For a fair comparison, the errors from our sample are also calculated using the
bootstrap method. In general the errors of the FPDF, FPS and CDDF statistics
from our sample are small thanks to the large sample size.  A comparison of
errors of the  observed flux statistics from this work with the literature is
discussed in detail in online supplementary appendix \ref{app:obs-flux-stat-error}.
\citet{kim2007,calura2012,rollinde2013} calculated the FPDF at $2.71 \leq z
\leq 3.21$ using 8, 2 and  5 high-resolution UVES spectra, respectively, while
our sample consists of $\sim 25$ high-resolution spectra. Panels A1 and A2 show
that the maximum difference between our FPDF (in the range $0.1 \leq F \leq
0.9$) and that of \citet{kim2007,calura2012,rollinde2013} is $\sim 13$ percent.
Panel B1 shows the comparison of our FPS with that from \citet{walther2018} at
$2.9 \leq z \leq 3.1$. Unlike \citet{walther2018}, we neither subtract noise
power from the FPS nor do we account for window effects in the FPS calculation.
Note that the effect of noise and finite spectral resolution play a
dominant role in the  FPS measurement at $k>0.15 \; {\rm km^{-1} \: s}$. 
Therefore, we use the FPS in the range \krange{0.01}{0.1} to
measure thermal parameters.  In this range our FPS is in good agreement
(maximum difference is 19 percent) with that from \citet{walther2018}.  Panels
C1 and C2 show that our CDDF is within 3.75 percent of that from
\citet{kim2013} at \zrange{1.9}{3.2}.  In panel C1, we define the CDDF similar
to \citet{kim2013}. However, for rest of this paper we define  the CDDF in
terms of \logNHI and $dz$.  
}{\label{fig:stat-comparison-literature}}

In this section,  we derive the statistics of the transmitted \lya flux for
our sample and compare them with results in the
literature.  In particular, we compare the evolution of mean flux, flux
probability distribution function (FPDF), flux power-spectrum (FPS) and \HI
column density distribution function (CDDF) from our sample with other
measurements in the literature. In \S \ref{sec:method}  we describe in detail
our method of calculating these flux statistics (along with other statistics)
from observations and simulations.

The last column in \tabref{tab:observations} provides the mean observed
transmitted flux measured in each redshift bin.  The error on the mean flux
given also accounts for the systematic uncertainty due to continuum placement.
As expected, the mean flux decreases monotonically with redshift due to the
increase in  the opacity of the IGM at higher redshifts. 
In \figref{fig:mean-flux-comparison-literature}, we compare the evolution of
the mean flux \Fmean of our sample (see \tabref{tab:observations}) with that
from \citet{kirkman2005,kim2007,faucher2008} and \citet{becker2013b}. The \Fmean evolution in \citet{kirkman2005,kim2007} and
\citet{faucher2008} is obtained using high-resolution QSO absorption spectra,
while  \citet{becker2013b} obtained their \Fmean evolution using very large
number of SDSS spectra  that have  lower resolution and low SNR compared  to
the typical high-resolution data used in the literature and our work here.

Our \Fmean evolution is broadly consistent with
\citet{kirkman2005,kim2007,faucher2008,becker2013b} with some notable
differences. \Fmean at $z=2.2, 3.0$ and $3.2$ in our sample is smaller by $\leq
2 \sigma$ than that of \citet{kirkman2005,kim2007}.
One possible reason could
be the number of QSO lines of sight used in their work, $\leq 10$ at those redshifts. Our \Fmean
measurement at \zrange{2.9}{3.5} (\zrange{2.1}{2.5}) is systematically larger
(smaller) than that of \citet{becker2013b}. The \Fmean evolution in our sample
is in good agreement with that from \citet{faucher2008} over the full redshift
range.  We see a slight enhancement in \Fmean at \zrange{2.9}{3.5} in the
\kodiaqdr sample (see online supplementary appendix \ref{appendix:excess-mean-flux} for more
details).  We have also analyzed the \squaddr QSO sample for the purpose of
calculating  the \Fmean evolution \citep[][]{murphy2019}. The \Fmean evolution from two samples are in good agreement (within $1\sigma$) with each other (except at $z=3.0,3.2$.
We find that the
enhancement in \Fmean at $z=3.2$ is less prominent in \squaddr, but  the \Fmean evolution
shows a change in slope at $z>3.2$.  It is noteworthy that the statistics
sensitive to thermal parameters are in good agreement with each other for the
\kodiaqdr and \squaddr samples.  Furthermore, note again that we rescale 
simulated optical depths to match the observed \Fmean which reduces the
effect of differences  in \Fmean on thermal parameters.  The presence or
absence of the slight excess of \Fmean discussed above has thus only a
marginal effect on measurements of $T_0-\gamma$ presented in this work.  

We compare the FPDF at \zrange{2.71}{3.21} from our sample with those from
\citet{kim2007,calura2012,rollinde2013} in
\figref{fig:stat-comparison-literature} (panel A1 and A2).  The number of
spectra used by \citet{calura2012,rollinde2013} and \citet{kim2007} is 2, 5
and 8 respectively,  while our sample contains $\sim 25$ spectra.
Despite this, our FPDF in the range $0.1 \leq F \leq 0.9$ is within $\sim 13$
percent of that from \citet{kim2007,calura2012,rollinde2013}. The FPDF
statistics appears reasonably well converged even for small number of spectra.
In this work, we focus on the FPDF in the flux range $0.1 \leq F \leq 0.9$
because the flux at $F < 0.1$ ($F>0.9$) could be dominated by sky subtraction
(continuum placement) uncertainty. 

In panels B1 and B2 of \figref{fig:stat-comparison-literature}, we show a
comparison of the FPS from our sample with that from \citet{walther2018}.  Even
though the sample used in \citet{walther2018} and our  work here is the same
(i.e, the \kodiaq DR2 sample), the number of QSOs per redshift bin is different
because of our selection criteria (see \S\ref{sec:observation}). Our method of
calculating the FPS is also different. Since our sample is selected such that
the spectra do not contain spectral gaps, we compute the power spectrum using
FFT. We also forward model the simulated \lya forest spectra  to mimic the
noise and  instrumental broadening properties of the observed spectra. Hence
when calculating the FPS, we neither subtract the noise power nor deconvolve
the instrumental broadening. We also replace metal lines with continuum and add
noise in the replaced regions, while \citet{walther2018} mask metal line
regions.  Despite these differences, when we compare our  FPS at
\krange{0.01}{0.1} with that from \citet{walther2018}, we find  good agreement.  Note that we use the FPS only  at
\krange{0.01}{0.1} to measure $T_0$ and $\gamma$, as these scales are most
sensitive to the variation of thermal parameters. 

In \figref{fig:stat-comparison-literature} (panel C1-C2),  we compare the CDDF
from \citet{kim2013} with that obtained from our sample using \viper.  For a
fair comparison, we restrict the observed spectra of our sample to the
redshift range \zrange{1.9}{3.2} and we use the same definition of the CDDF.
\citet{kim2013} present the CDDF for the range \logNHIrange{13}{18} since their
sample is incomplete at \logNHI $<13$. We account for the incompleteness in our
measurement by calculating the sensitivity curve hence our CDDF measurements
are shown from \logNHI $= 12.6$ to $15.8$.  Our observed CDDF is in good
agreement ($\sim 3.75$ percent) with that of \citet{kim2013}.  At \logNHI
$>15$,  we find more systems compared to \citet{kim2013}. This may be due to
the fact that \viper fits only \lya absorption while \citet{kim2013} fit \lya
and \lyb absorption simultaneously. Despite this,  the two CDDF in the range 
\logNHIrange{13}{15} are consistent with each other.  We refer the
reader to online supplementary appendix \ref{app:obs-flux-stat-error} for a discussion  of the
associated errors in these statistics for  our measurements and those in the
literature.

\section{Simulations}
\label{sec:simulation}
We evolve the cosmological density, velocity and temperature field using the
smooth particle hydrodynamic (SPH)
\gthree\footnote{\url{http://wwwmpa.mpa-garching.mpg.de/gadget/}} code.  The
initial conditions are generated at $z=99$ using \twolpt
\citep{2lpt2012}\footnote{\url{http://cosmo.nyu.edu/roman/2LPT/}}.  Our
fiducial simulations have a box size of $10 \: h^{-1}$ cMpc and have
$512^3$ baryon particles and an equal number of dark matter particles
corresponding to a gas particle mass of $\sim 10^5 \: {\rm M_{\odot}}$.  The
gravitational softening length is set to $\sim 0.65 \: h^{-1}$ ckpc.   We found
this to be the best compromise in terms of resolution, dynamic range and ability to
run a sufficiently fine grid of thermal parameters.  We have performed a
resolution study using the \sherwood simulation suite
\footnote{\url{https://www.nottingham.ac.uk/astronomy/sherwood/}} and show that
our  results are sufficiently well converged for our choice of simulation
\citep[see \S \ref{subsec:T0-gamma-recovery-sherwood}, ][]{bolton2017}. We also
use the Sherwood simulation suite to demonstrate that our measurements  are not
significantly affected by the somewhat limited box-size and thus  spatial
dynamical range.

\gthree is a modified version of the publicly available \gtwo code
\citep{springel2005} that incorporates the radiative heating and cooling by a
time varying, spatially uniform UV background.  For our fiducial simulations we
incorporate the \citetalias{khaire2019a} UVB model (QSO SED index
$\alpha=-1.4$) by modifying the TREECOOL file in \gthree.  We store the output
of the \gthree code at equal redshift intervals $z=6.0,5.9,\cdots,2.1,2.0$. We
employ a simplified star formation criteria that converts particles with
$\Delta > 1000$ and $T < 10^5$ into stars \citep[the so called quick \lya
option,][]{viel2004a} and do not include AGN or stellar feedback.

To generate physically motivated thermal histories one needs to perform 
computationally expensive cosmological simulations for a range of UVB models
\citep{becker2011,walther2019}.  In this work, we follow the approach laid out
by \citet{gaikwad2018} and explore the thermal parameter space efficiently. Our
procedure to simulate the thermal history is: (i) We perform a \gthree
simulation with the \citetalias{khaire2019a} UVB model using equilibrium
ionization evolution equation. The typical thermal parameters of such a
simulation  at $2.0 < z < 4.0$ are $T_0 \sim 9500$ K and $\gamma \sim 1.5$.
(ii) To obtain the variation in thermal parameters, we modify the photo-heating
rates in the UVB code (see online supplementary
appendix \ref{appendix:thermal-parameter-variation}).  We then solve the
ionization and thermal evolution equation for each particle on \gthree outputs
using our thermal evolution code \citefullform
\citep[\citecode,][]{gaikwad2017a} (iii) We apply pressure smoothing
corrections while extracting the $n_{\rm HI}, T$ and $v$ fields along a
sightline by convolving the SPH kernel with the pressure smoothing Gaussian
kernel \citep{gaikwad2018}.  (iv) Finally, we compute the \lya optical depth
accounting for thermal broadening, natural line width broadening and peculiar
velocity effects.  In \citet{gaikwad2017a,gaikwad2019}, we showed that thermal
parameters can be constrained within $1\sigma$ uncertainty provided the mean
flux is matched between simulation and observations.  However unlike
\citet{gaikwad2019}, in this work, we solve for the ionization evolution
assuming equilibrium while solving the thermal evolution using the
non-equilibrium equations.

We vary the thermal parameter evolution by scaling the \HI, \HeI and \HeII
photo-heating rates of the \citetalias{khaire2019a} UVB \citep[see][for a similar
approach]{becker2011}.  We generate $T_0-\gamma$ combinations for  a finely
sampled grid such that $T_0$ is varied from $\sim 6000$ K to $\sim 24000$ K  in
steps of 500 K while $\gamma$ is varied from $\sim 0.7$ to $\sim 2.0$ in steps
of 0.05 at $z=3$. The corresponding $T_0$ and $\gamma$ values at other
redshifts differ by $\leq 7$ percent. We have thus simulated the \lya forest
for $37 \times 27 = 999$ different thermal histories.  The computational time
required to run and extract \lya forest spectra for 999 UVB models is around
$\sim 2.25$ million cpu hours.  One can perform $\sim 50$ self-consistent L10N512 simulations in similar time. 
Note that the number of thermal parameters
probed in this work is larger by a factor of $\sim 13$ and $\sim 50$ than those
by \citet[][$N_{\rm thermal} = 76$]{walther2019} and \citet[][$N_{\rm
thermal} = 18$]{becker2011}, respectively. We refer the reader to 
online supplementary appendix
\ref{appendix:thermal-parameter-variation} for more details.  

We generate the overdensity ($\Delta$), neutral fraction (\fHI), temperature ($T$) and peculiar velocity $(v)$ fields along 20000 random skewers through our
simulation box at $z=1.9,2.0,2.1,\cdots,3.7,3.8~{\rm and}~3.9$. In each redshift bin we
concatenate the fields to match the observed redshift path length ($\Delta z =
0.2$). 
However, while computing the FPS, we do not
concatenate spectra as there is no correlation beyond the length of the
simulation box.  
The \lya optical depth is generated from the $\Delta$, \fHI, $T$ and
$v$ fields accounting for  Doppler broadening, natural line broadening and
peculiar velocity effects \citep{trc2001,hamsa2014}.  For each redshift bin, we
construct a single simulated mock sample with the same number of spectra as
observed in that redshift bin \citep[see][for a similar method at
$z<0.5$]{gaikwad2017a}.  The simulated spectra in a mock sample mimic the
observed spectra by 
(i) convolving the flux field with  Gaussian instrumental
broadening (FWHM $\sim 6$ \kmps) 
(iii) resampling the pixel distribution similar to the
observed data, 
and (iii) by adding Gaussian random noise
generated from the observed \SNR per pixel array.  We generate 100 such mock
samples to constitute a mock suite for each redshift bin.  For example, in the
redshift range \zrange{2.9}{3.1}, each mock sample consists of 26 simulated
spectra and a mock suite consists of $26 \times 100 = 2600$ simulated spectra.
We use these 100 mock samples to estimate the errors on different flux
statistics.  In order to compare the simulations with observations, we
calculate  the flux statistics that we will  show to be sensitive to thermal
parameters in the next section.


\section{Method}
\label{sec:method}
\InputFigCombine{Spectra_Example.pdf}{170}%
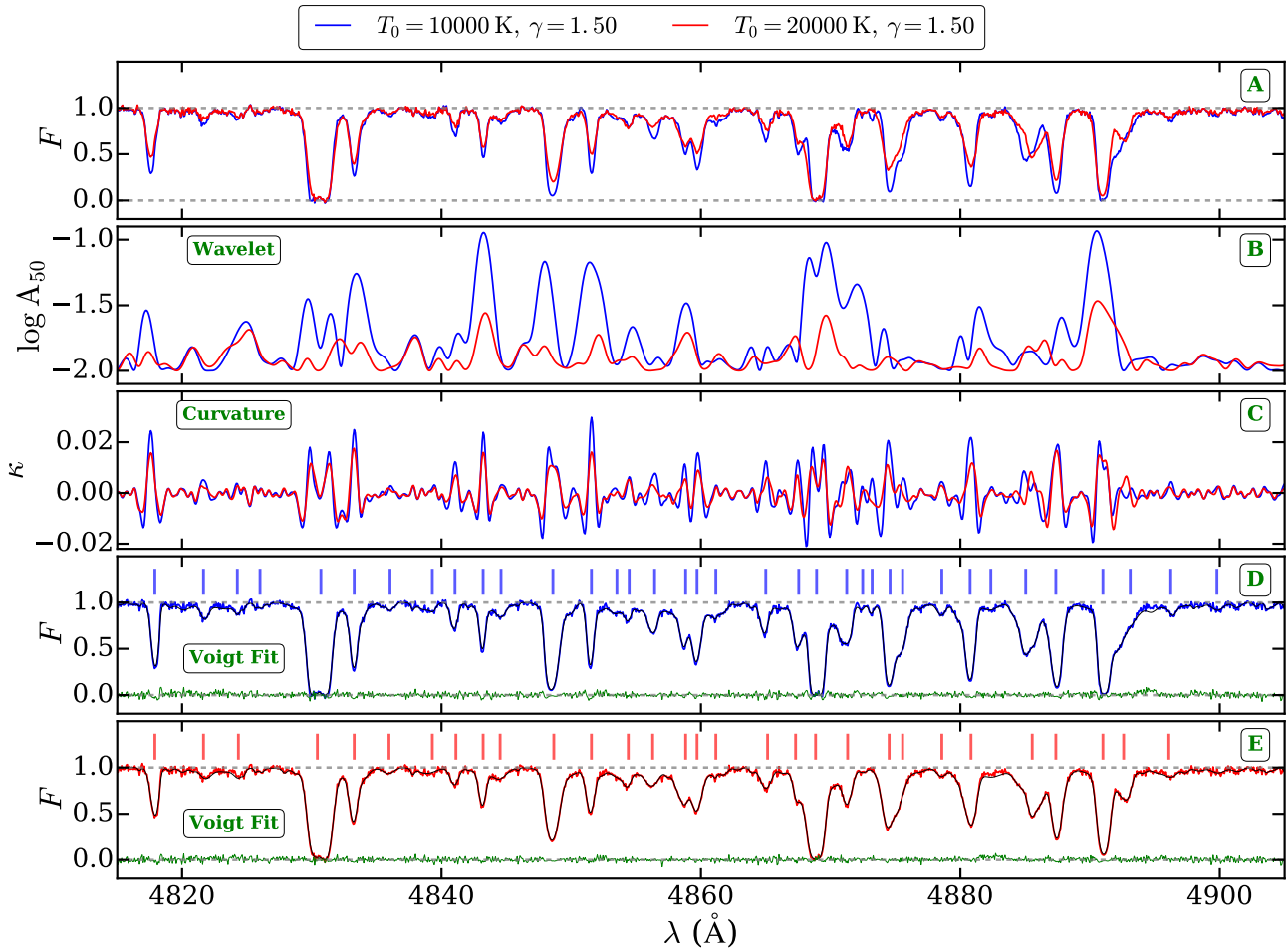
{Effects of changing the temperature field on the properties of the \lya
    forest.  Panel A shows the simulated \lya flux from a hot ($T_0 \sim 20000$
    K ) and a cold ($T_0 \sim 10000$ K) model accounting for instrumental
    broadening and finite \SNR. As expected, the absorption features  are
    systematically broader and shallower in the hot model than in the cold
    model. Panel B shows the wavelet amplitude from the two models obtained by
    convolving the \lya flux field from panel A with a wavelet of scale 50
    \kmps (see \S \ref{app:ws}).  Panel C shows the corresponding
    curvature amplitudes (see \S \ref{app:cs}).  The wavelet and
    curvature amplitudes are systematically smaller for the hot model as
    compared to those from the cold model.  Panels D and E show the Voigt
    profile decomposition of spectra for  the cold and hot model obtained using \viper respectively.  The
    Doppler parameter $b$ is systematically larger for the hot model  compared
    to the cold model (see \S \ref{app:bpdf}).  The black curve in panels D
    and E show the best-fit Voigt profiles.  The green curve near $F=0$ in
    these panels shows the residual between input and fitted flux. Vertical
    ticks in panel D and E show the locations of statistically significant Voigt
    components identified and fitted by \viper automatically.
}{\label{fig:spectra-example}}

The increase in temperature of the IGM due to \HeII reionization has the
following main effects on the  \HI \lya forest (i) broadening of the absorption
features \citep{schaye2000,lidz2010,becker2011}, (ii) decrease in neutral
fraction due to the temperature dependence of the recombination rate
\citep{rauch1997,bolton2007,becker2013,viel2016,khaire2019b}  and (iii) pressure smoothing
of the density and flux  fields
\citep{gnedin1998,peeples2010,kulkarni2015,lukic2015,rorai2017,maitra2019}. 
Since increasing the temperature of the IGM leads to smoothing of the flux, the
\lya forest statistics used in the past for measuring thermal parameters are based on either extracting small
scale features or measuring the broadening of the absorption lines. In what follows,
we describe the statistics used in this work to measure $T_0$ and $\gamma$.

\subsection{\lya forest statistics}
In the literature, $T_0$ and $\gamma$ have been measured by two different kind
of statistics, namely, those derived from the transmitted \lya flux directly and
those derived by fitting the transmitted \lya flux with multi-component Voigt
profiles. We use here four statistics, namely, the flux power spectrum (FPS), the
wavelet statistics (WS), the curvature statistics (CS) and the line width $(b)$
distribution (hereafter BPDF) of Voigt profile components to measure the
thermal parameters. 
%

\subsubsection{Flux power spectrum (FPS)}
\label{app:fps}
\InputFig{FPS_Simulation.pdf}{85}%
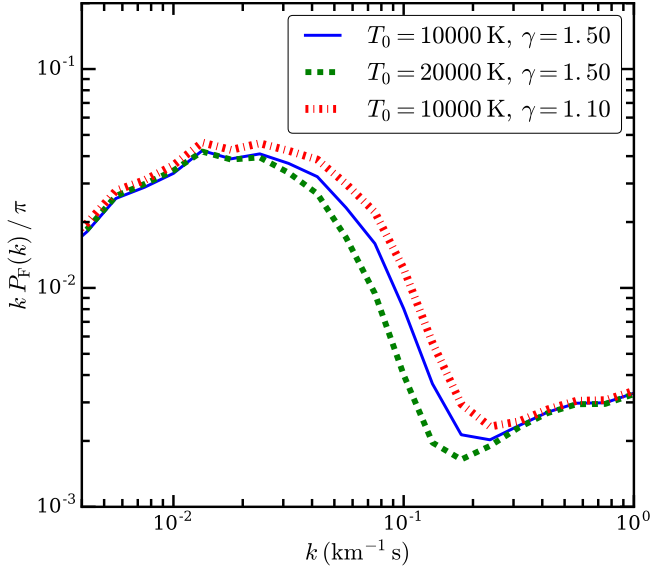
{The figure shows the sensitivity of the dimensionless FPS to $T_0$ and $\gamma$ for three
    models (i) $T_0=10000$ K, $\gamma=1.50$ (blue solid line), (ii)
    $T_0=20000$ K, $\gamma=1.50$ (green dashed line) and (iii) $T_0=10000$ K, $\gamma=1.10$
    (red dotted line) at \zrange{2.9}{3.1}.  The FPS is systematically lower 
    for models with higher $T_0$ at $0.02 \lesssim k$ (\spkm) $\lesssim 0.2$,
    while the FPS is systematically higher for models with smaller $\gamma$ at
    $0.02 \lesssim k$ (\spkm) $\lesssim 0.2$. The FPS at scales  $k \lesssim
    0.02$ is similar in all models because the underlying dark matter density
    field is the same in all the models, while the FPS at $0.2 \geq k$ is
    dominated by observational systematics such as noise and instrumental
    broadening.
}{\label{fig:T0-gamma-sensitivity-fps}}
The FPS is sensitive to a wide range of parameters: thermal parameters, \GHI,
cosmological parameters ($\Omega_b, \Omega_m, n_s$),   the free-streaming length of
dark matter particles and neutrinos etc
\citep{viel2004b,garzilli2015,irsic2017a,irsic2017b,gaikwad2017a,khaire2019b,gaikwad2020}.
For a given set of cosmological parameters in the $\Lambda$CDM framework, the
FPS is sensitive to small scale smoothing of the \lya flux arising from heating
of the IGM.  \figref{fig:spectra-example} (panel A) illustrates the effect of
increased temperature on the \lya flux. The flux in the higher-$T_0$ model (for
brevity, let us call it the hot model) is somewhat smoother and shows less
peaked absorption as compared to the lower-$T_0$ (the cold model) model. 

To calculate the FPS, we take the Fourier transform $\mathscr{F}$ of the flux
contrast  $\delta_f = (F / \langle F \rangle) - 1$ where $\langle F \rangle$ is
the mean flux in a given redshift bin. We bin the Fourier power
$|\mathscr{F}|^2$ in logarithmic  bins  centered on $k$ (\spkm) $=
0.005,\cdots,10.0$ and with bin width $\delta  \log k = 0.125$.  This probes
scales from $\sim 1250$ \kmps to $\sim 0.628$ \kmps.  The lower and upper
limits of $k$ are motivated by the size of our simulation box and the resolution of
the observed spectra, respectively. While computing the FPS, we do not
concatenate spectra as there is no correlation beyond the length of the
simulation box.  

\figref{fig:T0-gamma-sensitivity-fps} shows the sensitivity of the FPS to the
thermal parameters $T_0$ and $\gamma$. The power at \krange{0.01}{0.1} for a
model with higher $T_0$ (smaller $\gamma$) is smaller due to more smoothing of
the flux. The model predicted FPS at scales $k < 0.01$ \spkm  are
    nearly same as the dark matter density field  is the same for all 
models. On the other hand, the FPS at $k>0.3$ \spkm is dominated by observational systematic
effects such as instrumental broadening and finite \SNR per pixel. The FPS at
\krange{0.02}{0.3} (\vrange{20}{300}) is clearly sensitive to thermal
parameters.

\subsubsection{Wavelet Statistics (WS)}
\label{app:ws}
\InputFigCombine{WS_Simulation.pdf}{175}%
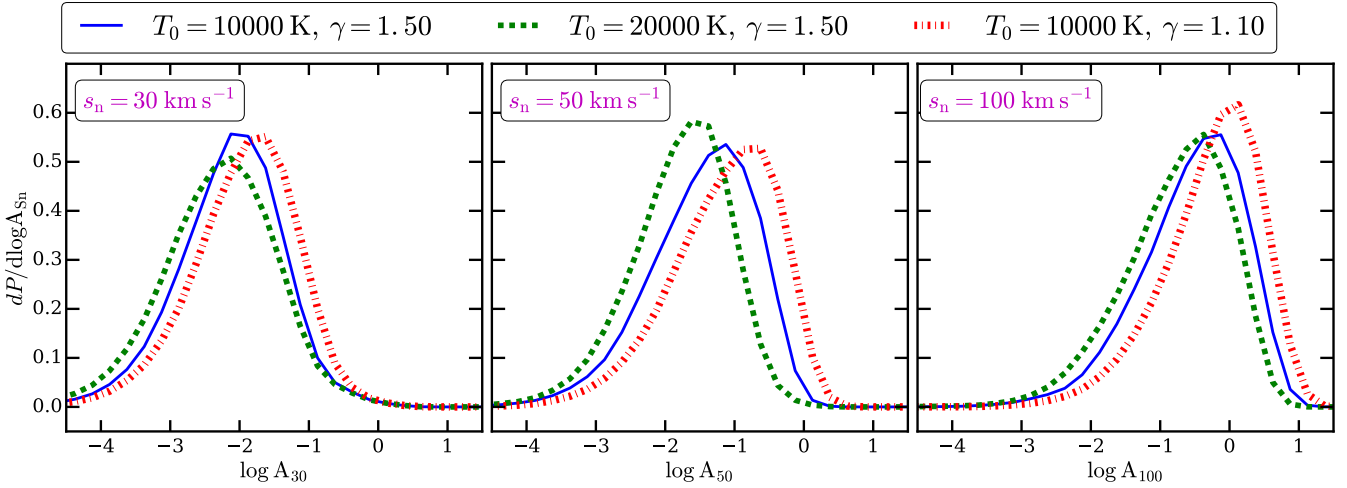
{The left, middle and right panels show the  sensitivity of the wavelet PDF to $T_0$ and
    $\gamma$ for a wavelet scale of $s_n = 30, \: 50$ and $100$ \kmps, 
    respectively (for \zrange{2.9}{3.1}). Each panel shows the wavelet PDF for
    three models (i) $T_0=10000$ K, $\gamma=1.50$ (blue solid line), (ii)
    $T_0=20000$ K, $\gamma=1.50$ (green dashed line) and (iii) $T_0=10000$ K, $\gamma=1.10$
    (red dotted line).  The wavelet PDF is systematically shifted to higher
    \logasn for models with smaller $T_0$,   while the wavelet PDF is
    systematically shifted to smaller \logasn for model with larger $\gamma$.
    The effect of $T_0$ and $\gamma$ variation is smaller at small wavelet
    scales, i.e., $s_n=30$ \kmps (panel A). The effect of $T_0$ on the wavelet PDF is
    larger at intermediate wavelet scale \snval{50},  whereas the effect of
    $\gamma$ is larger at higher wavelet scales \snval{100}.  We show in \figref{fig:T0-gamma-constraints-single-z-bin} that
    the simultaneous constraints using wavelet PDFs at scales $30 \leq s_{\rm n}$
    (\kmps) $\leq 100$ yield better constraints on $T_0$ and $\gamma$. All the
    wavelet PDFs are calculated using KDE (see online supplementary
    appendix \ref{app:kde}).
}{\label{fig:T0-gamma-sensitivity-wavelet-pdf}}
Wavelets are wave-like oscillations that are localized in both the real space
and frequency domains.  One can use a suitable wavelet to extract the scales in
transmitted \lya flux that are sensitive to $T_0$ and $\gamma$
\citep{zaldarriaga2002,theuns2002b,lidz2010,garzilli2012}.  Mathematically,
this is equivalent to a convolution of the transmitted \lya flux with the
wavelets.  Throughout this work, we use Morlet wavelets that look like a Gaussian
in Fourier space\footnote{We have also performed analysis with Mexican hat and
Shannon wavelets and found that the results are insensitive to the choice of
wavelet.}. The oscillation frequency of the wavelet depends on the center of
the Gaussian in Fourier space. While the width of the wavelet is related to
the width of the Gaussian in Fourier space.  The Morlet wavelets have the
functional form,
\begin{equation}
\begin{aligned}
    \Psi(v,s_{\rm n}) &= C \: \exp \bigg[ -\frac{2  \pi v i}{s_{\rm n}} \bigg] \:  \exp \bigg[ \frac{-v^2}{2 s^2_{\rm n}} \bigg] \\
\end{aligned}
\end{equation}
where $v$ is velocity, $s_{\rm n}$ is wavelet scale, $C$ is a normalization
constant obtained by demanding that wavelets are square integrable i.e., 
$\int \: |\Psi(v,s_{\rm n})|^2 \: dv = 1$.  The wavelet transform of the flux
(hereafter the wavelet field $A_{\rm Sn}$) is the convolution of the wavelet $\Psi(v,
s_{\rm n})$ with the transmitted \lya flux ($F$),
\begin{equation}
\begin{aligned}
    A_{S{\rm n}} &=  \int \limits_{v_{\rm min}}^{v_{\rm max}} \: dv^{\prime} \: F(v^{\prime})  \: \Psi(v-v^{\prime},s_{\rm n}) .
\end{aligned}
\end{equation}
\figref{fig:spectra-example} (panel B) demonstrates the sensitivity of the
wavelet amplitude (for the wavelet scale \snval{50}) to the temperature of the
IGM\footnote{Wavelet amplitudes can be negative hence we plot logarithm of
absolute values of wavelet amplitude.}. The wavelet amplitudes
\logasnbracket{50} in the hot model are systematically smaller than those from
the cold model. It is also evident that the wavelet amplitude picks out small
scale variations in the flux field.  To quantify the effect of $T_0$ and
$\gamma$ on the wavelet amplitude, we focus on the probability distribution
function of the wavelet field.  Usually, the wavelet field needs to be smoothed
\citep[e.g., using a boxcar filter of $\sim 1000$ \kmps, see][]{lidz2010} to
reduce the effect of flux noise on the wavelet PDF. However, we find that such
a boxcar smoothing reduces the sensitivity of the wavelet PDF to $T_0$ and
$\gamma$.  Instead we use  kernel density estimation to calculate the wavelet
PDF.  This  reduces the effect of noise while preserving the sensitivity of the
wavelet PDF to $T_0$ and $\gamma$ (see online supplementary
    appendix \ref{app:kde} for
details).  

\figref{fig:T0-gamma-sensitivity-wavelet-pdf} shows the comparison of the
wavelet PDF for 3 models
(i) $T_0=10000$ K, $\gamma=1.5$, (ii) $T_0=20000$ K, $\gamma=1.5$, and (iii) $T_0=10000$ K, $\gamma=1.1$.  The wavelet PDF
is systematically shifting to lower values for higher $T_0$ at all the three
wavelet scales \snval{30,50,100}.
\figref{fig:T0-gamma-sensitivity-wavelet-pdf} also shows that the wavelet scale
\snval{50} is most sensitive to variation in $T_0$. At smaller wavelet scales
\snval{30} it is less sensitive to $T_0$ due to observational effects
(noise and resolution) while larger wavelet scales \snval{100} trace the
underlying dark matter density field which is the same for the three models.  The
wavelet amplitudes are systematically lower for higher $\gamma$ values. This is
expected because for the same normalization $T_0$, a higher $\gamma$ value
corresponds to larger temperature at $\Delta>1$.  Since a significant fraction
of the \lya absorption at $z \sim 3$ comes from $\Delta > 1$, the \lya
absorption lines in models with higher $\gamma$ are expected to be broader. But
it is interesting to note that the wavelet scale \snval{100} is slightly more
sensitive to $\gamma$ than to  $T_0$. Thus, if we simultaneously use wavelet
PDFs at scales $30 \leq s_{\rm n}$ (\kmps) $\leq 100$, the constrains on $T_0$
and $\gamma$ should  be better than using the wavelet PDF for the individual
scales.

\subsubsection{Curvature Statistics}
\label{app:cs}
\InputFig{CS_Simulation.pdf}{83}%
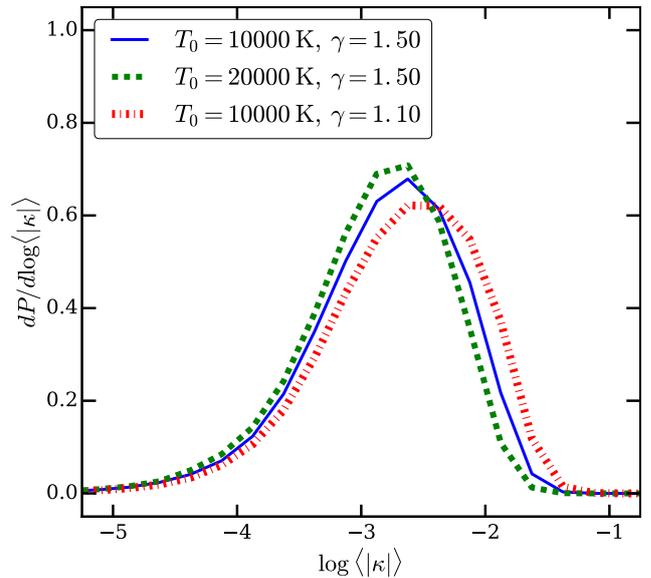
{The figure shows the sensitivity of the curvature PDF to $T_0$ and $\gamma$ at
    \zrange{2.9}{3.1}.  The curvature PDF is systematically shifted to higher
    (smaller) \logavgabskappa for smaller $T_0$ (larger $\gamma$). The
    curvature PDFs are calculated using KDE  (see online supplementary
    appendix \ref{app:kde}). The curvature PDFs show a similar behaviour as the 
    wavelet PDF except that it is less sensitive to thermal parameters than
    the wavelet PDF (at \snval{50}).
}{\label{fig:T0-gamma-sensitivity-curvature-pdf}}
The thermal broadening of the absorption lines can also be characterized by the
curvature of the absorption features \citep{becker2011}.  The curvature
($\kappa$) is essentially a measure of the rate of change of direction of a point
that moves on a curve and is defined as
\begin{equation}
\begin{aligned}
    \kappa &\equiv \frac{F''}{[1 + (F')^2]^{3/2}} \;\;,\\
\end{aligned}
\end{equation}
where $F'=dF/dv$ and $F''= d^2F/dv^2$ are the first and second derivatives of the
flux field (with respect to velocity), respectively.  Similar to wavelet
amplitudes, the curvature can also be affected by the finite \SNR of the spectra.
To circumvent this difficulty, \citet{becker2011} fitted a spline curve to the
transmitted \lya flux field and used that to compute the curvature. We,
however, use a simpler approach of smoothing the flux field by a Gaussian
filter of FWHM $\sim 10$ \kmps \citep[for a similar approach,
see][]{hamsa2015}.  
\figref{fig:spectra-example} (panel C) shows the comparison of the curvature
field for the hot and cold models.  Similar to the wavelet field, the curvature
amplitude is higher for the cold model as compared to the hot model.  In this
work, we use the curvature PDF to measure $T_0$ and $\gamma$.
\figref{fig:T0-gamma-sensitivity-curvature-pdf} shows the sensitivity of
the curvature PDF to $T_0$ and $\gamma$.  The curvature PDF is shifted to lower
values of \logavgabskappa for hot models as compared to cold models. The
curvature PDF is also sensitive to variations in $\gamma$.  However, note that 
the curvature PDF shows less sensitivity to $T_0$ and $\gamma$ than the 
wavelet PDF.


\subsubsection{The $b$-parameter  probability distribution function (BPDF)}
\label{app:bpdf}

\InputFigCombine{BNHI_Simulation.pdf}{175}%
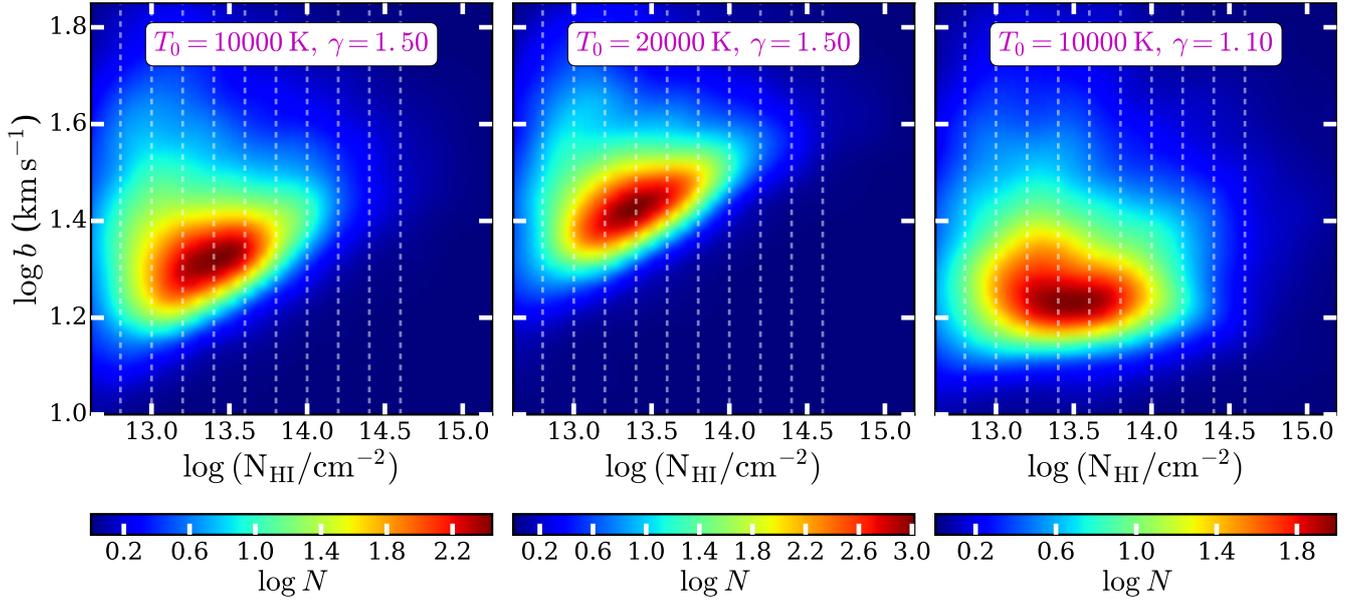
{Left, middle and right panels show the \blogNHI distribution for models (i)
    $T_0=10000$ K, $\gamma=1.50$, (ii) $T_0=20000$ K, $\gamma=1.50$ and (iii)
    $T_0=10000$ K, $\gamma=1.10$, respectively.  The \blogNHI distribution shows a
    systematic increase in $b$ values for the higher $T_0$ model while the slope 
    of the correlation remains approximately the same.  For higher 
    values of $\gamma$, the
    \blogNHI distribution becomes steeper.  The above distributions are
    calculated from $\sim 1300$ simulated spectra at \zrange{2.9}{3.1}.  Due to
    the limited number of observed spectra in any given redshift bins,  the
    \blogNHI plane is not sampled densely enough to be directly used for constraining the thermal parameters. 
    To circumvent this problem, we calculate 1D $b$ parameter
    probability distribution functions (BPDFs) in different \logNHI bins which 
    are individually sensitive to the thermal parameters (see
    \figref{fig:T0-gamma-sensitivity-BPDF}).  Vertical lines demarcate the \logNHI
    bins used in this work to compute 1D BPDFs. 
}{\label{fig:T0-gamma-sensitivity-BNHI}}

\InputFigCombine{BPDF_Simulation.pdf}{175}%
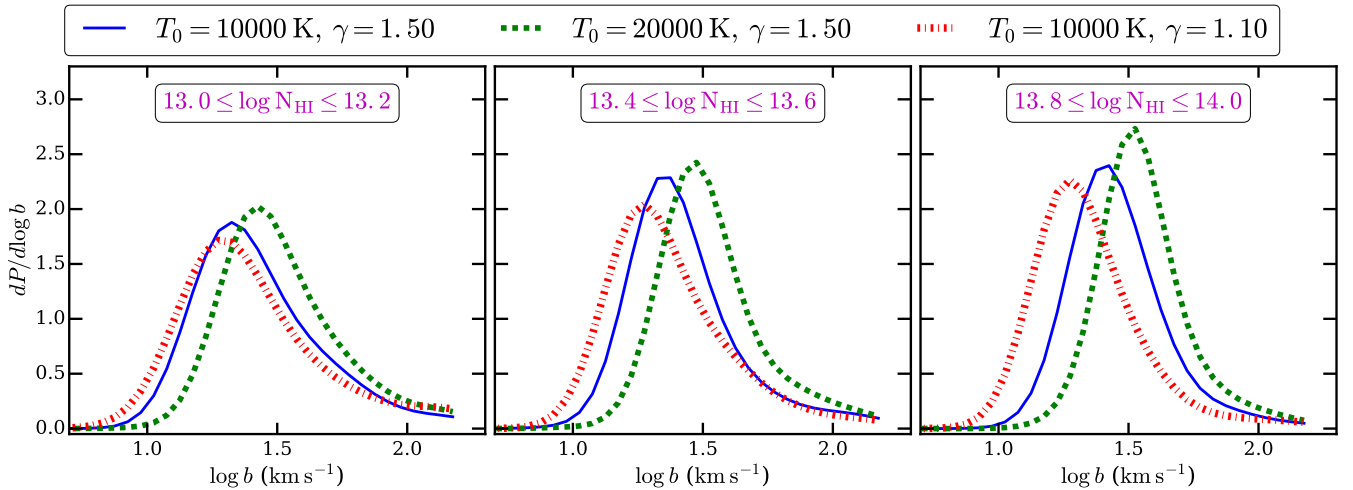
{Left, middle and right panels show the sensitivity of the  BPDF to $T_0$ and $\gamma$
    in \logNHI bins  [13.0,13.2], [13.4,13.6] and [13.8,14.0],  respectively.
    Each panel shows the BPDF for three models (i) $T_0=10000$ K, $\gamma=1.50$ (blue
    solid line), (ii) $T_0=20000$ K, $\gamma=1.50$ (green dashed line) and (iii)
    $T_0=10000$ K,$\gamma=1.10$ (red dotted line).  The $b$ values are systematically
    smaller for models with smaller $T_0$ and $\gamma$. The effect of $T_0$ and
    $\gamma$ variation is smaller for the  \logNHI bin [13.0,13.2]. The effect of
    $T_0$ on the BPDF is larger in all \logNHI bins, whereas the effect of $\gamma$
    is larger for higher \logNHI bins [13.8,14.0].  We show in \S
    \ref{sec:results} that the simultaneous constraints using the BPDF in different
    \logNHI bins yields better constraints on $T_0$ and $\gamma$.  All the
    BPDFs are calculated using KDE (see online supplementary appendix \ref{app:kde}).
}{\label{fig:T0-gamma-sensitivity-BPDF}}

Variations in the thermal parameters change the broadening of the \lya
absorption features. One can quantify this by
fitting a multi-component Voigt profile to \lya absorption spectra. The lower
envelope of the \blogNHI distribution has been used to measure the 
thermal parameters
\citep{schaye1999,bolton2014,gaikwad2017b,rorai2018,hiss2018,hiss2019,telikova2019}.
Fitting Voigt profiles to \lya forest spectra is challenging and somewhat
subjective.  Furthermore the complexity of \lya absorption increases with increasing
redshift as the opacity of the IGM increases.  We further need to fit a large
number of simulated spectra when measuring $T_0-\gamma$ model in each redshift
bin.  To facilitate this, we use our \viperfullform (\viper) that automatically
fits \lya forest spectra with multi-component Voigt profiles \citep[see][for
details]{gaikwad2017b}. The output of \viper consists of best fit values (along
with 1$\sigma$ uncertainty) of line center ($\lambda_{\rm c}$), H~{\sc i} column
density (\NHI), $b$ parameter and significance level of detection for each component. 
We used \viper to fit $N_{\rm los} \times 50$ (see
\tabref{tab:observations}) mock \lya forest spectra in each redshift bin. The
total time taken by \viper to fit all the spectra in all redshift bins is $\sim
1.5$ million cpu hours.  Panels D and E in \figref{fig:spectra-example} show
examples of the \viper fits (along with residuals) to the cold  and hot model,
respectively.

Neither the gas temperature nor the over-density are directly observable
quantities. However, the b-parameter of an absorption line is related to the
gas temperature while the measured column density(\NHI) is related to the
overdensity. The \blogNHI distribution can thus nevertheless be used to measure
the thermal parameters. 
The lower envelope in the \blogNHI distribution is shown to be sensitive to
thermal parameters \citep{webb1991,schaye2000,bolton2014,gaikwad2017b,rorai2018}.  
However, the lower envelope method utilizes fewer points that define the
lower cutoff in $b$ values at given \logNHI \citep[but see][]{hiss2019,telikova2019}.  Recently, \citet{hiss2019} have
proposed to use the full $b$-\logNHI 2D distribution to simultaneously measure
$T_0$ and $\gamma$. However, this requires to sample the \blogNHI plane densely
enough to reduce any statistical fluctuations introduced by binning.  This
becomes important when there are  few observed sightlines in a redshift bin.  

To circumvent this difficulty, we have used a modified approach to fit the
\blogNHI
2D distribution as shown in \figref{fig:T0-gamma-sensitivity-BNHI} and
\ref{fig:T0-gamma-sensitivity-BPDF}.  We divide the \blogNHI plane in different
\logNHI bins centered on $12.9,13.1, \cdots ~{\rm and}~14.5$ with bin width \dlogNHI$=0.2$. We
then calculate the 1D distribution of $b$ parameters in each \logNHI bin using
KDE.  

\figref{fig:T0-gamma-sensitivity-BPDF} shows the sensitivity of the $b$
distribution to $T_0$ and $\gamma$ in the three \logNHI bins
\logNHIrange{13.0}{13.2}, \logNHIrange{13.4}{13.6} and
\logNHIrange{13.8}{14.0}. The $b$ distribution shows a  systematic shift
towards higher $b$ values for higher $T_0$ models.  The  $b$ distribution is less
sensitive to $\gamma$ in the range\logNHIrange{13.0}{13.2},  but it  is more
sensitive in the range \logNHIrange{13.8}{14.0}.  Thus, the $b$ distribution at small
column density is  rather sensitive to $T_0$ but  less sensitive to $\gamma$.
This systematic variation in the $b$ distribution can change with redshift as 
the \lya forest is sensitive to different overdensities at different redshifts.
We simultaneously use the $b$ distribution in all the \logNHI bins to constrain the
thermal parameters.  Our method is similar to comparing the full 2D \blogNHI
distribution from simulations with observations  in the sense that we still use
all the data points in the \blogNHI distribution.  The only difference is that the size
of the \logNHI bin is larger than that used in the conventional 2D \blogNHI
distribution. This effectively reduces the statistical fluctuations due to
binning.


\subsection{Error estimation of \lya statistics}
\label{app:statistics-error-estimation}

The best fit values and the associated uncertainties on $T_0-\gamma$ crucially
depend on how accurately we can compute $\chi^2$ between data and model for a
given $T_0-\gamma$.  The accuracy of the $\chi^2$ estimation, in turn,
depends on (i) the estimated errors (covariance matrix) of the flux  statistic and
(ii) how finely we sample the $T_0-\gamma$ plane.  Since we use the
same method/code to derive the flux statistics from observations and
simulations,  numerical or computational systematics affecting  
the flux statistics should  cancel out in  the $\chi^2$ (and hence
$T_0-\gamma$) estimation.

We have considered two possible ways to calculate the error on each flux statistics: (i)
using bootstrap errors estimated from observed data and (ii) deriving the
flux statistics and computing the error from 100 simulated mock samples
\citep{rollinde2013}. The second method is not suitable in our case as these
errors depend on $T_0$ and $\gamma$. This is especially true for the flux statistics WS,
CS and BPDF because variation of  thermal parameters shifts the PDF to the right
or left i.e., along $x-$axis. We also estimate the covariance matrix using the 
bootstrap method. The bootstrap errors are found to be
slowly converging as the diagonal elements of the covariance matrix converge faster
than the off-diagonal ones. We find that the diagonal elements computed
from the bootstrap method are smaller by $\sim 25$ percent than those estimated
from the mock samples. This is expected as the bootstrap method usually tends
to underestimate the errors \citep{press1992}. To obtain the covariance matrix, we first inflate the boostrap error by $\sim 25$ percent. We then
calculate the correlation matrix from simulated mocks and rescale it using inflated
bootstrap errors. We
use this covariance matrix when calculating $\chi^2$.

Equally important for the $T_0-\gamma$ measurements and associated uncertainties is
the number of $T_0-\gamma$ grid points for which  $\chi^2$ is calculated (i.e., how finely we sample the parameter 
space).  In this work, $T_0$ is varied from $\sim 6000$ K to $\sim 24000$ K  in steps of $\delta T_0
\sim 500$ K while $\gamma$ is varied from $\sim 0.7$ to $\sim 2.0$ in steps of
$\delta \gamma = 0.05$ (at $z=3$). The \lya statistics are thus derived for $37
\times 27 = 999$ different $T_0-\gamma$ models.  However, we find that the sampling needs to be finer ($\delta T_0 \leq 200$ K
and $\delta \gamma \leq 0.02$) if we want the $1\sigma$
contours for $T_0$ and $\gamma$ to be converged. This is because the $T_0-\gamma$ uncertainty
from joint constraints can be smaller than that from individual
statistics. Due to limited computational resources, we populate the $\chi^2$
field by interpolating all the model statistics on to a finer
$T_0-\gamma$ grid with $\delta T_0 = 100$ K and $\delta \gamma = 0.01$. We
find that such a linear interpolation of statistics between different
$T_0-\gamma$ model is accurate to $\lesssim1.5$ percent.  Our
approach is similar to that in \citet{walther2019}, except that we use simple linear
interpolation instead of an emulator. Emulators are useful when the initial
$T_0$ and $\gamma$ grid is sparse. In our case, the initial $T_0-\gamma$ grid
is more densely sampled than previous works, hence the linear interpolation of
statistics between different $T_0-\gamma$ model is sufficient.

\subsection{Method of constraining thermal parameters}
\label{subsec:methodology}
In this work, we measure the thermal parameters by minimizing $\chi^2$
between the observed PDF/PS ($D$) and model PDF/PS ($M(T_0,\gamma)$),
\begin{equation}
\begin{aligned}
    \chi^2 = [D-M(T_0,\gamma)]\: C^{-1} \: [D-M(T_0,\gamma)]^{\rm T}  \;\;,
\end{aligned}
\end{equation}
where $C$ is the covariance matrix for each statistics is calculated as described in 
previous section. We compute $\chi^2$ for our $T_0-\gamma$ grid and find the best fit
model that corresponds to the minimum $\chi^2$ for each  
statistics. We then calculate the $1\sigma$
uncertainty on the parameters by marginalizing  over $\chi^2_{\rm min} \pm
\Delta \chi^2$ contours \citep[$\Delta \chi^2=3.50$ for 2 parameters + \Fmean
normalization,][]{avni1976}.  
For the wavelet statistics, the observed and model PDFs
depend on an additional parameter, the wavelet scale $s_n$. To perform
simultaneous measurements of $T_0$  and $\gamma$ using the  wavelet statistics,
we add the $\chi^2$ from different wavelet scales $s_{\rm n} = 30,40, \cdots,
90~{\rm and}~100$ \kmps i.e., $\chi^2_{\rm WS} = \chi^2_{30} + \chi^2_{40} 
+ \cdots + \chi^2_{90} + \chi^2_{100} $.
Similarly, the observed and model BPDF
depends on \logNHI bins. In this case, we add $\chi^2$ from different \logNHI
bins (\logNHIrange{12.8}{13.0}, $\cdots$, \logNHIrange{14.4}{14.6}).  When
performing the measurements from combined statistics, we add the $\chi^2$ from
all the statistics under consideration,
\begin{equation}
\begin{aligned}
    \chi_{\rm Joint}^2 = \chi_{\rm FPS}^2 + \chi_{\rm WS}^2 + \chi_{\rm CS}^2  + \chi_{\rm WS}^2 \;\; .
\end{aligned}
\end{equation}

In this work, we ignore the correlation among different statistics for
simplicity.  Hence the statistical uncertainty on $T_0-\gamma$ in joint
constraints may be somewhat underestimated.

\subsection{Validation of our approach using mock data}
\label{subsec:T0-gamma-recovery-sherwood}
\InputFigCombine{T0_gamma_Recovery_Sherwood_Multiple_z_bin.pdf}{170}%
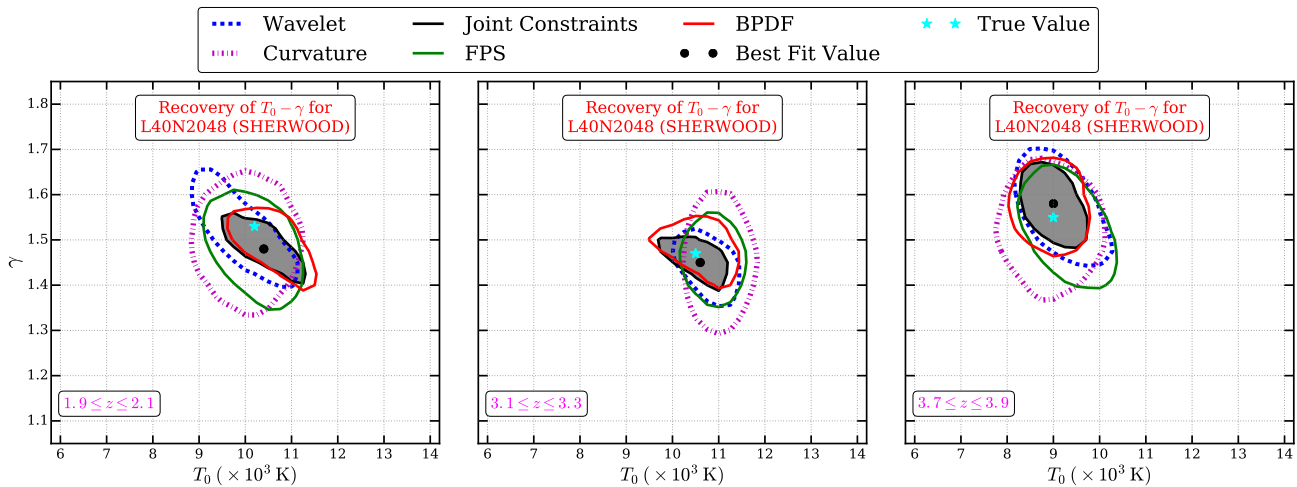
{ The left, middle and right panels show the recovery of thermal parameters
    using FPS, BPDF, wavelet and curvature statistics at $z=2,3.2~{\rm and}~3.8$
    respectively.  The $1\sigma$ joint constraints (using all statistics
    simultaneously) on the thermal parameters are shown by black contours and
    the  shaded regions. We here treat the \lya forest generated from the
    large dynamic range L40N2048 \sherwood simulation as mock data. We
    post-process our fiducial  L10N512 \gthree simulation  with 
    \citecode and generate model \lya forest spectra for 999 different thermal
    parameters by scaling the \citetalias{khaire2019a} photo-heating rates.
    The initial conditions of the L10N512 simulation used in this work are
    different from any of the ICs used in the \sherwood simulation suite. The
    mock and model \lya forest data  both mimic the observational properties
    at the respective redshifts. The true values (cyan stars) of the Sherwood
    simulation used to create the mock data sample lies 
    within the  gray shaded region at all three redshifts. 
    Thus our method successfully  recovers the thermal parameters
    despite the smaller spatial dynamic range of the simulations used 
    and the approximate nature of our modeling of the thermal
    history.
}{\label{fig:T0-gamma-recovery-sherwood}}
Before measuring the thermal parameters from observations, we demonstrate the
accuracy of our method in recovering the thermal parameters with an end-to-end
test on mock data created from  our fiducial hydro-simulation(s) and  from the
much larger dynamic range simulation from the \sherwood simulation suite (computed with cosmological parameters
similar to what we use in our models).  This
is with the aim to (i) test the sensitivity of statistics to thermal parameters,
(ii) study the degeneracy between thermal parameters, (iii) quantify the
accuracy of the method, (iv) check if there are any systematic effects between
true and recovered thermal parameters, (v)  check if the simulations are
sufficiently converged and (vi) test for the effect of Jeans smoothing.

The effect of box size and resolution on statistics of the \lya forest using a
range of \sherwood simulations are discussed in online supplementary appendix
\ref{app:resolution-study}.  As shown in \figref{fig:resolution-study-resolution}
to \ref{fig:resolution-study-box-size-high-z}, the
L10N512 simulation used in this work is sufficiently converged for the  corresponding
\sherwood model.  To test how well our models can recover the thermal
parameters, we generate \lya forest statistics in three  redshift bins
from  mock data based on the  L40N2048 simulations in the \sherwood suite which
has the same resolution and four times larger box size than  our fiducial 
hydro-simulation  The flux statistics  are generated from $\sim 20000$ skewers as
explained in \S \ref{sec:simulation}.  The mock data  and our model spectra
both closely  mimic  the properties of the observed sample e.g. number of
QSOs, \SNR, resolution and redshift path length in the corresponding redshift
bin.  The errors for each statistics are  calculated as discussed in \S
\ref{app:statistics-error-estimation}.  When  using FPS, BPDF, wavelet and
curvature PDF to characterise the thermal parameters we compute the $\chi^2$
between model and mock data (see \S \ref{app:statistics-error-estimation}
for details).

\figref{fig:T0-gamma-recovery-sherwood} shows the recovery of thermal
parameters using our models and the mock data generated from the L40N2048
\sherwood simulation. For all four  statistics, the true thermal parameters of
the mock data are within the  $1\sigma$ contours and are thus close to the best
fit measured thermal parameters. The true thermal parameters are also well
within the $1\sigma$ contours of the measured values if the four statistics are
combined (shown by the black contours and the  shaded region).  As expected the
joint analysis contours are narrower than those from individual statistics. The
best fit thermal parameters do not show any systematic deviation from the  true
thermal parameters indicating that any uncertainty in recovery is likely to be
statistical in nature.

\subsection{The effect of varying Jeans smoothing}
\label{subsec:Varying-Jeans-smoothing}
The width of absorption features depends on the instantaneous thermal state as
well as on the entire past thermal history due to the smoothing effect of the
thermal pressure  on the spatial distribution of the gas, an effect that has
been dubbed Jeans smoothing \citep{gnedin1998,peeples2010,kulkarni2015}. At the
redshifts considered here, well past \HI reionization and with the additional
heat input from \HeII reionization  ongoing, the memory of different possible
hydrogen reionization histories is expected to be modest. We have verified
this by repeating  our end-to-end test with mock data from an otherwise
identical simulation of the Sherwood suite which has the  same instantaneous
temperatures, but where hydrogen reionization occurs considerably later   (the
zr9 \sherwood simulation).  The difference in the measured thermal parameters
was hardly noticeable ($< 0.8$ percent).    The range of possible \HeII
reionization histories consistent with the \HeII opacity data  is also rather
small \citep{worseck2019}.  Hence the effect of the differences in Jeans
smoothing for the  same instantaneous temperature and  different \HeII
reionization  histories is again expected  to have a small effect on 
our measurements of thermal parameters. 

\subsection{The effect of spatial fluctuation of the temperature-density relation
due to inhomogeneous He~{\sc ii} reionization}
\label{subsec:spatial-TDR-fluctuations}

In reality, \HeII reionization will be spatially inhomogeneous and instead of a
well defined temperature density relation there will be a range of temperature
density  relations depending on when the \HeII in a particular region was
reionized \citep{rorai2017b,sanderbeck2020}.  In  \citet{gaikwad2020} we have shown that
for the equivalent situation during \HI reionization assuming a  homogeneous UVB
model  nevertheless recovers the median thermal state reasonably well.  We have
verified that this should also be the case for \HeII  reionization by producing
flux statistics for mock spectra obtained with a wide range  of temperature
density relations. As the box size of our simulations is much smaller than the
expected size of a region where \HeII is reionized simultaneously this should
mimic the effect of inhomogeneous \HeII reionization reasonably well.  The
thermal parameters measured from these samples of simulated mock spectra with a
range of temperature-density relations are indeed  within 1$\sigma$ of the
median  values. We are not sure if this can be interpreted as a systematic
bias, but we note that at and after the peak in temperature the inferred  $T_0$
values   were about 1$\sigma$ lower than the median while  at $z>3$  the
difference was hardly noticeable.

\section{Results}
\label{sec:results}

In this section, we present our  measurements of $T_0$ and  $\gamma$ from the
\kodiaq DR2 sample.  In order to make a fair comparison, we (i) derive the
statistics from simulations and observations in the same way, (ii) calculate
the error on each statistics from simulations and/or observations, (iii)
minimize the $\chi^2$ between data and model and (iv) find the best fit model
and the associated $1\sigma$ uncertainty on the parameters 
(see \S \ref{subsec:methodology})

\subsection{$T_0-\gamma$ constraints in the redshift range  \zrange{1.9}{3.9}}

\InputFig{T0_gamma_Constraints_Single_z_bin_Single_Panel.pdf}{83}%
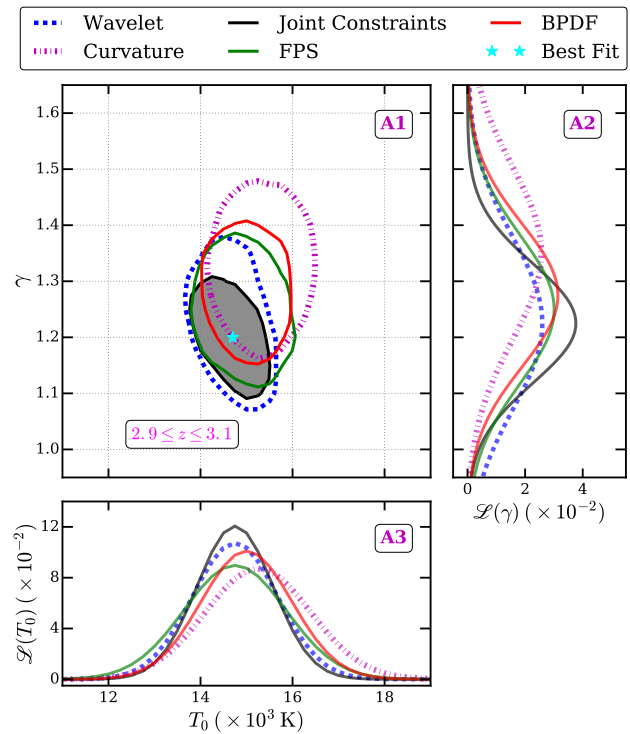
{Panel A1 shows $1\sigma$ constraints for $T_0$ and $\gamma$  from the wavelet
    PDF (blue dashed contour), curvature PDF (magenta dotted contour), BPDF
    (red solid contour) and FPS (green solid contour) statistics individually
    at \zrange{2.9}{3.1}. The joint constraints from all the four statistics is
    shown by the gray shaded region.  The constraints are obtained by
    minimizing the $\chi^2$ between data and model.  The wavelet constraints
    correspond to joint constraints from wavelet PDFs for wavelet scales
    $s_{\rm n} = 30, 40, \cdots, 100$ \kmps.  The BPDF constraints correspond
    to joint constraints from the BPDF calculated in 9 different \logNHI bins
    [13.0,13.2], [13.2,13.4], $\cdots$, [14.4,14.6].  
    The curvature PDF gives
    tighter constraints for $T_0$ as compared to $\gamma$. 
    The $1\sigma$
    contours from all the statistics are in good agreement with each other.
    Panels A2 and A3 show the marginalized $\gamma$ and $T_0$ distribution from
    all four statistics and our joint analysis.  
    The best
    fit $T_0$ and $\gamma$ values (cyan star in panel A) lie in the
overlapping region of  constraints from the the four flux statistics.  We show
the 1D reduced $\chi^2$ distribution in \figref{fig:T0-gamma-constraints-single-z-bin-chi-square} (also see \tabref{tab:chi-sq-dof}).
}{\label{fig:T0-gamma-constraints-single-z-bin}}

\InputFigCombine{{Observation_Simulation_Best_Fit_Model_With_Residuals_z-3.0}.pdf}{170}%
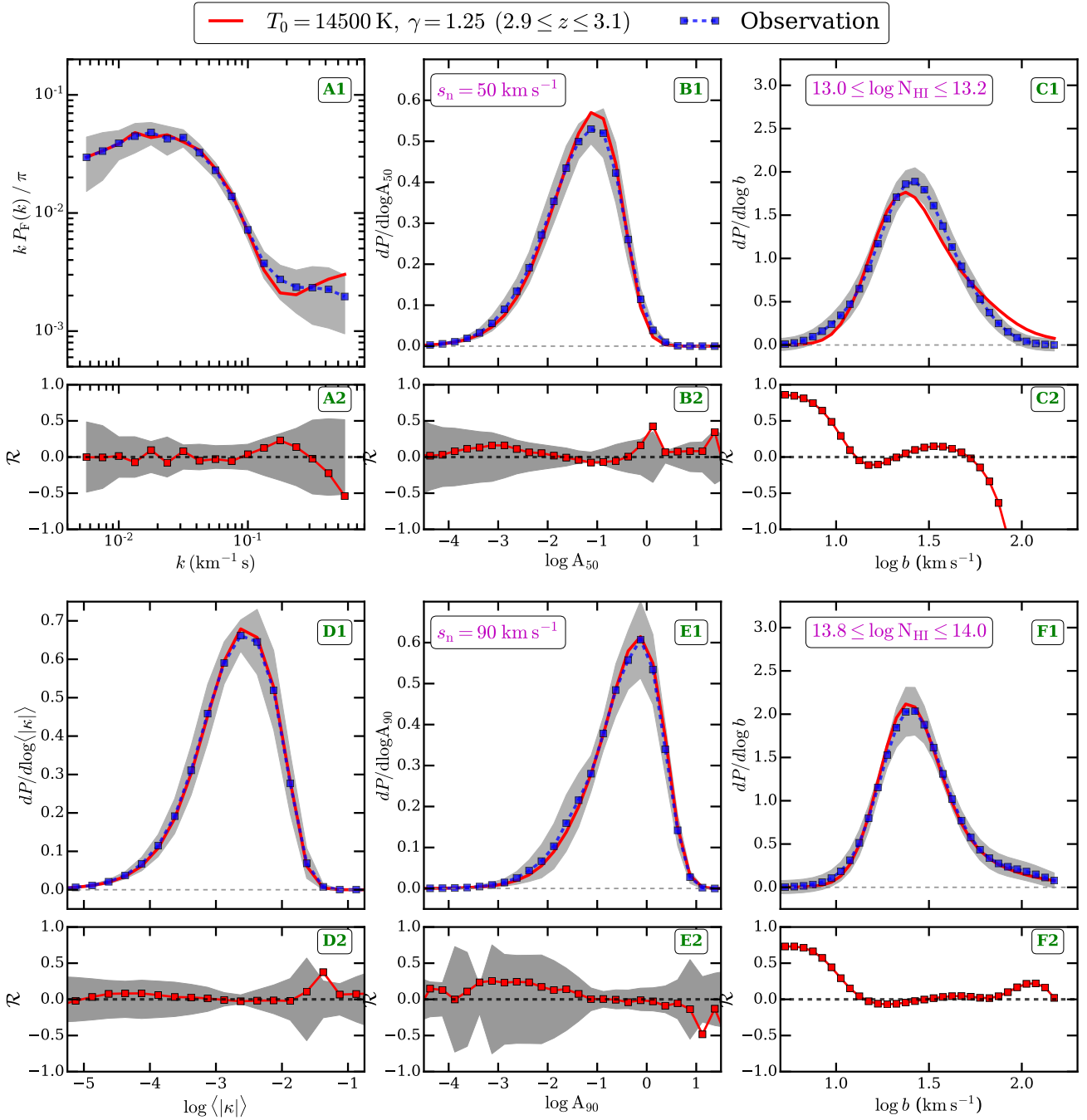
{Panel A1 shows a comparison of the FPS from the observations (blue squares
    with dashed line) with that of the best fit model [$T_0 = 14500 K $ and
    $\gamma = 1.25$] (red solid line) at \zrange{2.9}{3.1} generated by \citecode.  The  measurement errors are
    calculated from the observed data using the bootstrap method (gray shaded
    region).  Panel A2 shows the residuals between the observed distribution
    and the best fit model.  Panel B1, C1, D1, E1 and F1 are similar to panel
    A1 except the observations and predictions of the best fit
    models are shown for wavelet PDF \snvalbracket{50}, BPDF
    \logNHIrangebracket{13.4}{13.6}, curvature PDF, wavelet PDF
    \snvalbracket{70} and BPDF \logNHIrangebracket{13.8}{14.0}, respectively.
    Panel B2 to F2 are similar to panel A2 and show again the residuals between
    observations and best fit model.  For all the statistics, the best fit
    model is in good agreement ($\sim 1\sigma$) with observations.  Note that
    the best fit $T_0,\gamma$ given in \tabref{tab:T0-gamma-constraints} and
    \tabref{tab:T0-gamma-error-budget} corresponds to interpolated best fit
    values.  
}{\label{fig:T0-gamma-best-fit-model}}

\figref{fig:T0-gamma-constraints-single-z-bin} shows measurements of $T_0$ and
$\gamma$ in redshift bin \zrange{2.9}{3.1}. Panel A1 shows the measurements of
thermal parameters from individual as well as the combined statistics. 
The constraints from all  four
statistics agree with each other within  the $1 \sigma$ uncertainties.  The
constraints on $\gamma$ from the curvature statistics are poor compared to
those from the other statistics.  This is because the curvature PDF is less
sensitive to $\gamma$  compared to $T_0$. In previous measurements using
curvature statistics, $T_0$ is measured at a characteristic density
$\overline{\Delta}$ for an assumed value of $\gamma$.  
This does not allow for an independent measurement of $\gamma$.
It is also evident that
$T_0$ and $\gamma$ are anti-correlated for the wavelet statistics, BPDF and
FPS, while there is  little or no correlation between $T_0$ and $\gamma$ for
the curvature statistics.  Panels A2 and A3 show measurements of $\gamma$ and
$T_0$  marginalizing over $T_0$ and $\gamma$, respectively. The marginalized
distributions also show that the $T_0$ and $\gamma$ measurements using different statistics are consistent
with each other within $1\sigma$.  Note that the data and model \lya forest
spectra are the same in all cases only  the method/statistics used to measure
$T_0-\gamma$ is different.  In order to get tighter and more robust
measurements of $T_0$ and $\gamma$, we also plot the joint constraints for
$T_0$ and $\gamma$ from all the statistics in panels A1-A3 of
\figref{fig:T0-gamma-constraints-single-z-bin}. 
We calculate the joint constraints
by adding the $\chi^2$ between model and data for all the four statistics. For simplicity,
we ignore the correlation among different statistics. Hence the statistical 
uncertainty in joint constraints may be somewhat underestimated.
However, as shown in \figref{fig:T0-gamma-recovery-sherwood} 
when combining the statistics, the fiducial $T_0-\gamma$ is recovered 
well within $1\sigma$. This suggests that the statistical uncertainty for the
measurement from the combined statistics is realistic.
The joint constrains for $T_0$
and $\gamma$ are indeed tighter than the corresponding constraints from any
individual statistics.  The best fit value ($T_0 =14750$ K and $\gamma = 1.23$)
for the joint constraints from  all the four statistics  is shown as the cyan
star in panel A1 of \figref{fig:T0-gamma-constraints-single-z-bin}
(also see \figref{fig:T0-gamma-constraints-single-z-bin-chi-square}).

\figref{fig:T0-gamma-best-fit-model} shows the comparison of FPDF, BPDF,
wavelet PDF and curvature PDF from observations with the best fit model at
\zrange{2.9}{3.1}.  The corresponding residuals are shown in the lower panels.
The best fit models for all the statistics are in $1\sigma$ agreement with the
corresponding observations. The $\chi^2$ per degree of freedom for all the
statistics varies in the range  $0.7 \leq \chi^2_{\rm dof} \leq 1.35$ (see
\figref{fig:T0-gamma-constraints-single-z-bin-chi-square} and 
\tabref{tab:chi-sq-dof}).  The predictions of
our best fit model obtained from the joint analysis are thus in reasonably good
agreement with observations for all the statistics. 

\InputFigCombine{T0_gamma_Constraints_Joint_Analysis.pdf}{150}%
{Each panel is same as panel A1 in
    \figref{fig:T0-gamma-constraints-single-z-bin} except that the $1\sigma$
    contours for  $T_0$ and $\gamma$ from all the statistics are shown for different
    redshift bins.  The $1\sigma$ contours for   different
    statistics are in good agreement with each other at all redshifts. The
    joint constraints (the gray shaded regions) are more robust and tighter
    than the constraints from the individual statistics. The redshift evolution
    of $T_0$ and $\gamma$ is clearly evident from this figure. $T_0$ and
    $\gamma$ are anti-correlated  as expected. The degree
    of correlation decreases with increasing redshift as most of the \lya forest probes
    gas typically at cosmic mean density ($\Delta \sim 1$) at $z \sim 3.8$.
}{\label{fig:T0-gamma-constraints-joint}}
\InputFig{T0_gamma_Evolution_Statistics_Comparison_Single_Column.pdf}{85}
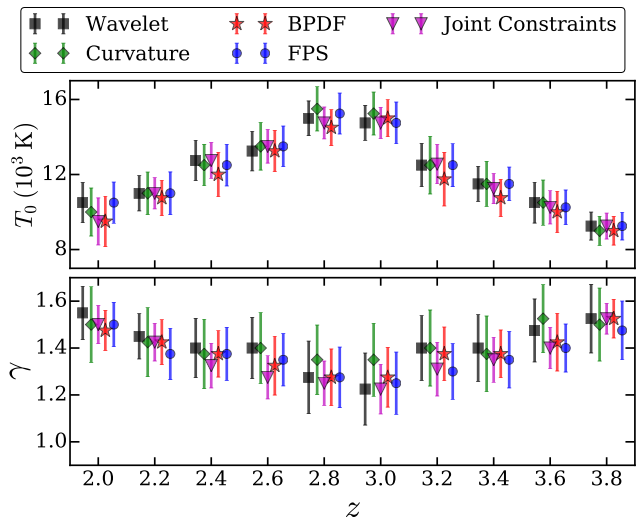
{The top and bottom panels show the evolution of $T_0$ and $\gamma$,
    respectively, obtained using five methods (i) Wavelet PDF (black squares),
    (ii) Curvature PDF (green diamonds), (iii) BPDF (red stars), (iv) FPS (blue
    circles) and (v) joint constraints using all the four statistics (magenta
    triangles). The errorbars show the $1\sigma$ statistical uncertainties on
    $T_0$ and $\gamma$ for all the 5 methods. All  $T_0$ and $\gamma$
    measurements  are consistent with each other in all redshift bins.  Note,
    however, that the errors for $T_0$ and $\gamma$  are different for
    different methods due to different sensitivities of these methods to
    variations of  these parameters.  The joint constraints minimize the
    scatter of the best fit values and the resultant $1\sigma$ uncertainty on
    $T_0$ and $\gamma$ is smaller than that for  the corresponding individual
    statistics.
}{\label{fig:T0-gamma-evolution-all-no-uvb}}
\figref{fig:T0-gamma-constraints-joint} shows measurements of $T_0$ and
$\gamma$ in the other redshift bins spanning \zrange{1.9}{3.9}. In all the
redshift bins, the $T_0$ and $\gamma$ measurements from  the individual
statistics are in good agreement with each other.  The $\gamma$ measurements
from the curvature statistics are less tight than those from the other statistics.
The uncertainty in $T_0-\gamma$ using the joint constraints is significantly
smaller than that from the individual statistics at all redshifts. 
The joint
$T_0-\gamma$ constraints are thereby in good agreement with the overlapping
regions of all the individual statistics.  
The variation of the  $T_0$ and
$\gamma$ measurements with redshift is also evident from
\figref{fig:T0-gamma-constraints-joint}.  As expected the $T_0$ and $\gamma$
constraints are anti-correlated in the lower redshift bins where the absorption
features probe densities above the mean. With increase in redshift the $T_0$
and $\gamma$ correlation becomes weaker, similar to the finding of
\citet{walther2019}.  This is because at $z \geq 3.5$ the \lya forest is mostly
sensitive to  approximately mean cosmic density $\Delta \sim 1$.

In \figref{fig:T0-gamma-evolution-all-no-uvb}, we compare the evolution of
$T_0$ and $\gamma$ from the individual statistics and joint analysis.
The $T_0$ ($\gamma$) measurements are marginalized over $\gamma$ ($T_0$). We
can clearly see an evolution of $T_0$ and $\gamma$ with redshift with a clearly
identifiable peak in $T_0$ that coincides with a minimum in $\gamma$ at $z\sim
3$. The $T_0$ and $\gamma$ measurements from each individual statistics as well as our joint constraints are
consistent within $1\sigma$ in all the redshift bins. 
As expected, the
$T_0-\gamma$ uncertainty for the  joint analysis is smaller than the
corresponding uncertainty from the individual flux statistics.  The best fit
$T_0$ and $\gamma$ values  show some scatter between neighboring redshift bins for
the individual statistics  due to the differences in sensitivity of individual
statistics to $T_0$ and $\gamma$ .  The scatter of the  best fit values is considerably smaller for the joint measurements.


\subsection{$T_0-\gamma$ Error Budget}
\InputFig{T0_gamma_Constraints_Continuum_Metal_Effect_Contour.pdf}{72}
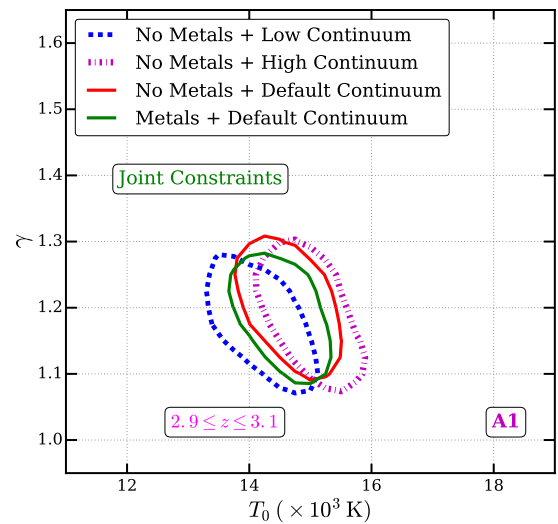
{The effect of continuum placement uncertainty and metal
    contamination on $T_0$ and $\gamma$  constraints from the joint
    analysis of data for \zrange{2.9}{3.1}.  Comparison of the green contours (with
    metals) with the red contour (without metals) shows that $T_0$ and $\gamma$
    are underpredicted by $\sim 2$ percent if  metal contamination is  not
    accounted for.  The effect of continuum placement uncertainty (assumed to
    be $\pm 5$ percent) biases the measurements  such that a low continuum
    predicts lower $T_0-\gamma$ measurements by 3\% or less.  We have accounted
    for both continuum placement and metal
    contamination uncertainties in our final measurements (see \tabref{tab:T0-gamma-constraints},  \tabref{tab:T0-gamma-error-budget} and online supplementary appendix
    \ref{app:error-analysis} for details).
}{\label{fig:T0-gamma-constraints-continuum-metal-single-zbin}}


\begin{table}
\centering
\caption{Measurements of $T_0$, $\gamma$ with total uncertainty (see 
\tabref{tab:T0-gamma-error-budget} for error budget).}
\begin{tabular}{ccc}
\hline  \hline 
$z \: \pm \: dz$  & $T_0 \pm \delta T_0$ (K) & $\gamma \pm \delta \gamma$ \\
\hline
2.0 $\pm$ 0.1 &  9500 $\pm$ 1393 & 1.500 $\pm$ 0.096 \\
2.2 $\pm$ 0.1 & 11000 $\pm$ 1028 & 1.425 $\pm$ 0.133 \\
2.4 $\pm$ 0.1 & 12750 $\pm$ 1132 & 1.325 $\pm$ 0.122 \\
2.6 $\pm$ 0.1 & 13500 $\pm$ 1390 & 1.275 $\pm$ 0.122 \\
2.8 $\pm$ 0.1 & 14750 $\pm$ 1341 & 1.250 $\pm$ 0.109 \\
3.0 $\pm$ 0.1 & 14750 $\pm$ 1322 & 1.225 $\pm$ 0.120 \\
3.2 $\pm$ 0.1 & 12750 $\pm$ 1493 & 1.275 $\pm$ 0.129 \\
3.4 $\pm$ 0.1 & 11250 $\pm$ 1125 & 1.350 $\pm$ 0.108 \\
3.6 $\pm$ 0.1 & 10250 $\pm$ 1070 & 1.400 $\pm$ 0.101 \\
3.8 $\pm$ 0.1 &  9250 $\pm$  876 & 1.525 $\pm$ 0.140 \\
\hline \hline
\end{tabular}
\\
\label{tab:T0-gamma-constraints}
\end{table}

We consider the following important 
uncertainties when measuring $T_0$ and $\gamma$:
(i) modeling uncertainty, (ii)
continuum placement uncertainty, (iii) uncertainty due to metal contamination
and (iv) cosmological parameter uncertainty.
\figref{fig:T0-gamma-constraints-continuum-metal-single-zbin} shows the effect
of continuum placement uncertainty and metal contamination uncertainty on our
measurements of  $T_0$ and $\gamma$ from our joint analysis of the four
different flux statistics (for $2.9\le z\le 3.1)$.  The uncertainty in $T_0$ and $\gamma$ due to
continuum placement uncertainty 
\citep[assumed to be $\pm 5$ percent, see][]{omeara2017} is less than 3 percent.  The continuum placement
uncertainty mainly affects  the mean flux of the observed sample. However, when
measuring $T_0$ and $\gamma$, we rescale the optical depth in the simulations
to match the observed mean flux. Contamination by narrow metal absorption lines in the observed \lya
forest 
can also  potentially bias our $T_0$ and $\gamma$ measurements.  We thus remove
the metal lines from the \lya forest to minimize their effect.  However, it is
still possible that some metal lines are not identified due to insufficient
wavelength coverage in the higher wavelength
side of the \lya emission 
and/or
blending effects.  By applying a cutoff in the fitted $b$ parameters, we show
in \figref{fig:T0-gamma-constraints-continuum-metal-single-zbin} that the
contribution of such metal lines to the $T_0-\gamma$ uncertainty is not larger
than $\sim 2$ percent.  Similarly, we find that the uncertainty in cosmological
parameters can lead to 0.5 percent uncertainty in $T_0$ and $\gamma$.  One of
the main uncertainties in the  $T_0$ and $\gamma$ measurements comes from our
modeling of the \lya forest. Since we vary the thermal state of the gas by
post-processing of \gthree simulations, the dynamical impact of pressure
smoothing may not have been captured accurately.  We find that the effect of
pressure smoothing on the $T_0$ and $\gamma$ uncertainty is below $2.5$
percent.  We refer the reader to online supplementary appendix \ref{app:error-analysis} for a
detailed discussion of the  effect of all the above uncertainties on our
$T_0-\gamma$ measurements.

\tabref{tab:T0-gamma-constraints} provides our $T_0$ and $\gamma$ measurements at different redshifts. In \tabref{tab:T0-gamma-error-budget} we give   $T_0$ and $\gamma$ errors contributed by various above listed uncertainties.
The
continuum placement uncertainty is systematic, while the other uncertainties
are statistical in nature.  We add the statistical uncertainties in quadrature
while the systematic uncertainty is additive in nature.  In the rest of the
paper, we show and discuss constraints on $T_0$ and $\gamma$ with total
uncertainties as given in Table~\ref{tab:T0-gamma-constraints}.

\subsection{Evolution of $T_0$ and $\gamma$}
\label{subsec:T0-gamma-evolution}
\InputFigCombine{T0_gamma_Evolution_Literature_Comparison_Separate_Statistics}{160}
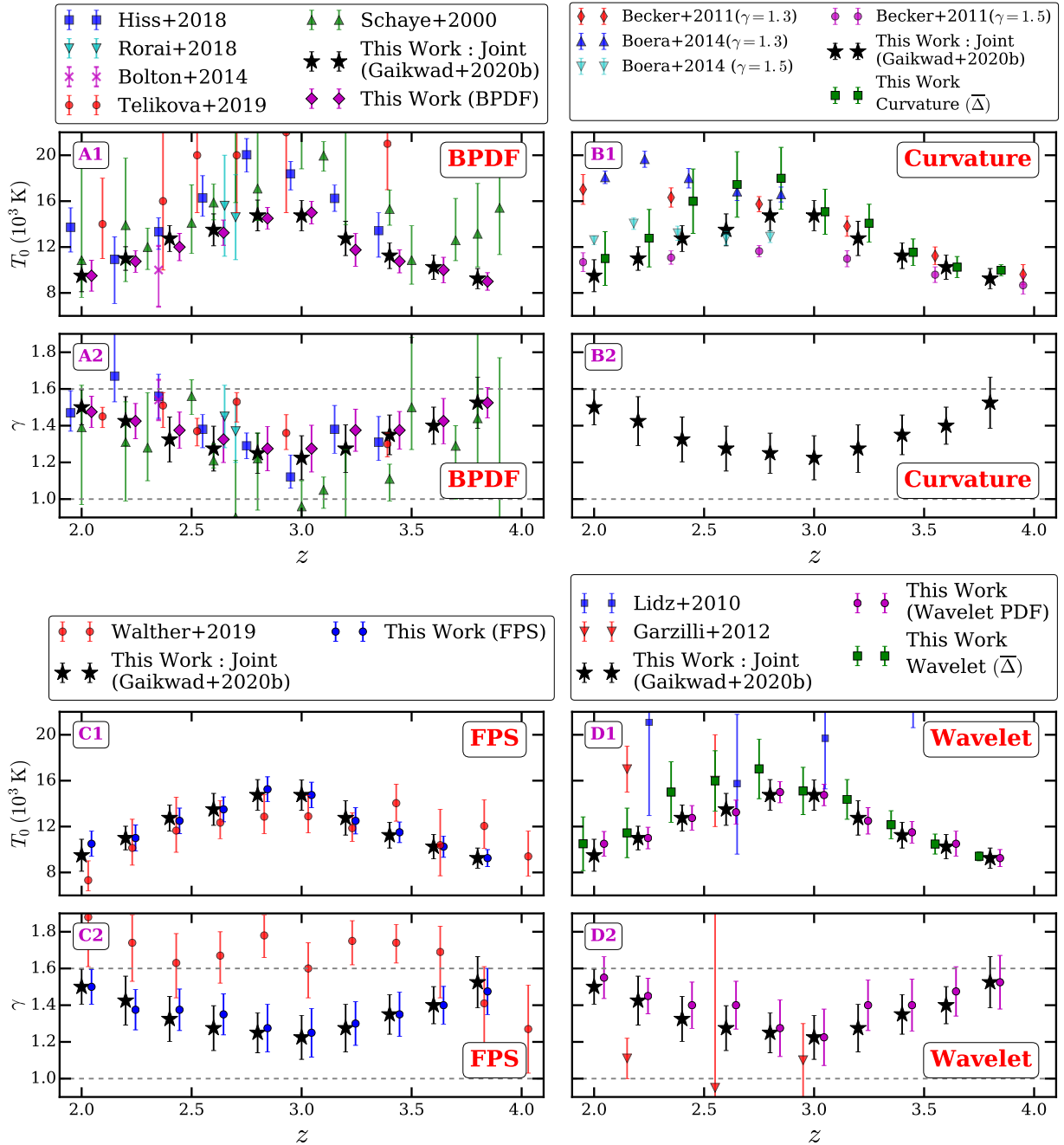
{The figure shows a comparison of our $T_0-\gamma$ measurements with
    measurements from the literature for the four flux statistics. The BPDF,
    Curvature, FPS and Wavelet statistics are shown in panel A1-A2, B1-B2,
    C1-C2 and D1-D2, respectively.  The joint constraints on thermal parameters
    from this work are shown by the black stars in each panel. For a fair
    comparison, we also show the measurements of thermal parameters obtained
    using individual statistics (i.e., same as
    \figref{fig:T0-gamma-evolution-all-no-uvb}, showing only statistical
uncertainty). The  uncertainty of our measurements for individual statistics
appears to be consistent with being statistical in nature and is slightly
bigger than that from the joint analysis.  For the BPDF statistics, our $T_0$
measurements are typically smaller than that obtained from the literature
(panel A1) while the $\gamma$ measurements  are similar to that  in the
literature.  For the curvature statistics we show the  $T_0$ measurements
obtained using the characteristic overdensity method \citep[\oDelta see \S
\ref{subsubsec:curvature-literature},][]{becker2011}. We use our $\gamma$
measurements obtained from the joint analysis as a prior to calculate the $T_0$
uncertainty for the \oDelta method in panel B1. The $T_0$ measurements using
the \oDelta method are systematically larger than that from the PDF method.
However, the $T_0$ in both measurements are in good agreement (maximum
difference $\sim 1.1\sigma$) with each other.  Our $T_0$ measurements at $z
\geq 2.6$ using the \oDelta method are within $1\sigma$ of  that from
\citet{becker2011,boera2014}. Panel C1 and C2 shows that the $T_0$ ($\gamma$)
measurements for the FPS statistics are systematically larger (smaller) than
those from \citet{walther2019}.  The systematic differences in $T_0$ and
$\gamma$ are likely due to constraints on pressure smoothing in \citet[][see \S
\ref{subsubsec:fps-literature}]{walther2019}.  However, our $T_0$ measurement
using the FPS statistics is within  $1 \sigma$ of  the corresponding
measurement by  \citet{walther2019}. Panel D1 and D2 show the comparison of
thermal parameters from this work with that from \citet{lidz2010,garzilli2012}
using wavelet statistics. The improvement in the uncertainty of $T_0$ and
$\gamma$ is due to our larger observational sample and the simultaneous
measurements from multiple wavelet scales (see \S
\ref{subsubsec:wavelet-literature}).  Similar to the curvature statistics, we
also show the $T_0$ measurements using the \oDelta method (see \S
\ref{subsubsec:wavelet-literature}).
}{\label{fig:T0-gamma-evolution-literature}}
\figref{fig:T0-gamma-evolution-literature} shows the evolution of $T_0$ and
$\gamma$ (as given in \tabref{tab:T0-gamma-constraints} and
\tabref{tab:T0-gamma-error-budget}) from the joint analysis with total
uncertainty in the redshift range \zrange{2}{4}.  Initially $T_0$ ($\gamma$) is
small (large) at \zrange{3.7}{3.9} then it increases (decreases) attaining a
maximum (minimum) at \zrange{2.7}{3.1}.  Subsequently $T_0$ ($\gamma$)
decreases (increases) at $z \leq 2.7$. As we will discuss in \S
\ref{sec:uvb-model-comparison}, the peak in the  evolution of $T_0$ and
$\gamma$ is due to the additional heating from \HeII reionization. Such a
peak in $T_0$ evolution has been suggested based on  other published measurements, 
but note the
rather large scatter between different measurements (see \figref{fig:T0-gamma-evolution-literature} and \figref{fig:T0-gamma-evolution-literature-combine}).  Our measurements show for
the first time a well defined peak and the expected smooth evolution of $T_0$
and $\gamma$ in the redshift range  \zrange{2.0}{3.8}, consistently for all four
different flux statistics.  We attribute this to the following main differences
of our study:  (i) our observed sample contains a larger number  of QSO sightlines than
previous studies, (ii) the simulated \lya forest in our analysis is generated
for  a finely sampled $T_0-\gamma$ grid than in previous studies, (iii) we treat
the data and simulations on an exact equal footing i.e., our simulated mock
spectra are mimicked to  match the observed sample, (iv) we apply the same
procedure/algorithm to calculate the \lya flux and Voigt  profile statistics to
observed and simulated spectra (v) by combining the $T_0-\gamma$ measurements
from different statistics, we mitigate possible biases of the  individual flux
statistics.

\subsection{Comparison with other measurements}
\label{subsec:T0-gamma-literature-comparison}


\begin{table}
\centering
\caption{Characteristic density probed by curvature and wavelet statistics in our simulations }
\begin{tabular}{cccc}
\hline  \hline 
$z$  & Curvature \oDelta & Wavelet \oDelta  & Wavelet \oDelta\\
  				  &  & \snval{50} & \snval{100} \\
\hline
2.0  & 4.85 & 4.95 & 5.37 \\
2.2  & 4.57 & 4.30 & 4.98 \\
2.4  & 3.90 & 3.89 & 4.37 \\
2.6  & 3.54 & 3.56 & 4.01 \\
2.8  & 3.19 & 3.26 & 3.50 \\
3.0  & 2.74 & 2.87 & 3.02 \\
3.2  & 2.48 & 2.53 & 2.67 \\
3.4  & 2.17 & 2.16 & 2.27 \\
3.6  & 2.00 & 1.90 & 2.25 \\
3.8  & 2.08 & 1.90 & 2.72 \\
\hline \hline
\end{tabular}
\\
\label{tab:Delta-c}
\end{table}

We now compare the $T_0$ and $\gamma$ measurements from this work
with that from the literature. \figref{fig:T0-gamma-evolution-literature} shows
a compilation of $T_0$ and $\gamma$ measurements from various papers. $T_0$
and $\gamma$ have been  measured in the past (with variants) of the same 
four different statistics we have used here.  In
order to do a fair comparison, we compare our measured $T_0$ and $\gamma$ from
the  joint analysis and the individual statistics 
with the
corresponding statistics from the literature.  

\subsubsection{Comparison with measurments from \blogNHI}
\label{subsubsec:bpdf-literature}
In panel A1 and A2 of \figref{fig:T0-gamma-evolution-literature}, we compare
our  thermal parameter evolution with that from
\citet{schaye2001,bolton2014,rorai2018,hiss2018,hiss2019,telikova2019}
obtained using the \blogNHI distribution. This  method is based on fitting the \lya
forest spectra with Voigt profiles and defining a lower envelope of the  2D
\blogNHI distribution. The motivation behind this is that for the
absorption components with low $b$  the broadening should be dominated by
thermal  broadening.

Our $T_0$ and $\gamma$ measurements from this statistics are consistent within
$1\sigma$ with that of \citet{bolton2014,rorai2018}. The $\gamma$ measurements
by \citet{hiss2018,telikova2019}  are in good agreement with our $\gamma$
measurements.  However, their $T_0$ measurements are significantly higher than
our measurements at \zrange{2.6}{3.6}.  Fitting a lower envelope to the
\blogNHI  distribution  although well motivated, is somewhat subjective.
Finding a lower envelope in an objective way is difficult \citep[but
see][]{telikova2019,hiss2019}. The lower envelope is found by iteratively
rejecting the lines with low $b$ parameter until  convergence is achieved.
More (less) rejection of low $b$ parameter lines can lead to systematically
higher (lower) inferred temperatures.  However, the high
values of  $T_0$ of \citet{hiss2018,telikova2019} obtained from the   \blogNHI
distribution  are  not consistent with the measurements  obtained using
e.g. the FPS by \citet{walther2019}.

Manual Voigt profile fitting is labour intensive and  the number of Voigt
components obtained from observations is rather limited in many analyses and
often the \blogNHI plane is sampled poorly. To avoid these systematic biases,
we use the 1D $b$ distribution (BPDF) computed in different \logNHI bins. The
BPDF does not rely on rejecting low $b$ parameters to find a lower envelope. At
the same time the BPDF statistics is sensitive to both $T_0$ and $\gamma$ (see
online supplementary appendix \ref{app:bpdf} for details).  Our
method of measuring $T_0$ and $\gamma$ using Voigt profile parameters is thus
significantly different from that used in the literature.  Unlike previous
work, our $T_0$ and $\gamma$ measurements using the BPDF statistics are
consistent with our measurements using the other statistics as shown in
\figref{fig:T0-gamma-evolution-all-no-uvb}.

\subsubsection{Comparison with measurments from curvature statistics}
\label{subsubsec:curvature-literature}
\InputFigCombine{Characteristic_Density_Illustration.pdf}{170}
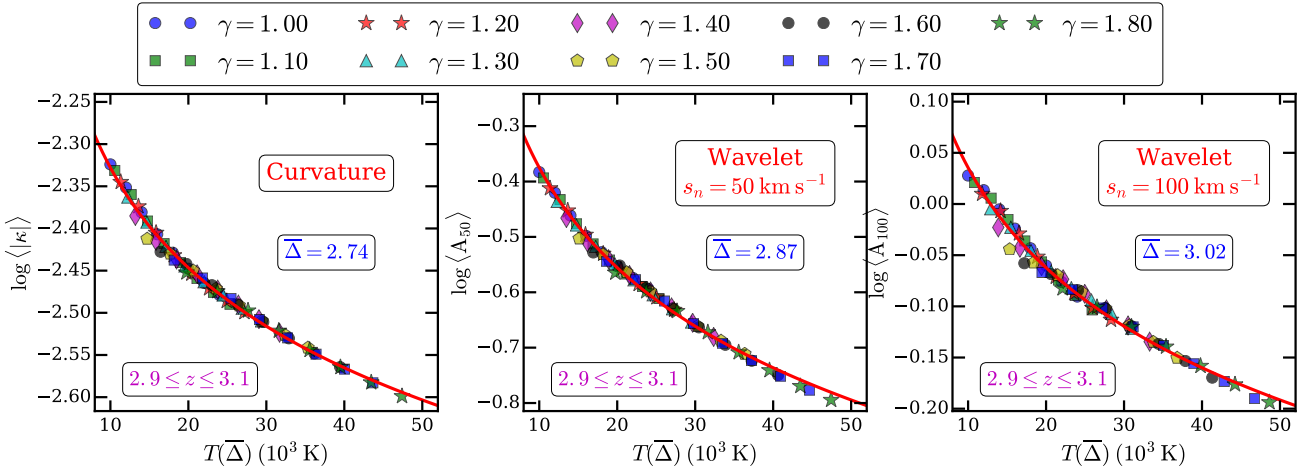
{ The figure shows the existence of a characteristic overdensity (\oDelta) in
    our simulations probed by the  curvature statistics (left panel) and
    wavelet statistics for two wavelet scales (50 \kmps and 100
    \kmps for middle and right panels respectively).  Simulations with different $\gamma$ are shown by different
    colors. For visualisation purposes, we show only points for selected
    $\gamma$ values. In the actual analysis, we use all the simulated $\gamma$
    values of our finely sampled $T_0-\gamma$ grids.  The red curve in each
    panel shows the  one-to-one relation between mean curvature (or wavelet)
    amplitude and temperature at \oDelta. The one-to-one relation is obtained
    by fitting a power-law.  The higher wavelet scale corresponds to slightly
    higher characteristic density indicating that \oDelta depends on the scales
    probed by the statistics.  The \oDelta probed by the curvature statistics
    is similar to that probed by the  wavelet statistics for \snval{50}.  The
    redshift evolution of \oDelta from this work (see \tabref{tab:Delta-c}) is
    in good agreement  with that obtained by  \citet{becker2011} for the
    curvature statistics.
}{\label{fig:characteristic-density}}
In panel B1 and B2 of \figref{fig:T0-gamma-evolution-literature}, we compare
our thermal parameter evolution with that from \citet{becker2011,boera2014}
obtained using the curvature statistics.  These authors measured $T_0$ by
measuring the temperature at a characteristic density \oDelta (\ToDelta) that is
a one-to-one empirical function of the
mean of the absolute curvature irrespective of $\gamma$. To map \ToDelta to the
temperature at cosmic mean density $T_0$, one needs to however assume a value of
$\gamma$. 

In \figref{fig:T0-gamma-evolution-all-no-uvb} we show the measurement of
thermal parameters using the PDF of the curvature statistics.  To make a more
direct  comparison between our measurement with that from
\citet{becker2011,boera2014}, we also calculate the $T_0$ using the
characteristic (over-)density method.  \figref{fig:characteristic-density} and
\tabref{tab:Delta-c} shows the  \oDelta probed by the  curvature statistics in
our simulations.  The evolution of \oDelta from this work is in good agreement
with that from \citet{becker2011}. We obtain the one-to-one empirical relation
between mean absolute curvature and temperature at \oDelta (as shown by the red
    curve in \figref{fig:characteristic-density}).  For a given observed mean
    absolute curvature, we measure the temperature at \oDelta.  We convert the
    temperature at \oDelta into $T_0$ using this empirical relation assuming a
    range of $\gamma$. For this we use the $\gamma$ values (with 1$\sigma$
    uncertainty)  from our joint constraints. The  green squares in panel B1 of
    \figref{fig:T0-gamma-evolution-literature} show our $T_0$ measurements
    using the \oDelta method.  The errors on the $T_0$ measurements for the
    \oDelta method account for the uncertainty in $\gamma$ only.

In \figref{fig:T0-gamma-evolution-literature}, we also show the $T_0$
measurements using the  \oDelta method and that using the joint analysis. The
best fit $T_0$ values from the \oDelta method are systematically higher than
that from the joint analysis as well as those obtained using the curvature PDF
(see \figref{fig:T0-gamma-evolution-all-no-uvb}) at $z<3.4$. The \oDelta
method assumes that most of the absorption in the \lya forest at any given
redshift is associated with densities $\sim$ \oDelta.  When converting
temperature at \oDelta to $T_0$, one assumes a single \oDelta value. However,
in a realistic scenario the \lya absorption will arise from a range in
densities. Unlike the \oDelta method, the $T_0$ measured in our joint analysis
(using PDFs and PS) is contributed by \lya absorption coming from all
densities. We suspect this to be the most likely  reason for the systematically
larger $T_0$ obtained with the \oDelta method.  Another interesting feature of
the  $T_0$ evolution obtained using the \oDelta method is that the uncertainty
of the  $T_0$ measurements increases with decreasing redshift.  This is mainly
because of the uncertainty in $\gamma$ and the increase of \oDelta at lower
redshifts.  

At $z>3.4$, our $T_0$ measurements from the  \oDelta method are consistent with
that from \citet{becker2011} irrespective of the assumed  value of $\gamma$.
This is because the \lya forest at these redshifts is sensitive to densities
close to the cosmic mean density $\Delta \sim 1$. This is reflected in the
smaller uncertainty of the  $T_0$ measurements at $z>3$. As $\gamma$ decreases
($1.2< \gamma < 1.5$) in our joint measurements at \zrange{2.6}{3.4}, $T_0$
increases and is in good agreement with that from \citet{becker2011,boera2014}
for $\gamma=1.3$.  At $z<2.6$, $\gamma$ increases from 1.2 to 1.5, and our
$T_0$ measurement using the \oDelta method is consistent with that from
\citet{becker2011,boera2014} for $\gamma=1.5$.

\subsubsection{Comparison with measurements from the FPS}
\label{subsubsec:fps-literature}
Panel C1 and C2 compares our $T_0$ measurements (FPS and joint analysis)  with
those from \citet{walther2019} using the FPS statistics.  Similar to this work,
\citet{walther2019} measured the $T_0$ and $\gamma$ evolution using the FPS
calculated from the \kodiaq DR2 sample.  However, their sample selection
criteria is different from ours.  Despite this the FPS from our work is in good
agreement with that from \citet[][]{walther2019}.
\figref{fig:T0-gamma-evolution-literature} shows that our $T_0$ measurements
are  also in  good agreement (within $1\sigma$ uncertainty) with their
measurements.  Note, however, that our $T_0$ measurements are systematically
larger (albeit within the errors) than their measurements at \zrange{2}{3}. The difference in the $\gamma$
evolution is more pronounced. Our $\gamma$ evolution shows a minimum at $z=3$
while their measurement shows a nearly flat $\gamma \sim 1.7$ at \zrange{2}{3.5}.  

Since the observed FPS are in good agreement with each other, the moderate
differences in the $T_0-\gamma$ measurements are likely to come from
differences in the simulations/analysis.  \citet{walther2019} vary  $T_0$,
$\gamma$ and the pressure smoothing scale $\lambda_p$ as a free parameter using
the \citetalias{onorbe2017} UVB.  They derive a single pressure smoothing scale
at a given redshift for the entire simulation box using a cutoff in the 3D real
space flux power spectrum that is calculated without accounting for peculiar
velocity and thermal broadening.  In reality, the pressure smoothing scale is
not an independent parameter.  This is because the pressure smoothing scale is
set by temperature and density of the gas particles i.e., $\lambda_p \propto (T
/ \Delta)^{1/2} \propto [T_0 \: \Delta^{\gamma-2}]^{1/2}$.  Simultaneous
fitting of  $T_0,\gamma$ and $\lambda_p$ can result in degeneracies of
$\lambda_p-T_0$ \citep[see Fig. 6, Fig. 6 in][respectively]{rorai2018,walther2019}. 
An overestimation of
$\lambda_p$ can result in a systematic decrease and increase in $T_0$ and
$\gamma$, respectively.  In our simulations, the pressure smoothing scale is
set by the density and temperature of each SPH particle which varies for each
particle.  Hence effectively we fit $T_0-\gamma$ as two free parameters to
match the simulations with observations.  It is also important to note that all
UVB models predict $\gamma < 1.55$ at \zrange{2}{4} (for equilibrium as well
as  non-equilibrium evolution of the ionization). As we show in \S
\ref{sec:uvb-model-comparison} the $\gamma$ evolution in \citet{walther2019}
is very difficult to reproduce with  any UVB model.


\subsubsection{Comparison of measurements from the wavelet statistics}
\label{subsubsec:wavelet-literature}
We compare our $T_0$ and $\gamma$ measurements using the wavelet statistics
with those from \citet{lidz2010} and \citet{garzilli2012} in panel D1 and D2 of
\figref{fig:T0-gamma-evolution-literature} respectively.  Our $T_0-\gamma$ measurements from
the wavelet analysis are more consistent with other flux statistics than those
of \citet{lidz2010} and \citet{garzilli2012}. The main difference between previous works
and this work are the observations and the methodology we use.  Our observed
sample size ($N_{\rm spectra} = 103$) is larger than that of previous studies using wavelet statistics  (40 and 20 for \citet{lidz2010} and \citet{garzilli2012} respectively).
%
%
To mitigate noise effects,
\citet{lidz2010} apply boxcar smoothing on the wavelet field
which somewhat reduces the sensitivity of the wavelets to $T_0$ and $\gamma$.
In this work, we compute the PDF from the unsmoothed  wavelet field using
kernel density estimation which converges faster than using a simple histogram
reducing the effect of noise on the PDF. In addition we simultaneously measure
$T_0$ and $\gamma$ at wavelet scales $s_{\rm n} = 30, 40, \cdots , 90, 100$
\kmps.  \citet{lidz2010} and \citet{garzilli2012} use only one wavelet scale at a time to
measure $T_0$ and $\gamma$. As a result, our $T_0$ and $\gamma$ constraints are
tighter than that of \citet{lidz2010} and \citet{garzilli2012}.

We find the  characteristic over-density  (\oDelta) probed by the  wavelet
statistics for \snval{50} and \snval{100}, respectively (see
\figref{fig:characteristic-density}). \tabref{tab:Delta-c} also summarizes the redshift
evolution of \oDelta for these two wavelet scales. The characteristic density
varies similarly  with other wavelet scales.  We find that the scatter in the
empirical relation increases with increasing  wavelet scales reducing the
sensitivity to $T_0$.  This is expected as  on large scales the baryons trace
the dark matter density field and large scales are  less sensitive to
variations in $T_0$. We also measure $T_0$ using the \oDelta method for the
wavelet statistics assuming the  $\gamma$ evolution from our joint analysis.
The $T_0$ values for all the wavelet scales are very similar. In
\figref{fig:T0-gamma-evolution-literature}, we show the $T_0$ values obtained using the \oDelta
method (green squares, for \snval{50}) and that obtained using the joint
analysis (black stars). The $T_0$ values from the two methods are in good
agreement with each other (within $1\sigma$).


\section{Discussion: Implication for UVB models and HeII reionization}
\label{sec:uvb-model-comparison}
\InputFigCombine{T0_gamma_Evolution_All_UVB_Models_Eqbm_Non_Eqbm.pdf}{170}
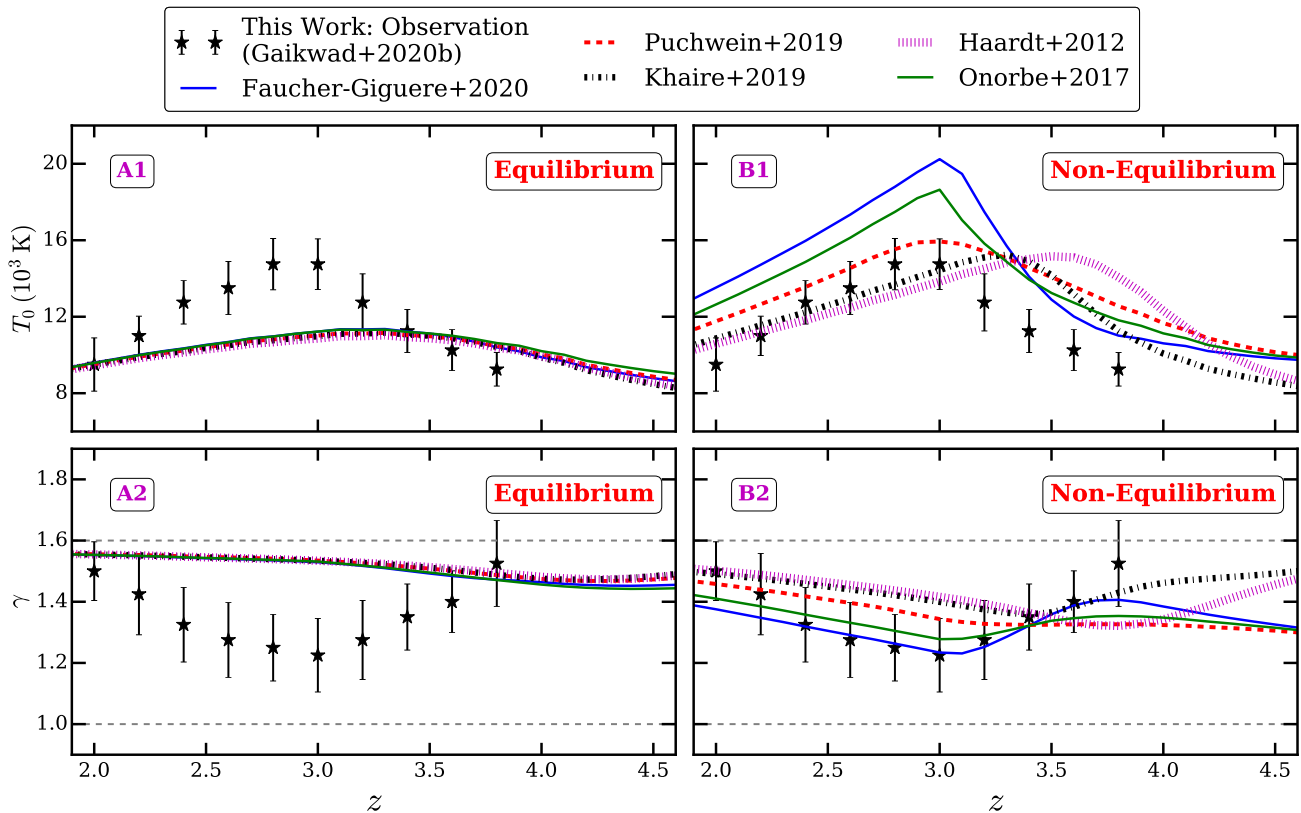
{   The figure  shows the evolution of thermal parameters obtained from
    our joint analysis (black star points)
    and the predictions of  the uniform UVB models of
    \citetalias{haardt2012,onorbe2017,puchwein2019,khaire2019a,faucher2020} 
    (different curves).  Panels A1 and A2 show the $T_0$, $\gamma$
    evolution obtained for UVB models assuming equilibrium ionization
    evolution.  The observed $T_0$ and $\gamma$ evolution is significantly
    different from those predicted by the  uniform UVB models assuming
    ionization equilibrium evolution.  Panels B1 and B2 show the predicted $T_0$,
    $\gamma$ evolution of the UVB models assuming  (physically very well
    motivated) non-equilibrium ionization evolution.  The $T_0$ evolution for
    the \citetalias{haardt2012,khaire2019a,puchwein2019} non-equilibrium models
    at $z<3$ are consistent  with that from observations. As we will discuss
    below the mismatch between observed and non-equilibrium UVB models at $z>3$
    is likely due to a somewhat wrong evolution of the assumed \HeII ionizing emissivity
    and/or (effective) \HeII photo-heating rates. Note also  that on the rising
    side of the temperature peak \HeII reionization will not yet be complete and
    the UVB is not expected to be spatially homogeneous.The $\gamma$ evolution
    of the \citetalias{onorbe2017,faucher2020} non-equilibrium UVB models is in
    good agreement with the observed values. However $T_0$ is systematically
    higher in these models. We have thus run the models  of
    \citetalias{onorbe2017,faucher2020} also with reduced  photo-heating rates.
    The observed $T_0$ evolution is well reproduced  if the photo-heating rate
    in these two models is lowered by factors of 0.8 and 0.7, respectively (see
    \figref{fig:T0-gamma-evolution-modified-FG18} and
\ref{fig:T0-gamma-evolution-modified-uvb}). The  good match between observation
and non-equilibrium \citetalias{onorbe2017,faucher2020} UVB models with
rescaled photo-heating rates is due to the rapid evolution of the \HeII ion
fraction in these two models.
}{\label{fig:T0-gamma-evolution-all-uvb}}
\InputFigCombine{T0_gamma_Evolution_Modified_UVB_Models_f_XI_Evolution.pdf}{170}
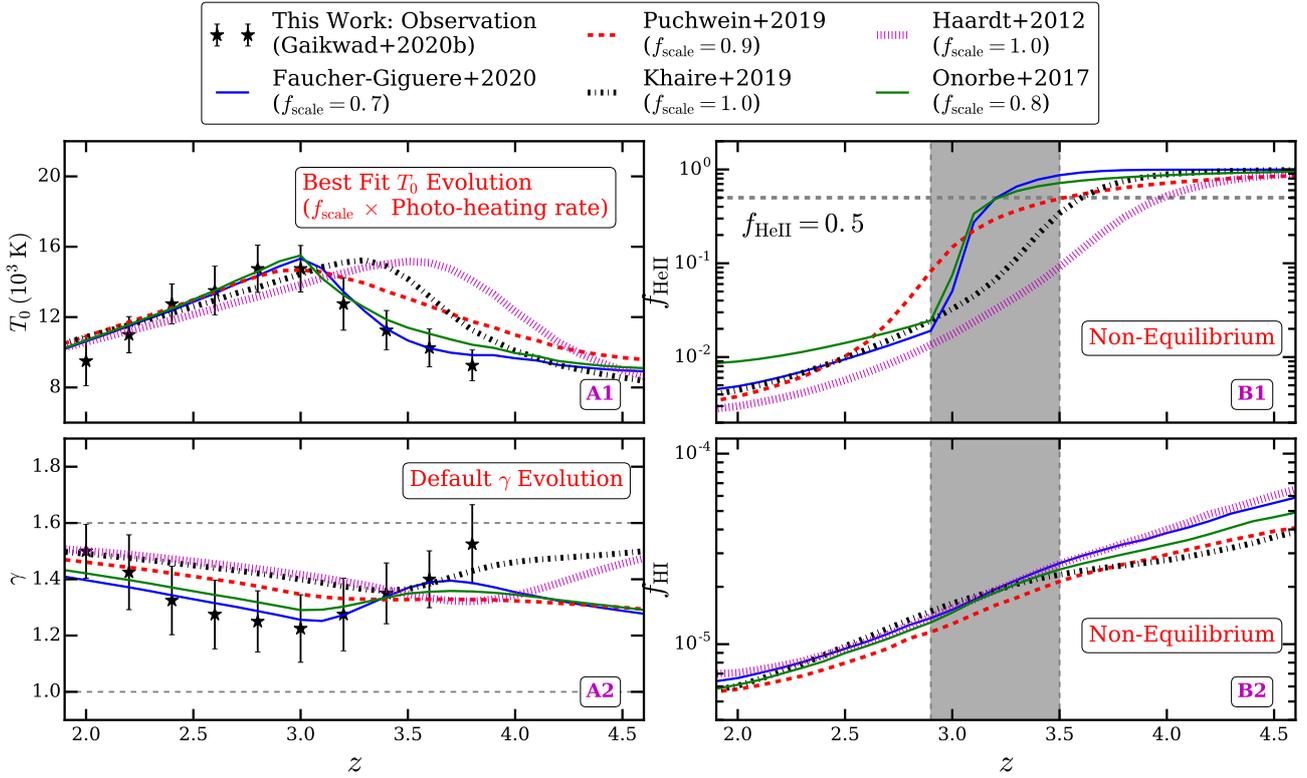
{  Panel A1 and A2 are similar to panel B1 and B2 in
    \figref{fig:T0-gamma-evolution-all-uvb} except  that the photo-heating
    rates of the \citetalias{onorbe2017,puchwein2019,faucher2020} UVB models
    are scaled by a factors of 0.8, 0.9 and 0.7, respectively. Note that such a
    scaling with a constant factor does not affect  $\gamma$.  Panel B1 and B2
    shows the evolution of the volume averaged \HeII and \HI fraction in our
    simulations for all the UVB models. \fHeII evolves rather rapidly at
    \zrange{2.9}{3.5} in the \citetalias{onorbe2017,faucher2020} models (shaded
    region in panel B1 and B2). Correspondingly \fHI in the same redshift range
    is similar for all UVB models (except \citetalias{puchwein2019}).
}{\label{fig:T0-gamma-evolution-modified-uvb-f-XI}}

In this section, we compare our measurements of
$T_0$ and $\gamma$ with predictions from published spatially homogeneous  UVB
models for the effect of \HeII reionization on the thermal evolution of the
IGM.  In \figref{fig:T0-gamma-evolution-all-uvb}, we compare the observed
$T_0-\gamma$ evolution with that predicted by  the UVB models of
\citetalias{haardt2012,onorbe2017,khaire2019a,puchwein2019} and
\citetalias{faucher2020}.  We use  \citecode to obtain  the evolution of $T_0$
and $\gamma$ for these UVB models, post-processing  \gthree outputs.

\subsection{Equilibrium vs Non-equilibrium UVB models}
\label{subsubsec:eqbm-non-eqbm}
Panels A1 and A2 in \figref{fig:T0-gamma-evolution-all-uvb} show that the
different UVB models with equilibrium ionization evolution show a very similar
$T_0-\gamma$ evolution in the redshift range \zrange{2}{4} irrespective of the
temperature after \HI reionization.  The observed $T_0$ ($\gamma$) is
systematically larger (smaller) than that predicted from equilibrium models at
\zrange{2.2}{3.4} (\zrange{2.2}{3.6}). It appears that it is not possible to
reproduce the higher $T_0$ (smaller $\gamma$) seen in the observed data with
UVB models assuming equilibrium ionization evolution. With equilibrium
ionization evolution, one assumes that the gas is ionized instantaneously as
the ionizing radiation field is turned on.  Because of this, the \HeII fraction
in equilibrium model is systematically underestimated
\citep{puchwein2015,gaikwad2019}.  Since the photo-heating is proportional to
the \HeII fraction, the temperature is underestimated. Similarly, the \HeII
fraction depends on density in photo-ionization equilibrium and as a result the
photo-heating of the gas is density dependent. As shown in \citet{gaikwad2019}
the density dependent photo-heating rates result in  $\gamma \sim 1.5$ in
equilibrium ionization models.

The \HeII reionization will in reality also be spatially inhomogeneous, and  spatially
homogeneous UVB models can only give an average evolution of the thermal
history. However,  in any given place the \HeIII ionization fronts will  move
too fast for equilibrium ionization evolution to be a good approximation.  We
have thus also implemented 
physically well
motivated non-equilibrium ionization evolution in \citecode.  Panels B1 and B2
in \figref{fig:T0-gamma-evolution-all-uvb} show the comparison of our  $T_0$
and $\gamma$ measurements with the prediction  from UVB models assuming
non-equilibrium evolution. The $T_0$ ($\gamma$) values  are systematically
larger (smaller) than in the corresponding equilibrium models and the
differences between the different UVB models are larger.  The non-equilibrium
predictions by all the  UVB models also show a more pronounced temperature
peak. This peak is generally somewhat higher than that in our measurements from
the observations and occurs in some of the models also somewhat earlier. 

\subsection{The temperature peak/rise due to HeII reionization predicted by  spatially homogeneous UVB models}
\label{subsubsec:reionization-history}
\begin{table}
\centering
\caption{Redshift and extent of \HeII reionization in UVB models}
\begin{threeparttable}
\begin{tabular}{ccccc}
\hline  \hline 
UVB Model  & $z_{\rm mid}$\tnote{[a]}  & $\Delta z$\tnote{[b]} & $f_{\rm scale}$\tnote{[c]} & $u_0(z=3)$\tnote{[d]} \\
\hline
\citetalias{faucher2020}   & 3.2  & 0.3 & 0.70 $\pm$ 0.10 & 2.97 \\
\citetalias{puchwein2019}  & 3.5  & 0.8 & 0.90 $\pm$ 0.20 & 3.05 \\
\citetalias{khaire2019a}   & 3.6  & 0.8 & 1.00 $\pm$ 0.20 & 2.43 \\
\citetalias{haardt2012}    & 4.0  & 1.0 & 1.00 $\pm$ 0.20 & 3.14 \\
\citetalias{onorbe2017}    & 3.2  & 0.4 & 0.75 $\pm$ 0.15 & 3.09 \\
\hline \hline
\end{tabular}
\begin{tablenotes}
\item[a] The redshift of reionization is defined as $z_{\rm mid}$  corresponding  to epoch when $f_{\rm HeII} = 0.5$ 
\item[b] The extent of reionization is defined as $\Delta z = z_{\rm start} - z_{\rm end}$. 
where $z_{\rm start}, \: z_{\rm end}$ corresponds to the redshift  when $f_{\rm HeII} = 0.75, 0.02$ 
respectively. 
\item[c] Factor by which photo-heating rates are scaled to better match the observed $T_0$
evolution at $z<3$. Errors represent the error in the scaling factor that corresponds to the
$1\sigma$ uncertainty for  $T_0$  
(see \figref{fig:T0-gamma-evolution-modified-FG18} and 
\figref{fig:T0-gamma-evolution-modified-uvb}).
\item[d] Cumulative energy deposited (in ${\rm eV} \: m^{-1}_p$) into the IGM  at
    $z=3$ for non-equilibrium UVB models without scaling the  photo-heating rates 
    (see online supplementary appendix \ref{app:cumulative-energy}).  
\end{tablenotes}
\end{threeparttable}
\\
\label{tab:early-late-reionization}
\end{table}

%
The basic concept  of the different UVB models is similar, but there are
considerable differences in the detailed assumptions.
\citetalias{onorbe2017,faucher2020} compute photo-ionization and photo-heating
rates during (\HI and \HeII) reionization differently than
\citetalias{haardt2012,khaire2019a,puchwein2019}.  To calculate
photo-ionization and photo-heating rates during \HeII reionization,
\citetalias{onorbe2017,faucher2020} assume the \HeII reionization history
(\fHeII evolution) to be  a free parametric function.  The reionization history
is calibrated to match the observed \taueffHeII evolution \citep{worseck2019}.
To compute the photo-heating rates during reionization,
\citetalias{onorbe2017,faucher2020} then use another free parameter namely the
total heat injected during \HeII reionization ($\Delta T_{\rm HeII}$).
\citetalias{haardt2012,khaire2019a,puchwein2019} on the other hand try to
predict photo-ionisation and  photo-heating rates based on observed source
luminosity functions and SEDs and a model of the \HeII opacity based on
observations.  There is thus considerable variety in the assumptions  of
spatially homogeneous UVB models.  As a result the exact time and height of
the temperature peak predicted by these models should be considered somewhat uncertain.  As can
be seen   in panel B1/B2 in \figref{fig:T0-gamma-evolution-all-uvb}, the timing
of the temperature peak in the \citetalias{onorbe2017,faucher2020} UVB models
agrees very well with the  timing of the peak in our measurements. The
evolution of $\gamma$ in these models is also  in better agreement with our
measurements than those from \citetalias{haardt2012,khaire2019a,puchwein2019}.
However, $T_0$ is systematically larger than our measured  values at all
redshifts.  Given the  uncertainty in the amplitude of the photo-heating rates
in all the UVB models, we have rescaled the  \HI, \HeI and \HeII photo-heating
rates in order to better match our measured evolution of the thermal parameters.
Such a rescaling changes  $T_0$ systematically, while  the $\gamma$ evolution
remains largely unaffected. 

Panels A1 and A2 in \figref{fig:T0-gamma-evolution-modified-uvb-f-XI} show the
evolution of $T_0$ and $\gamma$ for scaled photo-heating rates in the different UVB
models.  In \tabref{tab:early-late-reionization}, we summarize the
photo-heating rate scale factor required in different UVB models to match the
observed $T_0$ evolution at $z<3$ (see also
\figref{fig:T0-gamma-evolution-modified-FG18} and
\figref{fig:T0-gamma-evolution-modified-uvb} for details).  As expected, all
the rescaled UVB models show a better agreement with the $T_0$ evolution at $z<3$.
However, the $T_0$ evolution at $z>3$ in the rescaled
\citetalias{onorbe2017,faucher2020} UVB models is in much better agreement with
our measured values  than the \citetalias{haardt2012,khaire2019a,puchwein2019}
UVB models. The shape of the $T_0-\gamma$ evolution is  closely related to
the \HeII reionization history \citep[\fHeII evolution][]{gaikwad2019}.  It is noteworthy
here that the reionization history used in the \citetalias{onorbe2017,faucher2020} UVB
models is calibrated to match the observed \taueffHeII evolution from
\citet{worseck2019}. \textit{The good match between the shape of the observed
 $T_0-\gamma$ evolution and that predicted by the  \citet{} UVB models with scaled photo-heating rates 
at $z>3$ suggests that the $T_0-\gamma$ evolution from our work is  in
good agreement with the observed \taueffHeII  evolution from \citet{worseck2019}.}
Such consistency is important because the evolution of $T_0$ and $\gamma$ at
\zrange{2}{4} is mainly driven by the \fHeII evolution \citep[see Fig. 5
in][]{gaikwad2019}.  We should here emphasize that the observed dataset, methodology and
parameters that we used here  are completely complementary to those used to measure the  \taueffHeII
evolution in \citet{worseck2019}. 

Despite the  good match between observations and the predictions by
non-equilibrium \citetalias{onorbe2017,faucher2020} UVB models, we would like
to caution the reader that these UVB models do not account for inhomogeneous
\HeII reionization and the spatial fluctuations in the \HeII ionizing background
which are expected to be present.  To capture the effects of such spatial
fluctuations, will, however require rather challenging high dynamic range
radiative transfer simulations (ideally coupled to the hydrodynamics).  We hope
to do this in future work but caution that it  will be very expensive to
calibrate such simulations to match the observed data
\citep[see][]{gaikwad2020}.

\subsection{Late vs Early HeII reionization}
\label{subsubsec:late-early-reionization}
In \figref{fig:T0-gamma-evolution-modified-uvb-f-XI}, we compare the
evolution of the  \HeII and \HI fraction predicted by the different UVB models
with non-equilibrium ionisation evolution.  As already discussed the amount of
heat injected into the IGM is coupled to  the \fHeII evolution and we have
scaled  the photo-heating rates to match the observed $T_0$ evolution.  To
quantify the differences between the various UVB models, we define the mid
point of \HeII reionization as the redshift, where $f_{\rm HeII} = 0.5$. We
also define the extent of reionization ($\Delta z$) as corresponding to the
difference between the redshifts when $f_{\rm HeII}=0.75$ and $f_{\rm HeII} =
0.02$. These definitions facilitate the comparison of the UVB models
quantitatively, but note that they may differ from  definitions used in the
literature (see \tabref{tab:early-late-reionization}).

The midpoint of \HeII reionization in  the \citetalias{onorbe2017,faucher2020}
models ($z_{\rm mid} \sim 3.2$) is at lower redshift than that for the
models by \citetalias{haardt2012,khaire2019a,puchwein2019}. The observed
$T_0-\gamma$ evolution from our measurements is consistent with the relatively late \HeII
reionization in the \citetalias{onorbe2017,faucher2020} models.
\figref{fig:T0-gamma-evolution-modified-uvb-f-XI} also shows that the evolution
of  \fHeII in the \citetalias{haardt2012,khaire2019a,puchwein2019} models is
more gradual than that in the \citetalias{onorbe2017,faucher2020} UVB models.
The extent of \HeII reionization $\Delta z$ in the
\citetalias{onorbe2017,faucher2020} UVB models $(<0.4)$ is smaller than that
in the \citetalias{haardt2012,khaire2019a,puchwein2019} UVB models (see
\tabref{tab:early-late-reionization}).   We should note  here that the actual
duration and heating due to \HeII reionization may be different  if the
spatially inhomogeneous nature of \HeII reionization and the resulting
fluctuations in the \HeII ionizing background are taken into account.  With
this caveat in mind we note that for the spatially uniform UVB models considered
in this work, our measured  $T_0-\gamma$ evolution is consistent with late and
rapid \HeII reionization. 

We refer readers to online supplementary appendix \ref{app:T0-gamma-evolution-uvb-models} and
\ref{app:cumulative-energy} for more details on the effect of 
the UVB on the evolution of
thermal parameters and cumulative energy deposited in to the IGM.  In
particular, we show the effect of observational and modeling uncertainty in the 
UVB model on thermal parameters in online supplementary appendices 
\ref{app:initial-T0-gamma-effect} and
\ref{app:qso-sed-effect}. We discuss the likely modification needed for the
physically motivated non-equilibrium UVB models to reproduce our measured
thermal parameter evolution in online supplementary appendices \ref{app:uvb-modification} and
\ref{app:uvb-model-forward-way}. We discuss the evolution of the cumulative
energy deposited in IGM by various UVB models in online supplementary appendix
\ref{app:cumulative-energy}.  

Finally, we emphasize again that even though we use equilibrium
\citetalias{khaire2019a} UVB models to vary the thermal parameters in the
simulations, our final measured thermal parameter evolution is consistent with
the predictions by  the physically motivated non-equilibrium evolution for the
\citetalias{faucher2020,onorbe2017} UVB models with moderately reduced
photo-heating rates.  This gives us additional confidence that our measurements
of thermal parameters  are robust with regard to  observational and modeling
systematics.

\section{Summary}
\label{sec:summary}
We have measured the thermal parameters $T_0$ and $\gamma$  of the
IGM at \zrange{2}{4} using 103 high-resolution, high signal-to-noise 
QSO absorption spectra drawn from the  \kodiaq DR2  sample 
using four different \lya flux statistics.    The measurements are calibrated with a suite of  simulations based on 
a high-resolution smoothed particle hydrodynamics simulation post-processed
with our IGM thermal history code  ``\citefullform'' (\citecode) that finely samples 
the grid in  $T_0$ and $\gamma$. The salient points of our work are:
\begin{itemize}
\item We use a  sample of 103 QSO spectra satisfying the following criteria; (i) spectral \SNR $> 5$, (ii) no large spectral gaps and (iii) no identified
    DLAs or sub-DLAs to avoid systematic biases. We have divided our sample in
     redshift bins having $\Delta z =0.2$ with centers at $z=2.0,2.2, \cdots, 3.6~{\rm and}~3.8$. The
    typical number of lines of sight used per redshift bin is larger than what has been used in previous
    studies.  We have identified  metal lines contaminating the \lya forest and
    replaced them with continuum and added noise.
    Our simulations,  based on  L10N512 \gthree SPH simulation, sample $T_0$
    from $\sim 6000$ K to $24000$ K in bins of $\Delta T_0 = 500$ K and
    $\gamma$ from $\sim 0.7$ to $\sim 2.0$ in bins in of $\delta \gamma = 0.05$
    creating a densely sampled $T_0-\gamma$ grids with $37 \times 27 = 999$
    thermal parameter variations. The  mock spectra created from these
    simulations  account for the redshift path length, finite \SNR and
    instrumental broadening of the spectrograph of the observed sample.

\item Using our  code ``\viperfullform'' (\viper), we fit the observed and
    simulated \lya forest spectra with multi-component Voigt profiles. We
    derive 6 flux statistics for observed  and simulated spectra: (i) flux
    power spectrum (FPS), (ii) wavelet statistics (iii) curvature statistics,
    (iv) $b$ distribution function (BPDF), (v) transmitted flux probability
    distribution function (FPDF) and (iv) \HI column density distribution
    function (CDDF). The observed FPS, FPDF, CDDF and mean flux evolution from
    our sample is in good agreement with measurements in the  literature. By
    varying one parameter at a time, we show the sensitivity of these
    statistics to thermal parameters of the IGM.  

\item We measure the thermal parameters $T_0$ and $\gamma$ in the redshift
    range \zrange{2.0}{3.8} by quantitatively comparing four of the six flux
    statistic, namely  FPS, wavelet, curvature and BPDF statistics. The $T_0$
    and $\gamma$ measurements using different   statistics are consistent
    with each other within $1 \sigma$  at all redshifts.  By combining the four
    statistics we thus find joint    constraints on $T_0$ and $\gamma$ that are
    more robust and tightly constrained than the measured values from the individual
    statistics.  We estimate  the total uncertainty on $T_0$ and $\gamma$ by
    accounting for the uncertainty due to continuum fitting, metal line
    contamination, cosmological parameters and modeling uncertainties. 
    In this work, we ignore the correlation among different statistics for
    simplicity.  Hence the statistical uncertainty of $T_0-\gamma$ in joint
    constraints may be somewhat underestimated. However, our recovery using the Sherwood
    simulation suite suggests that the $T_0-\gamma$ uncertainty is not seviorly
    underestimated. We find
    that our measured $T_0$ varies from $9250$ K at $z=3.8$ to $14750$ K at
    $z=3$ to $9500$ K at $z=2$,  while  $\gamma$ evolves from $1.525$ at
    $z=3.8$ to $1.225$ at $z=3$ to $1.500$ at $z=2$. We robustly detect a
    maximum in $T_0$ and minimum in $\gamma$ that occur simultaneously at
    $z\simeq3$. 

\item We compare our measurements to the evolution of $T_0$ and $\gamma$
    predicted by  spatially homogeneous UVB background models of
    \citetalias{haardt2012,onorbe2017,khaire2019a,puchwein2019,faucher2020}
    assuming  equilibrium and non-equilibrium ionization evolution.
    Our temperature measurements are difficult  to reproduce assuming
    equilibrium ionization  evolution. Our measurements are, however,  in good
    agreement with the thermal evolution predicted by the UVB models that have
    been calibrated to match the observed \HeII opacity evolution
    (\citetalias{onorbe2017,faucher2020}) if we assume non-equilibrium
    ionization evolution and reduce the photo-heating rates in these models by a
    moderate factor  0.7-0.8. The timing of the peak  in our measured
    temperatures due to the extra heat input during \HeII reionization appears
    thus to be nicely consistent with the timing of \HeII reionization inferred
    from the \HeII opacity evolution and appears to be rather rapid.  Our
    measured temperatures are also (to varying degree) systematically lower
    than those in the other three models. These three models  predict \HeII
    reionization to be more extended  and either an onset of the temperature
    rise (\citetalias{puchwein2019})  or a timing of the peak  that is too
    early  (\citetalias{haardt2012,khaire2019a}) to  be consistent with our
    temperature  measurements. 
 \end{itemize}

The simulations used to calibrate our temperature measurements assume
time  varying but spatially uniform UVB models.  While spatial fluctuations in
the \HeII ionizing background are clearly important during \HeII reionization
($z>3$), simulating these self-consistently  is  beyond the scope of the
current work. Large suites of high dynamic range, large scale, multi-frequency
cosmological radiative transfer simulations will be  required to capture this
effect accurately and to sample the observational uncertainties regarding
relevant properties of the ionizing sources.  We hope  to perform such
simulations in future. The tests we have performed and our experience with
hydrogen reionization  suggest that modeling the thermal evolution with
homogeneous UVB models should nevertheless  recover the evolution of the
(median) thermal state of  the IGM  not only after but also before overlap of
\HeIII regions reasonably well.


\section*{Acknowledgments}
We thank the DiRAC@Durham facility managed by the Institute for Computational
Cosmology on behalf of the STFC DiRAC HPC Facility (www.dirac.ac.uk). The
equipment was funded by BEIS capital funding via STFC capital grants
ST/P002293/1, ST/R002371/1 and ST/S002502/1, Durham University and STFC
operations grant ST/R000832/1. DiRAC is part of the National e-Infrastructure.
Computations in this work were also  performed using the Perseus cluster at
IUCAA, the HPC cluster at NCRA and the CALX machines at IoA.  Support by ERC
Advanced Grant 320596 `The Emergence of Structure During the Epoch of
reionization' is gratefully acknowledged. PG and MGH acknowledge the support of
the UK Science and Technology Facilities Council (STFC). TRC acknowledges
support of the Department of Atomic Energy, Government of India, under project
no.~12-R\&D-TFR-5.02-0700.  We thank J. Bolton for making the \sherwood simulation
suite available to our work.
The \sherwood simulations were performed using the Curie supercomputer at the
Tre Grand Centre de Calcul (TGCC), and the DiRAC Data Analytic system at the
University of Cambridge, operated by the University of Cambridge High
Performance Computing Service on behalf of the STFC DiRAC HPC Facility
(www.dirac.ac.uk). This equipment was funded by BIS National E-infrastructure
capital grant (ST/K001590/1), STFC capital grants ST/H008861/1 and
ST/H00887X/1, and STFC DiRAC Operations grant ST/K00333X/1. DiRAC is part of
the National E- Infrastructure.
We thank Vid Irsic, Ewald Puchwein and Jamie Bolton for useful comments on the
manuscript.

\section*{Data Availability}
The data underlying this article are available in the article and in its online
supplementary material.


\bibliographystyle{mnras}
\bibliography{igm_temperature} 

\begin{thebibliography}{}
\makeatletter
\relax
\def\mn@urlcharsother{\let\do\@makeother \do\$\do\&\do\#\do\^\do\_\do\%\do\~}
\def\mn@doi{\begingroup\mn@urlcharsother \@ifnextchar [ {\mn@doi@}
  {\mn@doi@[]}}
\def\mn@doi@[#1]#2{\def\@tempa{#1}\ifx\@tempa\@empty \href
  {http://dx.doi.org/#2} {doi:#2}\else \href {http://dx.doi.org/#2} {#1}\fi
  \endgroup}
\def\mn@eprint#1#2{\mn@eprint@#1:#2::\@nil}
\def\mn@eprint@arXiv#1{\href {http://arxiv.org/abs/#1} {{\tt arXiv:#1}}}
\def\mn@eprint@dblp#1{\href {http://dblp.uni-trier.de/rec/bibtex/#1.xml}
  {dblp:#1}}
\def\mn@eprint@#1:#2:#3:#4\@nil{\def\@tempa {#1}\def\@tempb {#2}\def\@tempc
  {#3}\ifx \@tempc \@empty \let \@tempc \@tempb \let \@tempb \@tempa \fi \ifx
  \@tempb \@empty \def\@tempb {arXiv}\fi \@ifundefined
  {mn@eprint@\@tempb}{\@tempb:\@tempc}{\expandafter \expandafter \csname
  mn@eprint@\@tempb\endcsname \expandafter{\@tempc}}}

\bibitem[\protect\citeauthoryear{{Almgren}, {Bell}, {Lijewski}, {Luki{\'c}}  \&
  {Van Andel}}{{Almgren} et~al.}{2013}]{almgren2013}
{Almgren} A.~S.,  {Bell} J.~B.,  {Lijewski} M.~J.,  {Luki{\'c}} Z.,   {Van
  Andel} E.,  2013, \mn@doi [\apj] {10.1088/0004-637X/765/1/39}, \href
  {https://ui.adsabs.harvard.edu/abs/2013ApJ...765...39A} {765, 39}

\bibitem[\protect\citeauthoryear{{Avni}}{{Avni}}{1976}]{avni1976}
{Avni} Y.,  1976, \mn@doi [\apj] {10.1086/154870}, \href
  {https://ui.adsabs.harvard.edu/abs/1976ApJ...210..642A} {210, 642}

\bibitem[\protect\citeauthoryear{{Becker} \& {Bolton}}{{Becker} \&
  {Bolton}}{2013}]{becker2013}
{Becker} G.~D.,  {Bolton} J.~S.,  2013, \mn@doi [\mnras]
  {10.1093/mnras/stt1610}, \href
  {https://ui.adsabs.harvard.edu/abs/2013MNRAS.436.1023B} {436, 1023}

\bibitem[\protect\citeauthoryear{{Becker}, {Bolton}, {Haehnelt}  \&
  {Sargent}}{{Becker} et~al.}{2011}]{becker2011}
{Becker} G.~D.,  {Bolton} J.~S.,  {Haehnelt} M.~G.,   {Sargent} W. L.~W.,
  2011, \mn@doi [\mnras] {10.1111/j.1365-2966.2010.17507.x}, \href
  {https://ui.adsabs.harvard.edu/abs/2011MNRAS.410.1096B} {410, 1096}

\bibitem[\protect\citeauthoryear{{Becker}, {Hewett}, {Worseck}  \&
  {Prochaska}}{{Becker} et~al.}{2013}]{becker2013b}
{Becker} G.~D.,  {Hewett} P.~C.,  {Worseck} G.,   {Prochaska} J.~X.,  2013,
  \mn@doi [\mnras] {10.1093/mnras/stt031}, \href
  {https://ui.adsabs.harvard.edu/abs/2013MNRAS.430.2067B} {430, 2067}

\bibitem[\protect\citeauthoryear{{Boera}, {Murphy}, {Becker}  \&
  {Bolton}}{{Boera} et~al.}{2014}]{boera2014}
{Boera} E.,  {Murphy} M.~T.,  {Becker} G.~D.,   {Bolton} J.~S.,  2014, \mn@doi
  [\mnras] {10.1093/mnras/stu660}, \href
  {https://ui.adsabs.harvard.edu/abs/2014MNRAS.441.1916B} {441, 1916}

\bibitem[\protect\citeauthoryear{{Boera}, {Becker}, {Bolton}  \&
  {Nasir}}{{Boera} et~al.}{2019}]{boera2019}
{Boera} E.,  {Becker} G.~D.,  {Bolton} J.~S.,   {Nasir} F.,  2019, \mn@doi
  [\apj] {10.3847/1538-4357/aafee4}, \href
  {https://ui.adsabs.harvard.edu/abs/2019ApJ...872..101B} {872, 101}

\bibitem[\protect\citeauthoryear{{Bolton} \& {Haehnelt}}{{Bolton} \&
  {Haehnelt}}{2007}]{bolton2007}
{Bolton} J.~S.,  {Haehnelt} M.~G.,  2007, \mn@doi [\mnras]
  {10.1111/j.1365-2966.2007.12372.x}, \href
  {https://ui.adsabs.harvard.edu/abs/2007MNRAS.382..325B} {382, 325}

\bibitem[\protect\citeauthoryear{{Bolton}, {Haehnelt}, {Viel}  \&
  {Springel}}{{Bolton} et~al.}{2005}]{bolton2005}
{Bolton} J.~S.,  {Haehnelt} M.~G.,  {Viel} M.,   {Springel} V.,  2005, \mn@doi
  [\mnras] {10.1111/j.1365-2966.2005.08704.x}, \href
  {https://ui.adsabs.harvard.edu/abs/2005MNRAS.357.1178B} {357, 1178}

\bibitem[\protect\citeauthoryear{{Bolton}, {Becker}, {Raskutti}, {Wyithe},
  {Haehnelt}  \& {Sargent}}{{Bolton} et~al.}{2012}]{bolton2012}
{Bolton} J.~S.,  {Becker} G.~D.,  {Raskutti} S.,  {Wyithe} J. S.~B.,
  {Haehnelt} M.~G.,   {Sargent} W. L.~W.,  2012, \mn@doi [\mnras]
  {10.1111/j.1365-2966.2011.19929.x}, \href
  {https://ui.adsabs.harvard.edu/abs/2012MNRAS.419.2880B} {419, 2880}

\bibitem[\protect\citeauthoryear{{Bolton}, {Becker}, {Haehnelt}  \&
  {Viel}}{{Bolton} et~al.}{2014}]{bolton2014}
{Bolton} J.~S.,  {Becker} G.~D.,  {Haehnelt} M.~G.,   {Viel} M.,  2014, \mn@doi
  [\mnras] {10.1093/mnras/stt2374}, \href
  {https://ui.adsabs.harvard.edu/abs/2014MNRAS.438.2499B} {438, 2499}

\bibitem[\protect\citeauthoryear{{Bolton}, {Puchwein}, {Sijacki}, {Haehnelt},
  {Kim}, {Meiksin}, {Regan}  \& {Viel}}{{Bolton} et~al.}{2017}]{bolton2017}
{Bolton} J.~S.,  {Puchwein} E.,  {Sijacki} D.,  {Haehnelt} M.~G.,  {Kim} T.-S.,
   {Meiksin} A.,  {Regan} J.~A.,   {Viel} M.,  2017, \mn@doi [\mnras]
  {10.1093/mnras/stw2397}, \href
  {https://ui.adsabs.harvard.edu/abs/2017MNRAS.464..897B} {464, 897}

\bibitem[\protect\citeauthoryear{{Calura}, {Tescari}, {D'Odorico}, {Viel},
  {Cristiani}, {Kim}  \& {Bolton}}{{Calura} et~al.}{2012}]{calura2012}
{Calura} F.,  {Tescari} E.,  {D'Odorico} V.,  {Viel} M.,  {Cristiani} S.,
  {Kim} T.~S.,   {Bolton} J.~S.,  2012, \mn@doi [\mnras]
  {10.1111/j.1365-2966.2012.20811.x}, \href
  {https://ui.adsabs.harvard.edu/abs/2012MNRAS.422.3019C} {422, 3019}

\bibitem[\protect\citeauthoryear{{Calverley}, {Becker}, {Haehnelt}  \&
  {Bolton}}{{Calverley} et~al.}{2011}]{calverley2011}
{Calverley} A.~P.,  {Becker} G.~D.,  {Haehnelt} M.~G.,   {Bolton} J.~S.,  2011,
  \mn@doi [\mnras] {10.1111/j.1365-2966.2010.18072.x}, \href
  {https://ui.adsabs.harvard.edu/abs/2011MNRAS.412.2543C} {412, 2543}

\bibitem[\protect\citeauthoryear{{Carswell}, {Whelan}, {Smith}, {Boksenberg}
  \& {Tytler}}{{Carswell} et~al.}{1982}]{carswell1982}
{Carswell} R.~F.,  {Whelan} J.~A.~J.,  {Smith} M.~G.,  {Boksenberg} A.,
  {Tytler} D.,  1982, \mn@doi [\mnras] {10.1093/mnras/198.1.91}, \href
  {https://ui.adsabs.harvard.edu/abs/1982MNRAS.198...91C} {198, 91}

\bibitem[\protect\citeauthoryear{{Chang}, {Broderick}  \& {Pfrommer}}{{Chang}
  et~al.}{2012}]{Chang2012}
{Chang} P.,  {Broderick} A.~E.,   {Pfrommer} C.,  2012, \mn@doi [\apj]
  {10.1088/0004-637X/752/1/23}, \href
  {https://ui.adsabs.harvard.edu/abs/2012ApJ...752...23C} {752, 23}

\bibitem[\protect\citeauthoryear{{Choudhury}, {Srianand}  \&
  {Padmanabhan}}{{Choudhury} et~al.}{2001}]{trc2001}
{Choudhury} T.~R.,  {Srianand} R.,   {Padmanabhan} T.,  2001, \mn@doi [\apj]
  {10.1086/322327}, \href
  {https://ui.adsabs.harvard.edu/abs/2001ApJ...559...29C} {559, 29}

\bibitem[\protect\citeauthoryear{{Cirelli}, {Iocco}  \& {Panci}}{{Cirelli}
  et~al.}{2009}]{Cirelli2009}
{Cirelli} M.,  {Iocco} F.,   {Panci} P.,  2009, \mn@doi [\jcap]
  {10.1088/1475-7516/2009/10/009}, \href
  {https://ui.adsabs.harvard.edu/abs/2009JCAP...10..009C} {2009, 009}

\bibitem[\protect\citeauthoryear{{Croft}, {Weinberg}, {Bolte}, {Burles},
  {Hernquist}, {Katz}, {Kirkman}  \& {Tytler}}{{Croft}
  et~al.}{2002}]{croft2002}
{Croft} R. A.~C.,  {Weinberg} D.~H.,  {Bolte} M.,  {Burles} S.,  {Hernquist}
  L.,  {Katz} N.,  {Kirkman} D.,   {Tytler} D.,  2002, \mn@doi [\apj]
  {10.1086/344099}, \href
  {https://ui.adsabs.harvard.edu/abs/2002ApJ...581...20C} {581, 20}

\bibitem[\protect\citeauthoryear{{D'Aloisio}, {McQuinn}, {Trac}, {Cain}  \&
  {Mesinger}}{{D'Aloisio} et~al.}{2020}]{daloisio2020}
{D'Aloisio} A.,  {McQuinn} M.,  {Trac} H.,  {Cain} C.,   {Mesinger} A.,  2020,
  arXiv e-prints, \href {https://ui.adsabs.harvard.edu/abs/2020arXiv200202467D}
  {p. arXiv:2002.02467}

\bibitem[\protect\citeauthoryear{{Day}, {Tytler}  \& {Kambalur}}{{Day}
  et~al.}{2019}]{day2019}
{Day} A.,  {Tytler} D.,   {Kambalur} B.,  2019, \mn@doi [\mnras]
  {10.1093/mnras/stz2214}, \href
  {https://ui.adsabs.harvard.edu/abs/2019MNRAS.489.2536D} {489, 2536}

\bibitem[\protect\citeauthoryear{{Faucher-Gigu{\`e}re}}{{Faucher-Gigu{\`e}re}}{2020}]{faucher2020}
{Faucher-Gigu{\`e}re} C.-A.,  2020, \mn@doi [\mnras] {10.1093/mnras/staa302},
  \href {https://ui.adsabs.harvard.edu/abs/2020MNRAS.493.1614F} {493, 1614}

\bibitem[\protect\citeauthoryear{{Faucher-Gigu{\`e}re}, {Lidz}  \&
  {Hernquist}}{{Faucher-Gigu{\`e}re} et~al.}{2008}]{faucher2008}
{Faucher-Gigu{\`e}re} C.-A.,  {Lidz} A.,   {Hernquist} L.,  2008, \mn@doi
  [Science] {10.1126/science.1151476}, \href
  {https://ui.adsabs.harvard.edu/abs/2008Sci...319...52F} {319, 52}

\bibitem[\protect\citeauthoryear{{Faucher-Gigu{\`e}re}, {Lidz}, {Zaldarriaga}
  \& {Hernquist}}{{Faucher-Gigu{\`e}re} et~al.}{2009}]{faucher2009}
{Faucher-Gigu{\`e}re} C.-A.,  {Lidz} A.,  {Zaldarriaga} M.,   {Hernquist} L.,
  2009, \mn@doi [\apj] {10.1088/0004-637X/703/2/1416}, \href
  {https://ui.adsabs.harvard.edu/abs/2009ApJ...703.1416F} {703, 1416}

\bibitem[\protect\citeauthoryear{{Gaikwad}, {Khaire}, {Choudhury}  \&
  {Srianand}}{{Gaikwad} et~al.}{2017a}]{gaikwad2017a}
{Gaikwad} P.,  {Khaire} V.,  {Choudhury} T.~R.,   {Srianand} R.,  2017a,
  \mn@doi [\mnras] {10.1093/mnras/stw3086}, \href
  {https://ui.adsabs.harvard.edu/abs/2017MNRAS.466..838G} {466, 838}

\bibitem[\protect\citeauthoryear{{Gaikwad}, {Srianand}, {Choudhury}  \&
  {Khaire}}{{Gaikwad} et~al.}{2017b}]{gaikwad2017b}
{Gaikwad} P.,  {Srianand} R.,  {Choudhury} T.~R.,   {Khaire} V.,  2017b,
  \mn@doi [\mnras] {10.1093/mnras/stx248}, \href
  {https://ui.adsabs.harvard.edu/abs/2017MNRAS.467.3172G} {467, 3172}

\bibitem[\protect\citeauthoryear{{Gaikwad}, {Choudhury}, {Srianand }  \&
  {Khaire}}{{Gaikwad} et~al.}{2018}]{gaikwad2018}
{Gaikwad} P.,  {Choudhury} T.~R.,  {Srianand } R.,   {Khaire} V.,  2018,
  \mn@doi [\mnras] {10.1093/mnras/stx2859}, \href
  {https://ui.adsabs.harvard.edu/abs/2018MNRAS.474.2233G} {474, 2233}

\bibitem[\protect\citeauthoryear{{Gaikwad}, {Srianand}, {Khaire}  \&
  {Choudhury}}{{Gaikwad} et~al.}{2019}]{gaikwad2019}
{Gaikwad} P.,  {Srianand} R.,  {Khaire} V.,   {Choudhury} T.~R.,  2019, \mn@doi
  [\mnras] {10.1093/mnras/stz2692}, \href
  {https://ui.adsabs.harvard.edu/abs/2019MNRAS.490.1588G} {490, 1588}

\bibitem[\protect\citeauthoryear{{Gaikwad} et~al.,}{{Gaikwad}
  et~al.}{2020}]{gaikwad2020}
{Gaikwad} P.,  et~al., 2020, arXiv e-prints, \href
  {https://ui.adsabs.harvard.edu/abs/2020arXiv200110018G} {p. arXiv:2001.10018}

\bibitem[\protect\citeauthoryear{{Garzilli}, {Bolton}, {Kim}, {Leach}  \&
  {Viel}}{{Garzilli} et~al.}{2012}]{garzilli2012}
{Garzilli} A.,  {Bolton} J.~S.,  {Kim} T.~S.,  {Leach} S.,   {Viel} M.,  2012,
  \mn@doi [\mnras] {10.1111/j.1365-2966.2012.21223.x}, \href
  {https://ui.adsabs.harvard.edu/abs/2012MNRAS.424.1723G} {424, 1723}

\bibitem[\protect\citeauthoryear{{Garzilli}, {Boyarsky}  \&
  {Ruchayskiy}}{{Garzilli} et~al.}{2017}]{garzilli2015}
{Garzilli} A.,  {Boyarsky} A.,   {Ruchayskiy} O.,  2017, \mn@doi [Physics
  Letters B] {10.1016/j.physletb.2017.08.022}, \href
  {https://ui.adsabs.harvard.edu/abs/2017PhLB..773..258G} {773, 258}

\bibitem[\protect\citeauthoryear{{Gnedin} \& {Hui}}{{Gnedin} \&
  {Hui}}{1998}]{gnedin1998}
{Gnedin} N.~Y.,  {Hui} L.,  1998, \mn@doi [\mnras]
  {10.1046/j.1365-8711.1998.01249.x}, \href
  {https://ui.adsabs.harvard.edu/abs/1998MNRAS.296...44G} {296, 44}

\bibitem[\protect\citeauthoryear{{Haardt} \& {Madau}}{{Haardt} \&
  {Madau}}{1996}]{haardt1996}
{Haardt} F.,  {Madau} P.,  1996, \mn@doi [\apj] {10.1086/177035}, \href
  {https://ui.adsabs.harvard.edu/abs/1996ApJ...461...20H} {461, 20}

\bibitem[\protect\citeauthoryear{{Haardt} \& {Madau}}{{Haardt} \&
  {Madau}}{2012}]{haardt2012}
{Haardt} F.,  {Madau} P.,  2012, \mn@doi [\apj] {10.1088/0004-637X/746/2/125},
  \href {https://ui.adsabs.harvard.edu/abs/2012ApJ...746..125H} {746, 125}

\bibitem[\protect\citeauthoryear{{Hiss}, {Walther}, {Hennawi}, {O{\~n}orbe},
  {O'Meara}, {Rorai}  \& {Luki{\'c}}}{{Hiss} et~al.}{2018}]{hiss2018}
{Hiss} H.,  {Walther} M.,  {Hennawi} J.~F.,  {O{\~n}orbe} J.,  {O'Meara} J.~M.,
   {Rorai} A.,   {Luki{\'c}} Z.,  2018, \mn@doi [\apj]
  {10.3847/1538-4357/aada86}, \href
  {https://ui.adsabs.harvard.edu/abs/2018ApJ...865...42H} {865, 42}

\bibitem[\protect\citeauthoryear{{Hiss}, {Walther}, {O{\~n}orbe}  \&
  {Hennawi}}{{Hiss} et~al.}{2019}]{hiss2019}
{Hiss} H.,  {Walther} M.,  {O{\~n}orbe} J.,   {Hennawi} J.~F.,  2019, \mn@doi
  [\apj] {10.3847/1538-4357/ab1418}, \href
  {https://ui.adsabs.harvard.edu/abs/2019ApJ...876...71H} {876, 71}

\bibitem[\protect\citeauthoryear{{Hui} \& {Gnedin}}{{Hui} \&
  {Gnedin}}{1997}]{hui1997}
{Hui} L.,  {Gnedin} N.~Y.,  1997, \mn@doi [\mnras] {10.1093/mnras/292.1.27},
  \href {https://ui.adsabs.harvard.edu/abs/1997MNRAS.292...27H} {292, 27}

\bibitem[\protect\citeauthoryear{{Ir{\v{s}}i{\v{c}}}
  et~al.,}{{Ir{\v{s}}i{\v{c}}} et~al.}{2017a}]{irsic2017a}
{Ir{\v{s}}i{\v{c}}} V.,  et~al., 2017a, \mn@doi [\prd]
  {10.1103/PhysRevD.96.023522}, \href
  {https://ui.adsabs.harvard.edu/abs/2017PhRvD..96b3522I} {96, 023522}

\bibitem[\protect\citeauthoryear{{Ir{\v{s}}i{\v{c}}}, {Viel}, {Haehnelt},
  {Bolton}  \& {Becker}}{{Ir{\v{s}}i{\v{c}}} et~al.}{2017b}]{irsic2017b}
{Ir{\v{s}}i{\v{c}}} V.,  {Viel} M.,  {Haehnelt} M.~G.,  {Bolton} J.~S.,
  {Becker} G.~D.,  2017b, \mn@doi [\prl] {10.1103/PhysRevLett.119.031302},
  \href {https://ui.adsabs.harvard.edu/abs/2017PhRvL.119c1302I} {119, 031302}

\bibitem[\protect\citeauthoryear{{Khaire}}{{Khaire}}{2017}]{khaire2017}
{Khaire} V.,  2017, \mn@doi [\mnras] {10.1093/mnras/stx1487}, \href
  {https://ui.adsabs.harvard.edu/abs/2017MNRAS.471..255K} {471, 255}

\bibitem[\protect\citeauthoryear{{Khaire} \& {Srianand}}{{Khaire} \&
  {Srianand}}{2015}]{khaire2015a}
{Khaire} V.,  {Srianand} R.,  2015, \mn@doi [\apj]
  {10.1088/0004-637X/805/1/33}, \href
  {https://ui.adsabs.harvard.edu/abs/2015ApJ...805...33K} {805, 33}

\bibitem[\protect\citeauthoryear{{Khaire} \& {Srianand}}{{Khaire} \&
  {Srianand}}{2019}]{khaire2019a}
{Khaire} V.,  {Srianand} R.,  2019, \mn@doi [\mnras] {10.1093/mnras/stz174},
  \href {https://ui.adsabs.harvard.edu/abs/2019MNRAS.484.4174K} {484, 4174}

\bibitem[\protect\citeauthoryear{{Khaire}, {Srianand}, {Choudhury}  \&
  {Gaikwad}}{{Khaire} et~al.}{2016}]{khaire2016}
{Khaire} V.,  {Srianand} R.,  {Choudhury} T.~R.,   {Gaikwad} P.,  2016, \mn@doi
  [\mnras] {10.1093/mnras/stw192}, \href
  {https://ui.adsabs.harvard.edu/abs/2016MNRAS.457.4051K} {457, 4051}

\bibitem[\protect\citeauthoryear{{Khaire} et~al.,}{{Khaire}
  et~al.}{2019}]{khaire2019b}
{Khaire} V.,  et~al., 2019, \mn@doi [\mnras] {10.1093/mnras/stz344}, \href
  {https://ui.adsabs.harvard.edu/abs/2019MNRAS.486..769K} {486, 769}

\bibitem[\protect\citeauthoryear{{Kim}, {Viel}, {Haehnelt}, {Carswell}  \&
  {Cristiani}}{{Kim} et~al.}{2004}]{kim2004}
{Kim} T.~S.,  {Viel} M.,  {Haehnelt} M.~G.,  {Carswell} R.~F.,   {Cristiani}
  S.,  2004, \mn@doi [\mnras] {10.1111/j.1365-2966.2004.07221.x}, \href
  {https://ui.adsabs.harvard.edu/abs/2004MNRAS.347..355K} {347, 355}

\bibitem[\protect\citeauthoryear{{Kim}, {Bolton}, {Viel}, {Haehnelt}  \&
  {Carswell}}{{Kim} et~al.}{2007}]{kim2007}
{Kim} T.~S.,  {Bolton} J.~S.,  {Viel} M.,  {Haehnelt} M.~G.,   {Carswell}
  R.~F.,  2007, \mn@doi [\mnras] {10.1111/j.1365-2966.2007.12406.x}, \href
  {https://ui.adsabs.harvard.edu/abs/2007MNRAS.382.1657K} {382, 1657}

\bibitem[\protect\citeauthoryear{{Kim}, {Partl}, {Carswell}  \&
  {M{\"u}ller}}{{Kim} et~al.}{2013}]{kim2013}
{Kim} T.~S.,  {Partl} A.~M.,  {Carswell} R.~F.,   {M{\"u}ller} V.,  2013,
  \mn@doi [\aap] {10.1051/0004-6361/201220042}, \href
  {https://ui.adsabs.harvard.edu/abs/2013A&amp;A...552A..77K} {552, A77}

\bibitem[\protect\citeauthoryear{{Kirkman} et~al.,}{{Kirkman}
  et~al.}{2005}]{kirkman2005}
{Kirkman} D.,  et~al., 2005, \mn@doi [\mnras]
  {10.1111/j.1365-2966.2005.09126.x}, \href
  {https://ui.adsabs.harvard.edu/abs/2005MNRAS.360.1373K} {360, 1373}

\bibitem[\protect\citeauthoryear{{Kulkarni} \& {Fall}}{{Kulkarni} \&
  {Fall}}{1993}]{kulkarni1993}
{Kulkarni} V.~P.,  {Fall} S.~M.,  1993, \mn@doi [\apjl] {10.1086/186960}, \href
  {https://ui.adsabs.harvard.edu/abs/1993ApJ...413L..63K} {413, L63}

\bibitem[\protect\citeauthoryear{{Kulkarni}, {Hennawi}, {O{\~n}orbe}, {Rorai}
  \& {Springel}}{{Kulkarni} et~al.}{2015}]{kulkarni2015}
{Kulkarni} G.,  {Hennawi} J.~F.,  {O{\~n}orbe} J.,  {Rorai} A.,   {Springel}
  V.,  2015, \mn@doi [\apj] {10.1088/0004-637X/812/1/30}, \href
  {https://ui.adsabs.harvard.edu/abs/2015ApJ...812...30K} {812, 30}

\bibitem[\protect\citeauthoryear{{Kulkarni}, {Worseck}  \&
  {Hennawi}}{{Kulkarni} et~al.}{2019}]{kulkarni2019b}
{Kulkarni} G.,  {Worseck} G.,   {Hennawi} J.~F.,  2019, \mn@doi [\mnras]
  {10.1093/mnras/stz1493}, \href
  {https://ui.adsabs.harvard.edu/abs/2019MNRAS.488.1035K} {488, 1035}

\bibitem[\protect\citeauthoryear{{Lidz}, {McQuinn}, {Zaldarriaga}, {Hernquist}
  \& {Dutta}}{{Lidz} et~al.}{2007}]{lidz2007}
{Lidz} A.,  {McQuinn} M.,  {Zaldarriaga} M.,  {Hernquist} L.,   {Dutta} S.,
  2007, \mn@doi [\apj] {10.1086/521974}, \href
  {https://ui.adsabs.harvard.edu/abs/2007ApJ...670...39L} {670, 39}

\bibitem[\protect\citeauthoryear{{Lidz}, {Faucher-Gigu{\`e}re}, {Dall'Aglio},
  {McQuinn}, {Fechner}, {Zaldarriaga}, {Hernquist}  \& {Dutta}}{{Lidz}
  et~al.}{2010}]{lidz2010}
{Lidz} A.,  {Faucher-Gigu{\`e}re} C.-A.,  {Dall'Aglio} A.,  {McQuinn} M.,
  {Fechner} C.,  {Zaldarriaga} M.,  {Hernquist} L.,   {Dutta} S.,  2010,
  \mn@doi [\apj] {10.1088/0004-637X/718/1/199}, \href
  {https://ui.adsabs.harvard.edu/abs/2010ApJ...718..199L} {718, 199}

\bibitem[\protect\citeauthoryear{{Liu}, {Qin}, {Ridgway}  \& {Slatyer}}{{Liu}
  et~al.}{2020}]{liu2020}
{Liu} H.,  {Qin} W.,  {Ridgway} G.~W.,   {Slatyer} T.~R.,  2020, arXiv
  e-prints, \href {https://ui.adsabs.harvard.edu/abs/2020arXiv200801084L} {p.
  arXiv:2008.01084}

\bibitem[\protect\citeauthoryear{{Luki{\'c}}, {Stark}, {Nugent}, {White},
  {Meiksin}  \& {Almgren}}{{Luki{\'c}} et~al.}{2015}]{lukic2015}
{Luki{\'c}} Z.,  {Stark} C.~W.,  {Nugent} P.,  {White} M.,  {Meiksin} A.~A.,
  {Almgren} A.,  2015, \mn@doi [\mnras] {10.1093/mnras/stu2377}, \href
  {https://ui.adsabs.harvard.edu/abs/2015MNRAS.446.3697L} {446, 3697}

\bibitem[\protect\citeauthoryear{{Lusso} et~al.,}{{Lusso}
  et~al.}{2014}]{lusso2014}
{Lusso} E.,  et~al., 2014, \mn@doi [\apj] {10.1088/0004-637X/784/2/176}, \href
  {https://ui.adsabs.harvard.edu/abs/2014ApJ...784..176L} {784, 176}

\bibitem[\protect\citeauthoryear{{Maitra}, {Srianand}, {Petitjean}, {Rahmani},
  {Gaikwad}, {Choudhury}  \& {Pichon}}{{Maitra} et~al.}{2019}]{maitra2019}
{Maitra} S.,  {Srianand} R.,  {Petitjean} P.,  {Rahmani} H.,  {Gaikwad} P.,
  {Choudhury} T.~R.,   {Pichon} C.,  2019, \mn@doi [\mnras]
  {10.1093/mnras/stz2828}, \href
  {https://ui.adsabs.harvard.edu/abs/2019MNRAS.tmp.2422M} {p.~2422}

\bibitem[\protect\citeauthoryear{{Maitra}, {Srianand}, {Gaikwad}, {Choudhury},
  {Paranjape}  \& {Petitjean}}{{Maitra} et~al.}{2020}]{maitra2020}
{Maitra} S.,  {Srianand} R.,  {Gaikwad} P.,  {Choudhury} T.~R.,  {Paranjape}
  A.,   {Petitjean} P.,  2020, arXiv e-prints, \href
  {https://ui.adsabs.harvard.edu/abs/2020arXiv200505346M} {p. arXiv:2005.05346}

\bibitem[\protect\citeauthoryear{{McDonald}}{{McDonald}}{2003}]{mcdonald2003}
{McDonald} P.,  2003, \mn@doi [\apj] {10.1086/345945}, \href
  {https://ui.adsabs.harvard.edu/abs/2003ApJ...585...34M} {585, 34}

\bibitem[\protect\citeauthoryear{{McDonald} \& {Miralda-Escud{\'e}}}{{McDonald}
  \& {Miralda-Escud{\'e}}}{2001}]{mcdonald2001b}
{McDonald} P.,  {Miralda-Escud{\'e}} J.,  2001, \mn@doi [\apjl]
  {10.1086/319123}, \href
  {https://ui.adsabs.harvard.edu/abs/2001ApJ...549L..11M} {549, L11}

\bibitem[\protect\citeauthoryear{{Miralda-Escud{\'e}} \&
  {Rees}}{{Miralda-Escud{\'e}} \& {Rees}}{1994}]{miralda1994}
{Miralda-Escud{\'e}} J.,  {Rees} M.~J.,  1994, \mn@doi [\mnras]
  {10.1093/mnras/266.2.343}, \href
  {https://ui.adsabs.harvard.edu/abs/1994MNRAS.266..343M} {266, 343}

\bibitem[\protect\citeauthoryear{{Murphy}, {Kacprzak}, {Savorgnan}  \&
  {Carswell}}{{Murphy} et~al.}{2019}]{murphy2019}
{Murphy} M.~T.,  {Kacprzak} G.~G.,  {Savorgnan} G. A.~D.,   {Carswell} R.~F.,
  2019, \mn@doi [\mnras] {10.1093/mnras/sty2834}, \href
  {https://ui.adsabs.harvard.edu/abs/2019MNRAS.482.3458M} {482, 3458}

\bibitem[\protect\citeauthoryear{{Nasir}, {Bolton}  \& {Becker}}{{Nasir}
  et~al.}{2016}]{nasir2016}
{Nasir} F.,  {Bolton} J.~S.,   {Becker} G.~D.,  2016, \mn@doi [\mnras]
  {10.1093/mnras/stw2147}, \href
  {https://ui.adsabs.harvard.edu/abs/2016MNRAS.463.2335N} {463, 2335}

\bibitem[\protect\citeauthoryear{{Nath} \& {Biermann}}{{Nath} \&
  {Biermann}}{1993}]{Nath1993}
{Nath} B.~B.,  {Biermann} P.~L.,  1993, \mn@doi [\mnras]
  {10.1093/mnras/265.1.241}, \href
  {https://ui.adsabs.harvard.edu/abs/1993MNRAS.265..241N} {265, 241}

\bibitem[\protect\citeauthoryear{{O{\~n}orbe}, {Hennawi}  \&
  {Luki{\'c}}}{{O{\~n}orbe} et~al.}{2017}]{onorbe2017}
{O{\~n}orbe} J.,  {Hennawi} J.~F.,   {Luki{\'c}} Z.,  2017, \mn@doi [\apj]
  {10.3847/1538-4357/aa6031}, \href
  {https://ui.adsabs.harvard.edu/abs/2017ApJ...837..106O} {837, 106}

\bibitem[\protect\citeauthoryear{{O{\~n}orbe}, {Davies}, {Luki{\'c}}, {},
  {Hennawi}  \& {Sorini}}{{O{\~n}orbe} et~al.}{2019}]{onorbe2019}
{O{\~n}orbe} J.,  {Davies} F.~B.,  {Luki{\'c}} {} Z.,  {Hennawi} J.~F.,
  {Sorini} D.,  2019, \mn@doi [\mnras] {10.1093/mnras/stz984}, \href
  {https://ui.adsabs.harvard.edu/abs/2019MNRAS.486.4075O} {486, 4075}

\bibitem[\protect\citeauthoryear{{O'Meara} et~al.,}{{O'Meara}
  et~al.}{2015}]{omeara2015}
{O'Meara} J.~M.,  et~al., 2015, \mn@doi [\aj] {10.1088/0004-6256/150/4/111},
  \href {https://ui.adsabs.harvard.edu/abs/2015AJ....150..111O} {150, 111}

\bibitem[\protect\citeauthoryear{{O'Meara}, {Lehner}, {Howk}, {Prochaska},
  {Fox}, {Peeples}, {Tumlinson}  \& {O'Shea}}{{O'Meara}
  et~al.}{2017}]{omeara2017}
{O'Meara} J.~M.,  {Lehner} N.,  {Howk} J.~C.,  {Prochaska} J.~X.,  {Fox} A.~J.,
   {Peeples} M.~S.,  {Tumlinson} J.,   {O'Shea} B.~W.,  2017, \mn@doi [\aj]
  {10.3847/1538-3881/aa82b8}, \href
  {https://ui.adsabs.harvard.edu/abs/2017AJ....154..114O} {154, 114}

\bibitem[\protect\citeauthoryear{{Padmanabhan}, {Choudhury}  \&
  {Srianand}}{{Padmanabhan} et~al.}{2014}]{hamsa2014}
{Padmanabhan} H.,  {Choudhury} T.~R.,   {Srianand} R.,  2014, \mn@doi [\mnras]
  {10.1093/mnras/stu1433}, \href
  {https://ui.adsabs.harvard.edu/abs/2014MNRAS.443.3761P} {443, 3761}

\bibitem[\protect\citeauthoryear{{Padmanabhan}, {Srianand}  \&
  {Choudhury}}{{Padmanabhan} et~al.}{2015}]{hamsa2015}
{Padmanabhan} H.,  {Srianand} R.,   {Choudhury} T.~R.,  2015, \mn@doi [\mnras]
  {10.1093/mnrasl/slv041}, \href
  {https://ui.adsabs.harvard.edu/abs/2015MNRAS.450L..29P} {450, L29}

\bibitem[\protect\citeauthoryear{{Park}, {Shapiro}, {Choi}, {Yoshida}, {Hirano}
   \& {Ahn}}{{Park} et~al.}{2016}]{park2016}
{Park} H.,  {Shapiro} P.~R.,  {Choi} J.-h.,  {Yoshida} N.,  {Hirano} S.,
  {Ahn} K.,  2016, \mn@doi [\apj] {10.3847/0004-637X/831/1/86}, \href
  {https://ui.adsabs.harvard.edu/abs/2016ApJ...831...86P} {831, 86}

\bibitem[\protect\citeauthoryear{{Peeples}, {Weinberg}, {Dav{\'e}}, {Fardal}
  \& {Katz}}{{Peeples} et~al.}{2010}]{peeples2010}
{Peeples} M.~S.,  {Weinberg} D.~H.,  {Dav{\'e}} R.,  {Fardal} M.~A.,   {Katz}
  N.,  2010, \mn@doi [\mnras] {10.1111/j.1365-2966.2010.16383.x}, \href
  {https://ui.adsabs.harvard.edu/abs/2010MNRAS.404.1281P} {404, 1281}

\bibitem[\protect\citeauthoryear{{Planck Collaboration} et~al.,}{{Planck
  Collaboration} et~al.}{2014}]{planck2014}
{Planck Collaboration} et~al., 2014, \mn@doi [\aap]
  {10.1051/0004-6361/201321591}, \href
  {https://ui.adsabs.harvard.edu/abs/2014A&amp;A...571A..16P} {571, A16}

\bibitem[\protect\citeauthoryear{{Press}, {Teukolsky}, {Vetterling}  \&
  {Flannery}}{{Press} et~al.}{1992}]{press1992}
{Press} W.~H.,  {Teukolsky} S.~A.,  {Vetterling} W.~T.,   {Flannery} B.~P.,
  1992, {Numerical recipes in FORTRAN. The art of scientific computing}

\bibitem[\protect\citeauthoryear{{Puchwein}, {Pfrommer}, {Springel},
  {Broderick}  \& {Chang}}{{Puchwein} et~al.}{2012}]{Puchwein2012}
{Puchwein} E.,  {Pfrommer} C.,  {Springel} V.,  {Broderick} A.~E.,   {Chang}
  P.,  2012, \mn@doi [\mnras] {10.1111/j.1365-2966.2012.20738.x}, \href
  {https://ui.adsabs.harvard.edu/abs/2012MNRAS.423..149P} {423, 149}

\bibitem[\protect\citeauthoryear{{Puchwein}, {Bolton}, {Haehnelt}, {Madau},
  {Becker}  \& {Haardt}}{{Puchwein} et~al.}{2015}]{puchwein2015}
{Puchwein} E.,  {Bolton} J.~S.,  {Haehnelt} M.~G.,  {Madau} P.,  {Becker}
  G.~D.,   {Haardt} F.,  2015, \mn@doi [\mnras] {10.1093/mnras/stv773}, \href
  {https://ui.adsabs.harvard.edu/abs/2015MNRAS.450.4081P} {450, 4081}

\bibitem[\protect\citeauthoryear{{Puchwein}, {Haardt}, {Haehnelt}  \&
  {Madau}}{{Puchwein} et~al.}{2019}]{puchwein2019}
{Puchwein} E.,  {Haardt} F.,  {Haehnelt} M.~G.,   {Madau} P.,  2019, \mn@doi
  [\mnras] {10.1093/mnras/stz222}, \href
  {https://ui.adsabs.harvard.edu/abs/2019MNRAS.485...47P} {485, 47}

\bibitem[\protect\citeauthoryear{{Rauch} et~al.,}{{Rauch}
  et~al.}{1997}]{rauch1997}
{Rauch} M.,  et~al., 1997, \mn@doi [\apj] {10.1086/304765}, \href
  {https://ui.adsabs.harvard.edu/abs/1997ApJ...489....7R} {489, 7}

\bibitem[\protect\citeauthoryear{{Rollinde}, {Theuns}, {Schaye}, {P{\^a}ris}
  \& {Petitjean}}{{Rollinde} et~al.}{2013}]{rollinde2013}
{Rollinde} E.,  {Theuns} T.,  {Schaye} J.,  {P{\^a}ris} I.,   {Petitjean} P.,
  2013, \mn@doi [\mnras] {10.1093/mnras/sts057}, \href
  {https://ui.adsabs.harvard.edu/abs/2013MNRAS.428..540R} {428, 540}

\bibitem[\protect\citeauthoryear{{Rorai} et~al.,}{{Rorai}
  et~al.}{2017a}]{rorai2017}
{Rorai} A.,  et~al., 2017a, \mn@doi [Science] {10.1126/science.aaf9346}, \href
  {https://ui.adsabs.harvard.edu/abs/2017Sci...356..418R} {356, 418}

\bibitem[\protect\citeauthoryear{{Rorai} et~al.,}{{Rorai}
  et~al.}{2017b}]{rorai2017b}
{Rorai} A.,  et~al., 2017b, \mn@doi [\mnras] {10.1093/mnras/stw2917}, \href
  {https://ui.adsabs.harvard.edu/abs/2017MNRAS.466.2690R} {466, 2690}

\bibitem[\protect\citeauthoryear{{Rorai}, {Carswell}, {Haehnelt}, {Becker},
  {Bolton}  \& {Murphy}}{{Rorai} et~al.}{2018}]{rorai2018}
{Rorai} A.,  {Carswell} R.~F.,  {Haehnelt} M.~G.,  {Becker} G.~D.,  {Bolton}
  J.~S.,   {Murphy} M.~T.,  2018, \mn@doi [\mnras] {10.1093/mnras/stx2862},
  \href {https://ui.adsabs.harvard.edu/abs/2018MNRAS.474.2871R} {474, 2871}

\bibitem[\protect\citeauthoryear{{Samui}, {Subramanian}  \& {Srianand
  }}{{Samui} et~al.}{2005}]{samui2005}
{Samui} S.,  {Subramanian} K.,   {Srianand } R.,  2005, in 29th International
  Cosmic Ray Conference (ICRC29), Volume 9. p.~215 (\mn@eprint {arXiv}
  {astro-ph/0505590})

\bibitem[\protect\citeauthoryear{{Schaye}}{{Schaye}}{2001}]{schaye2001}
{Schaye} J.,  2001, \mn@doi [\apj] {10.1086/322421}, \href
  {https://ui.adsabs.harvard.edu/abs/2001ApJ...559..507S} {559, 507}

\bibitem[\protect\citeauthoryear{{Schaye}, {Theuns}, {Leonard}  \&
  {Efstathiou}}{{Schaye} et~al.}{1999}]{schaye1999}
{Schaye} J.,  {Theuns} T.,  {Leonard} A.,   {Efstathiou} G.,  1999, \mn@doi
  [\mnras] {10.1046/j.1365-8711.1999.02956.x}, \href
  {https://ui.adsabs.harvard.edu/abs/1999MNRAS.310...57S} {310, 57}

\bibitem[\protect\citeauthoryear{{Schaye}, {Theuns}, {Rauch}, {Efstathiou}  \&
  {Sargent}}{{Schaye} et~al.}{2000}]{schaye2000}
{Schaye} J.,  {Theuns} T.,  {Rauch} M.,  {Efstathiou} G.,   {Sargent} W. L.~W.,
   2000, \mn@doi [\mnras] {10.1046/j.1365-8711.2000.03815.x}, \href
  {https://ui.adsabs.harvard.edu/abs/2000MNRAS.318..817S} {318, 817}

\bibitem[\protect\citeauthoryear{{Scoccimarro}, {Hui}, {Manera}  \&
  {Chan}}{{Scoccimarro} et~al.}{2012}]{2lpt2012}
{Scoccimarro} R.,  {Hui} L.,  {Manera} M.,   {Chan} K.~C.,  2012, \mn@doi
  [\prd] {10.1103/PhysRevD.85.083002}, \href
  {https://ui.adsabs.harvard.edu/abs/2012PhRvD..85h3002S} {85, 083002}

\bibitem[\protect\citeauthoryear{{Shull} \& {Danforth}}{{Shull} \&
  {Danforth}}{2020}]{shull2020}
{Shull} M.,  {Danforth} C.,  2020, arXiv e-prints, \href
  {https://ui.adsabs.harvard.edu/abs/2020arXiv200702948S} {p. arXiv:2007.02948}

\bibitem[\protect\citeauthoryear{{Springel}}{{Springel}}{2005}]{springel2005}
{Springel} V.,  2005, \mn@doi [\mnras] {10.1111/j.1365-2966.2005.09655.x},
  \href {https://ui.adsabs.harvard.edu/abs/2005MNRAS.364.1105S} {364, 1105}

\bibitem[\protect\citeauthoryear{{Srianand} \& {Khare}}{{Srianand} \&
  {Khare}}{1996}]{Srianand1996}
{Srianand} R.,  {Khare} P.,  1996, \mn@doi [\mnras] {10.1093/mnras/280.3.767},
  \href {https://ui.adsabs.harvard.edu/abs/1996MNRAS.280..767S} {280, 767}

\bibitem[\protect\citeauthoryear{{Stevans}, {Shull}, {Danforth}  \&
  {Tilton}}{{Stevans} et~al.}{2014}]{stevans2014}
{Stevans} M.~L.,  {Shull} J.~M.,  {Danforth} C.~W.,   {Tilton} E.~M.,  2014,
  \mn@doi [\apj] {10.1088/0004-637X/794/1/75}, \href
  {https://ui.adsabs.harvard.edu/abs/2014ApJ...794...75S} {794, 75}

\bibitem[\protect\citeauthoryear{{Telikova}, {Shternin}  \&
  {Balashev}}{{Telikova} et~al.}{2019}]{telikova2019}
{Telikova} K.~N.,  {Shternin} P.~S.,   {Balashev} S.~A.,  2019, \mn@doi [\apj]
  {10.3847/1538-4357/ab52fe}, \href
  {https://ui.adsabs.harvard.edu/abs/2019ApJ...887..205T} {887, 205}

\bibitem[\protect\citeauthoryear{{Theuns}, {Schaye}  \& {Haehnelt}}{{Theuns}
  et~al.}{2000}]{theuns2000a}
{Theuns} T.,  {Schaye} J.,   {Haehnelt} M.~G.,  2000, \mn@doi [\mnras]
  {10.1046/j.1365-8711.2000.03423.x}, \href
  {https://ui.adsabs.harvard.edu/abs/2000MNRAS.315..600T} {315, 600}

\bibitem[\protect\citeauthoryear{{Theuns}, {Zaroubi}, {Kim}, {Tzanavaris}  \&
  {Carswell}}{{Theuns} et~al.}{2002a}]{theuns2002b}
{Theuns} T.,  {Zaroubi} S.,  {Kim} T.-S.,  {Tzanavaris} P.,   {Carswell} R.~F.,
   2002a, \mn@doi [\mnras] {10.1046/j.1365-8711.2002.05316.x}, \href
  {https://ui.adsabs.harvard.edu/abs/2002MNRAS.332..367T} {332, 367}

\bibitem[\protect\citeauthoryear{{Theuns}, {Schaye}, {Zaroubi}, {Kim},
  {Tzanavaris}  \& {Carswell}}{{Theuns} et~al.}{2002b}]{theuns2002c}
{Theuns} T.,  {Schaye} J.,  {Zaroubi} S.,  {Kim} T.-S.,  {Tzanavaris} P.,
  {Carswell} B.,  2002b, \mn@doi [\apjl] {10.1086/339998}, \href
  {https://ui.adsabs.harvard.edu/abs/2002ApJ...567L.103T} {567, L103}

\bibitem[\protect\citeauthoryear{{Upton Sanderbeck} \& {Bird}}{{Upton
  Sanderbeck} \& {Bird}}{2020}]{sanderbeck2020}
{Upton Sanderbeck} P.,  {Bird} S.,  2020, arXiv e-prints, \href
  {https://ui.adsabs.harvard.edu/abs/2020arXiv200205733U} {p. arXiv:2002.05733}

\bibitem[\protect\citeauthoryear{{Upton Sanderbeck}, {D'Aloisio}  \&
  {McQuinn}}{{Upton Sanderbeck} et~al.}{2016}]{sanderbeck2016}
{Upton Sanderbeck} P.~R.,  {D'Aloisio} A.,   {McQuinn} M.~J.,  2016, \mn@doi
  [\mnras] {10.1093/mnras/stw1117}, \href
  {https://ui.adsabs.harvard.edu/abs/2016MNRAS.460.1885U} {460, 1885}

\bibitem[\protect\citeauthoryear{{Viel}, {Haehnelt}  \& {Springel}}{{Viel}
  et~al.}{2004a}]{viel2004a}
{Viel} M.,  {Haehnelt} M.~G.,   {Springel} V.,  2004a, \mn@doi [\mnras]
  {10.1111/j.1365-2966.2004.08224.x}, \href
  {https://ui.adsabs.harvard.edu/abs/2004MNRAS.354..684V} {354, 684}

\bibitem[\protect\citeauthoryear{{Viel}, {Weller}  \& {Haehnelt}}{{Viel}
  et~al.}{2004b}]{viel2004b}
{Viel} M.,  {Weller} J.,   {Haehnelt} M.~G.,  2004b, \mn@doi [\mnras]
  {10.1111/j.1365-2966.2004.08498.x}, \href
  {https://ui.adsabs.harvard.edu/abs/2004MNRAS.355L..23V} {355, L23}

\bibitem[\protect\citeauthoryear{{Viel}, {Haehnelt}, {Bolton}, {Kim},
  {Puchwein}, {Nasir}  \& {Wakker}}{{Viel} et~al.}{2017}]{viel2016}
{Viel} M.,  {Haehnelt} M.~G.,  {Bolton} J.~S.,  {Kim} T.-S.,  {Puchwein} E.,
  {Nasir} F.,   {Wakker} B.~P.,  2017, \mn@doi [\mnras]
  {10.1093/mnrasl/slx004}, \href
  {https://ui.adsabs.harvard.edu/abs/2017MNRAS.467L..86V} {467, L86}

\bibitem[\protect\citeauthoryear{{Walther}, {Hennawi}, {Hiss}, {O{\~n}orbe},
  {Lee}, {Rorai}  \& {O'Meara}}{{Walther} et~al.}{2018}]{walther2018}
{Walther} M.,  {Hennawi} J.~F.,  {Hiss} H.,  {O{\~n}orbe} J.,  {Lee} K.-G.,
  {Rorai} A.,   {O'Meara} J.,  2018, \mn@doi [\apj] {10.3847/1538-4357/aa9c81},
  \href {https://ui.adsabs.harvard.edu/abs/2018ApJ...852...22W} {852, 22}

\bibitem[\protect\citeauthoryear{{Walther}, {O{\~n}orbe}, {Hennawi}  \&
  {Luki{\'c}}}{{Walther} et~al.}{2019}]{walther2019}
{Walther} M.,  {O{\~n}orbe} J.,  {Hennawi} J.~F.,   {Luki{\'c}} Z.,  2019,
  \mn@doi [\apj] {10.3847/1538-4357/aafad1}, \href
  {https://ui.adsabs.harvard.edu/abs/2019ApJ...872...13W} {872, 13}

\bibitem[\protect\citeauthoryear{{Webb} \& {Carswell}}{{Webb} \&
  {Carswell}}{1991}]{webb1991}
{Webb} J.~K.,  {Carswell} R.~F.,  1991, in {Shaver} P.~A.,  {Wampler} E.~J.,
  {Wolfe} A.~M.,  eds, Quasar Absorption Lines. p.~3

\bibitem[\protect\citeauthoryear{{Worseck} et~al.,}{{Worseck}
  et~al.}{2011}]{worseck2011}
{Worseck} G.,  et~al., 2011, \mn@doi [\apjl] {10.1088/2041-8205/733/2/L24},
  \href {https://ui.adsabs.harvard.edu/abs/2011ApJ...733L..24W} {733, L24}

\bibitem[\protect\citeauthoryear{{Worseck}, {Prochaska}, {Hennawi}  \&
  {McQuinn}}{{Worseck} et~al.}{2016}]{worseck2016}
{Worseck} G.,  {Prochaska} J.~X.,  {Hennawi} J.~F.,   {McQuinn} M.,  2016,
  \mn@doi [\apj] {10.3847/0004-637X/825/2/144}, \href
  {https://ui.adsabs.harvard.edu/abs/2016ApJ...825..144W} {825, 144}

\bibitem[\protect\citeauthoryear{{Worseck}, {Davies}, {Hennawi}  \&
  {Prochaska}}{{Worseck} et~al.}{2019}]{worseck2019}
{Worseck} G.,  {Davies} F.~B.,  {Hennawi} J.~F.,   {Prochaska} J.~X.,  2019,
  \mn@doi [\apj] {10.3847/1538-4357/ab0fa1}, \href
  {https://ui.adsabs.harvard.edu/abs/2019ApJ...875..111W} {875, 111}

\bibitem[\protect\citeauthoryear{{Zaldarriaga}}{{Zaldarriaga}}{2002}]{zaldarriaga2002}
{Zaldarriaga} M.,  2002, \mn@doi [\apj] {10.1086/324212}, \href
  {https://ui.adsabs.harvard.edu/abs/2002ApJ...564..153Z} {564, 153}

\makeatother
\end{thebibliography}

\appendix


\section{Coaddition of observed spectra}
\label{app:spectra-coaddition}
\InputFigCombine{Coadd_Spectra_Example.pdf}{170}%
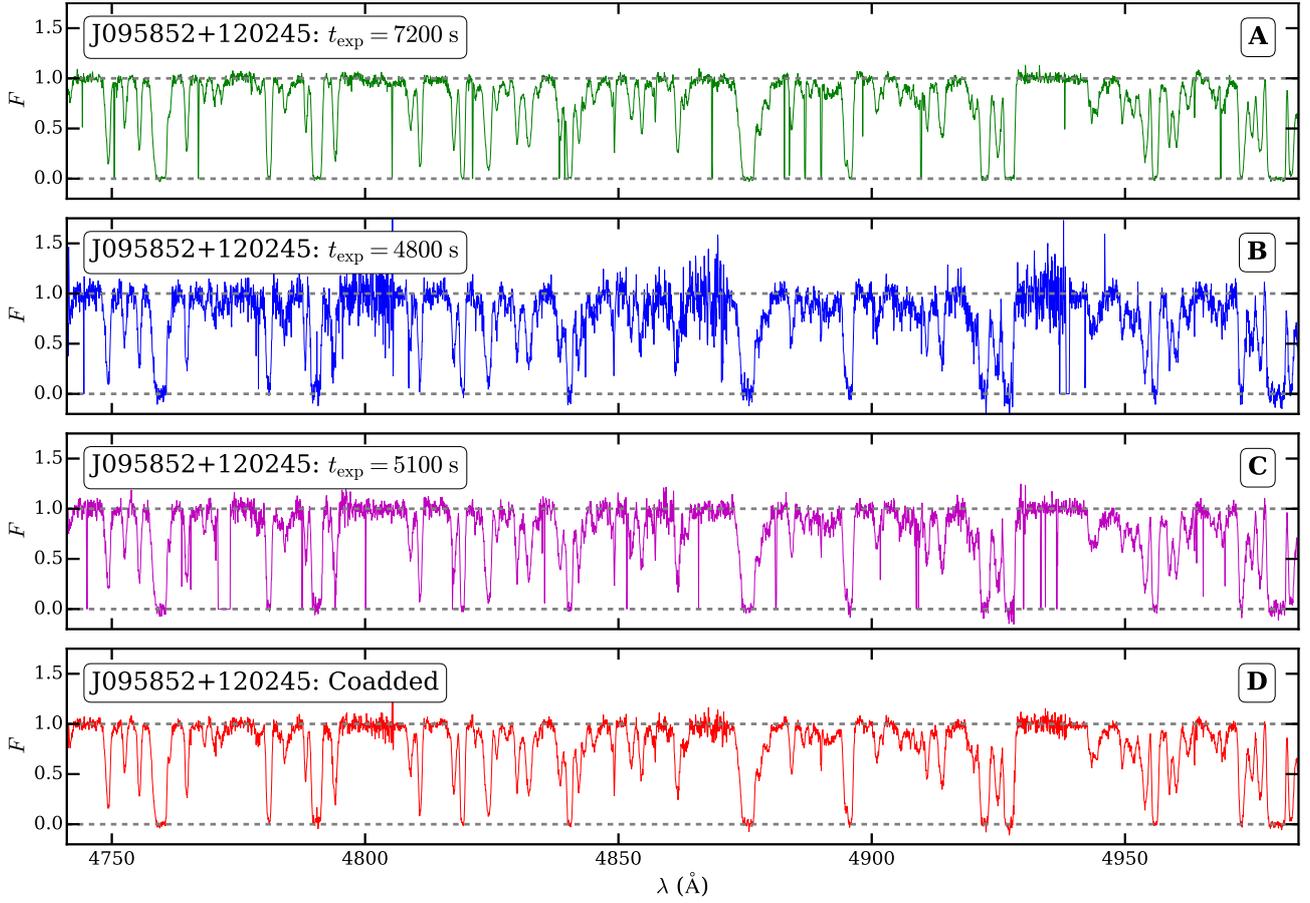
{ Panel A, B and C show the observed spectrum of QSO J095852+120245 at three different epochs having exposure times of 7200
    s, 4800 s and 5100 s, respectively.  Many pixels in the individual
    exposures are bad pixels due to observational systematics.  Panel D shows
    the coadded spectrum obtained by our method as discussed in online supplementary 
    appendix
    \ref{app:spectra-coaddition}. Our coaddition procedure takes care of bad pixels
    due to observational systematics.
}{\label{fig:coadd-spectra-example}}
Most of the QSOs in the \kodiaqdr sample were observed more than once with different exposure times.
In this section, we discuss our procedure to coadd the spectra.
Let a QSO be  observed $N$ times with exposure times
$t_1$,$t_2$,$\cdots$,$t_N$.  Let $F_{i}(\lambda)$, $\sigma_{i}(\lambda)$ be the
value of the normalized flux and associated error at wavelength $\lambda$ in
the $i^{\rm th}$ observation.  The coadded flux $F_{\rm coadd}(\lambda)$ and 
coadded variance $\sigma^2_{\rm coadd}$ of the  flux is given by,
\begin{equation}
\begin{aligned}
F_{\rm coadd}(\lambda) &= \frac{\sum \limits_{i=1}^{N} \; t_i \: w_i \: F_i(\lambda)}{\sum \limits_{i=1}^{N} \; t_i \: w_i} \\
\sigma^2_{\rm coadd}(\lambda) &= \frac{\sum \limits_{i=1}^{N} \; t_i \: w_i \: \sigma^2_i(\lambda)}{\sum \limits_{i=1}^{N} \; t_i \: w_i}
\end{aligned}
\end{equation}
where $w_i$ is weight for the $i^{\rm th}$ pixel. $w_i = 0$ for bad pixels and
$w_i=1$ for good pixels.  For the \kodiaqdr sample, we define a pixel as a bad pixel 
if it satisfies any of the following criteria, 
(i) pixels with negative error i.e., $\sigma_{i}(\lambda) \leq 0$, 
(ii) pixels with $\sigma_{i}(\lambda) \gg 1$ (i.e., SNR per pixel $\ll 1$) 
(iii) $F_{i}(\lambda) = 0$ (due to sky background $F_i(\lambda) \neq 0$ for good pixel), 
(iv)  negative outliers i.e., pixels with $F_{i}(\lambda) \leq -3 \sigma_i(\lambda)$ and 
(v) positive outliers i.e., pixels with $F_i(\lambda) \geq 1 + 3\sigma_i(\lambda)$, 

\figref{fig:coadd-spectra-example} shows an example of coadded spectrum obtained using our method.
Spectrum obtained in individual exposures show many bad pixels due to observational
systematics. Our coaddition method accounts for these effects as shown in panel D
of \figref{fig:coadd-spectra-example}.

In our analysis, we also exclude sightlines that contain large spectral gaps.  We decide whether a sightline contains a large spectral gap or not in
the following way.  
For a given emission redshift of QSO, we identify redshift bins
that are spanned by QSO absorption spectra excluding proximity regions and
\lyb, O~{\sc vi} emission lines.
The choice of QSO proximity region size corresponds to $\sim 1166$  \AA~at $z=3$. To avoid possible contamination from intrinsic O~{\sc iv} absorption we consider only regions with rest wavelength greater than 1050\AA\ at the quasar emission redshift.
If the spectral gap is more than 25 percent of the redshift bin we mark it as a large gap and exclude the sightline from the analysis for that redshift bin.
We note that the number of such
sightlines are small $(\leq  5)$.


\section{Excess mean flux \Fmean at $z=3.2$}
\label{appendix:excess-mean-flux}
\InputFigCombine{FMEAN_Scatter_FPDF_Observations.pdf}{175}%
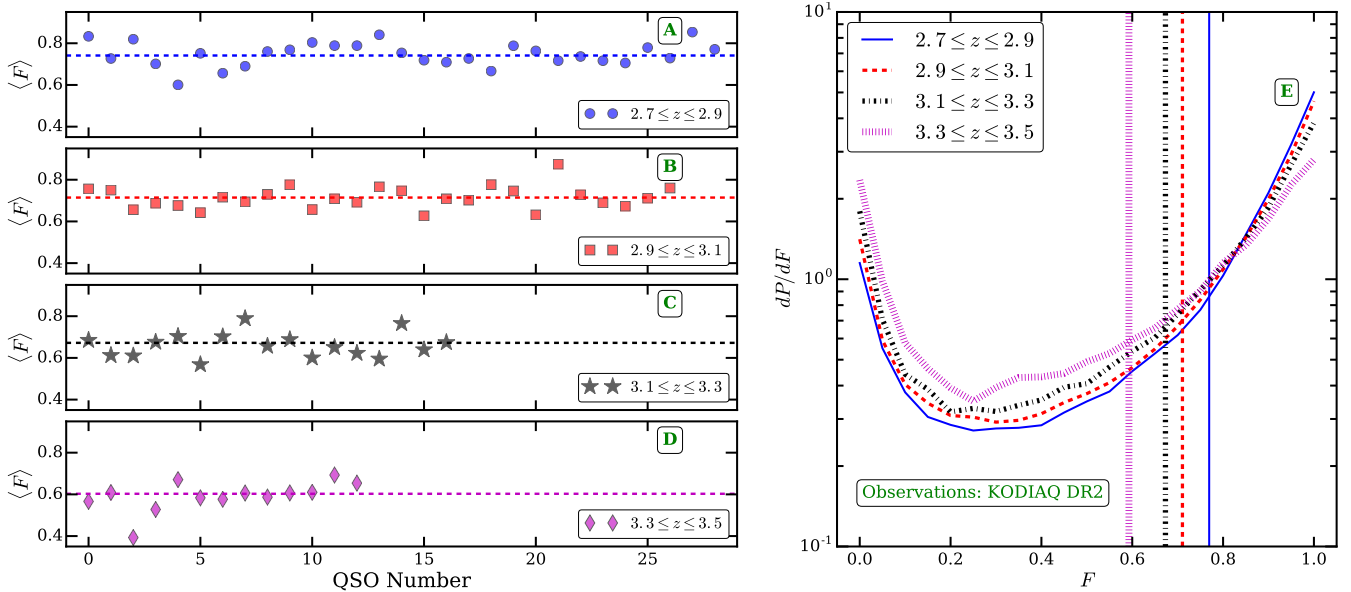
{ Panels A to D show the observed mean flux along individual QSO
sightlines in different redshift bins. The dashed line in each panel shows the
mean flux of the sample at the given redshift. We exclude the possibility of
larger mean flux at \zrange{3.1}{3.2} due to any outliers. Panel 
E shows the observed FPDF in four redshift bins. The vertical 
lines in panel E show the observed mean flux in the corresponding redshift bins.
Similar to panel A-D, the FPDF comparison in panel E does not show any unusual
evolution.
}{\label{fig:excess-mean-flux-test}}
In this section we test if the observed excess in mean flux at
\zrange{3.1}{3.3} in the \kodiaqdr sample is due to the presence of any unusual
QSO sightlines.  To check if there are any outliers in this particular redshift
bin, we plot the mean flux along individual sightlines in
\figref{fig:excess-mean-flux-test} for four redshift bins.
\figref{fig:excess-mean-flux-test} illustrates that the distribution of mean
flux along sightlines is random around the mean flux (dashed line in panel A to
D) of the entire sample for all redshift bins. We also do not see any outlier
at \zrange{3.1}{3.3}. Since the mean flux along a sightline is an average
quantity, we also check if there is any significant deviation in the observed
FPDF evolution in panel E of \figref{fig:excess-mean-flux-test}. The observed
FPDF at \zrange{3.1}{3.3} also does not show any unusual deviation from other
redshift bins. The excess in \Fmean at \zrange{3.1}{3.3} may thus be real in
the \kodiaqdr sample.  Another reason for the excess \Fmean could be that the
number of sightlines at \zrange{3.1}{3.3} are limited in our sample. Our preliminary analysis using UVES \squaddr \citep[][]{murphy2019} while showing distinct change in the slope at z$\sim3.2$ does not show any prominent excess.
We plan to
investigate this further in our future work. However,
we emphasize that the effect of excess in \Fmean at \zrange{3.1}{3.3} has a
marginal effect on our measurements of thermal parameters.  This is because for
any thermal parameter variation we rescale the optical depth in our simulations
to match the mean flux.  Thus, even if there are any systematics in the \Fmean
evolution in the \kodiaqdr sample, this does not significantly change the main
results of this work.

\section{Comparison of observed flux statistics errors}
\label{app:obs-flux-stat-error}
\InputFigCombine{FPDF_FPS_CDDF_Literature_Comparison_With_Errors.pdf}{175}%
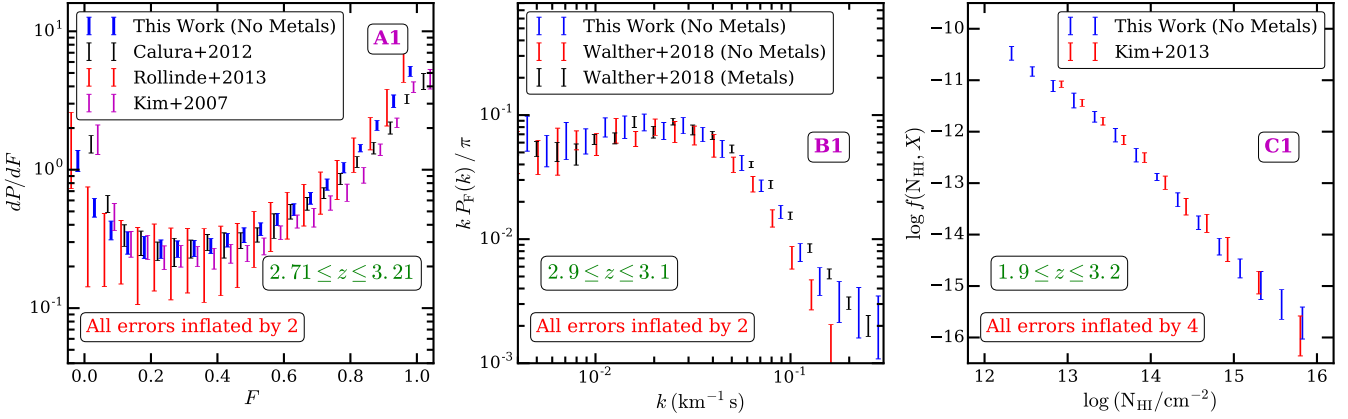
{Each panel is same as the corresponding panel in 
    \figref{fig:stat-comparison-literature}
    except that the errors of the observed FPDF, FPS and CDDF statistics
    are shown. For visual purposes, the errors of the FPDF, FPS  and CDDF
    statistics are inflated by a factor of 2, 2 and 4 respectively. For a fair
    comparison, the errors from our sample are calculated using the bootstrap
    method. In general the errors of the FPDF, FPS and CDDF statistics from our
    sample are smaller because of the large sample size.  
    The errors of the FPDF in \citet{rollinde2013} are
    computed from simulated mock spectra and hence are larger than that of the 
    other measurements.  The errors of the FPS statistics from this work are 
    similar to those in \citet[][for the no metal case]{walther2018} due to 
    similar sample size and data quality. The errors of our CDDF
    at \logNHI$<13.6$ are larger due to the somewhat smaller \SNR of the 
    spectra in our
    sample.  At \logNHI$>13.6$, the errors of our CDDF are smaller than
    that in \citet{kim2013}.
}{\label{fig:stat-comparison-literature-error}}
\InputFigCombine{FPS_Literature_Comparison_With_Errors_2.pdf}{175}%
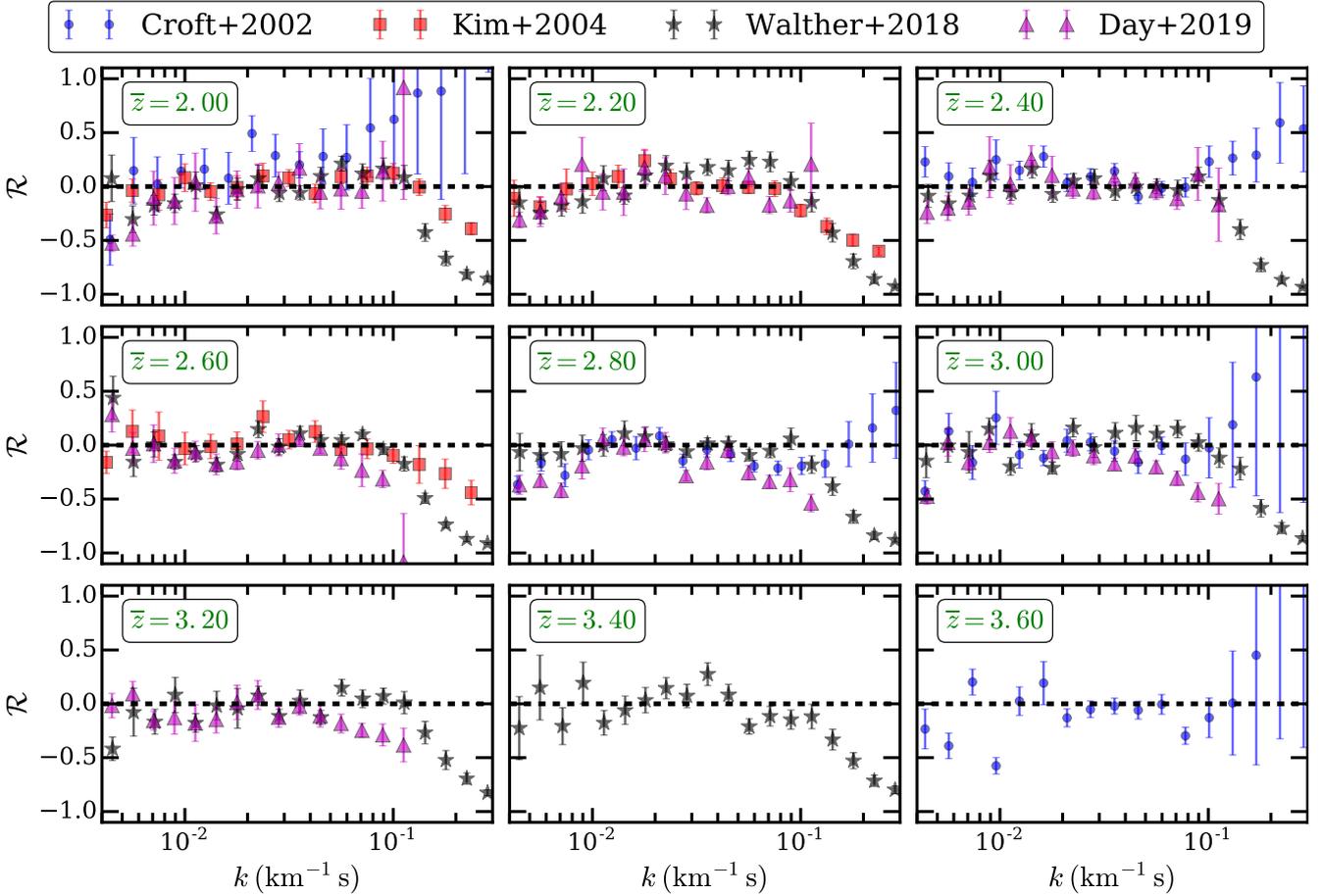
{Each panel shows residuals between observed FPS calculated in this work and
corresponding observed FPS in the literature 
\citep{croft2002,kim2004,walther2018,day2019}. The observed number of spectra,
redshift path length, quality of spectra and methods of computing flux power 
spectrum are different in each work. Even with such differences the measured FPS
from this work is generally in good agreement with the measurements in the literature. 
}{\label{fig:fps-comparison-literature-residual}}

In \S \ref{subsec:stat-comparison} we discuss the consistency of observed flux
statistics from this work with that in the literature. In this section, we compare
the errors of the observed FPDF, FPS and CDDF from this work with  
measurements in the literature. 
The errors for observed statistics are normally determined using the
bootstrap method \citep[except in][]{rollinde2013}. For a fair comparison, we
also compute the errors using the bootstrap method.
\figref{fig:stat-comparison-literature-error} shows a comparison of the errors for
different statistics.  In general, the errors in different statistics from this 
work are either consistent or slightly smaller than 
those in the literature.

Panel A1 in \figref{fig:stat-comparison-literature-error} shows that the error
in our observed FPDF is smaller than that in \citet{kim2007,calura2012,rollinde2013}.
The smaller errors are due to the large number of QSOs in our sample. 
The large error in the FPDF of \citet{rollinde2013} is due to the difference in their
method of calculating the errors. \citet{rollinde2013} argue that the errors
computed using the bootstrap method are underestimated and may not account for 
the cosmic variance accurately. Hence they derive the errors from large number
of simulated mocks. Furthermore, their sample size was limited to $<5$ at the
redshift of interest. We have calculated the errors on our other observed 
statistics using both methods and accounted for cosmic variance (see \S \ref{app:statistics-error-estimation} for details).

Panel B1 in \figref{fig:stat-comparison-literature-error} shows a comparison 
of the error of the FPS from this work with that from \citet{walther2018}. 
The errors (for the no metal case) are similar in the two works. This is expected
because the number of QSOs per redshift bin are similar in both cases.
Similarly, the data quality is the same because the subsets of QSOs are 
both drawn from the \kodiaqdr survey. 
The consistency in FPS and its error of this work with that
from \citet{walther2018} reflects the consistency of both the data and our 
method of calculating the FPS. This consistency is important because the FPS is 
one of the statistics used in this work to measure thermal parameters.

In \figref{fig:fps-comparison-literature-residual}, we compare the observed
FPS measurements from this work with those from
\citet{croft2002,kim2004,walther2018,day2019}.
We find that our FPS
measurements are generally in good agreement (maximum deviation is $\sim 1.8\sigma$) 
with those from the literature at $k \leq 0.1 \: {\rm s\: km^{-1}}$. The large differences
at $k>0.1~ {\rm s\: km^{-1}}$ are due to differences in the method of computing the 
FPS. In particular \citet{kim2004,walther2018} subtract power due to white noise 
in the spectra, whereas we do not subtract the noise power. Furthermore the 
number of observed spectra, quality of spectra (\SNR) and total observed 
redshift  path length is quite different in \citet{croft2002,kim2004} from that in our analysis.
Even with such differences the FPS in these work seem to be in good agreement with our measurements.
We also like to emphasize that any systematic differences in the method of computing
the FPS does not significantly affect the measurements of thermal parameters. This 
is because we use the same method to compute the FPS statistics  in simulations and
observations.

In panel C1 of \figref{fig:stat-comparison-literature-error}, we compare the
errors of our CDDF with that from \citet{kim2013}. \citet{kim2013} measured
the CDDF from 18 high resolution high \SNR (usually $>50$) \lya forest spectra.
On the other hand we measure the CDDF from a sample of $\sim 35$ high resolution
but moderate \SNR spectra. Thus the data size and data quality is different
in both the works. These differences of the observed dataset are reflected in 
the CDDF errors. For example the errors in our CDDF at \logNHI$<13.6$ is larger 
than that from \citet{kim2013}. This is because low \NHI systems are
affected more by the \SNR of the spectra. Since \SNR of our sample is smaller
than in \citet{kim2013}, our CDDF error at small \logNHI is correspondingly large.
At high \logNHI $>13.6$, \SNR is no longer an issue since lines are detected 
with relatively high significance level. The errors in our CDDF at \logNHI$>13.6$
are smaller than that from \citet{kim2013} because the number of spectra in 
our sample are nearly twice that in \citet{kim2013}. Even though there 
are differences in data quality and sample size, the CDDF and the associated 
errors from the two works are in good agreement with each other. The
agreement between the CDDF demonstrates the consistency of \viper in fitting the
\lya forest with Voigt profiles. This consistency is important because we use 
the $b$ parameter of Voigt profile fits to constrain thermal parameters.
In summary, the observed statistics with associated errors from this work are 
in good agreement with that from the literature.

\section{Variation in thermal parameters}
\label{appendix:thermal-parameter-variation}
\InputFigCombine{Thermal_history_in_simulation.pdf}{160}%
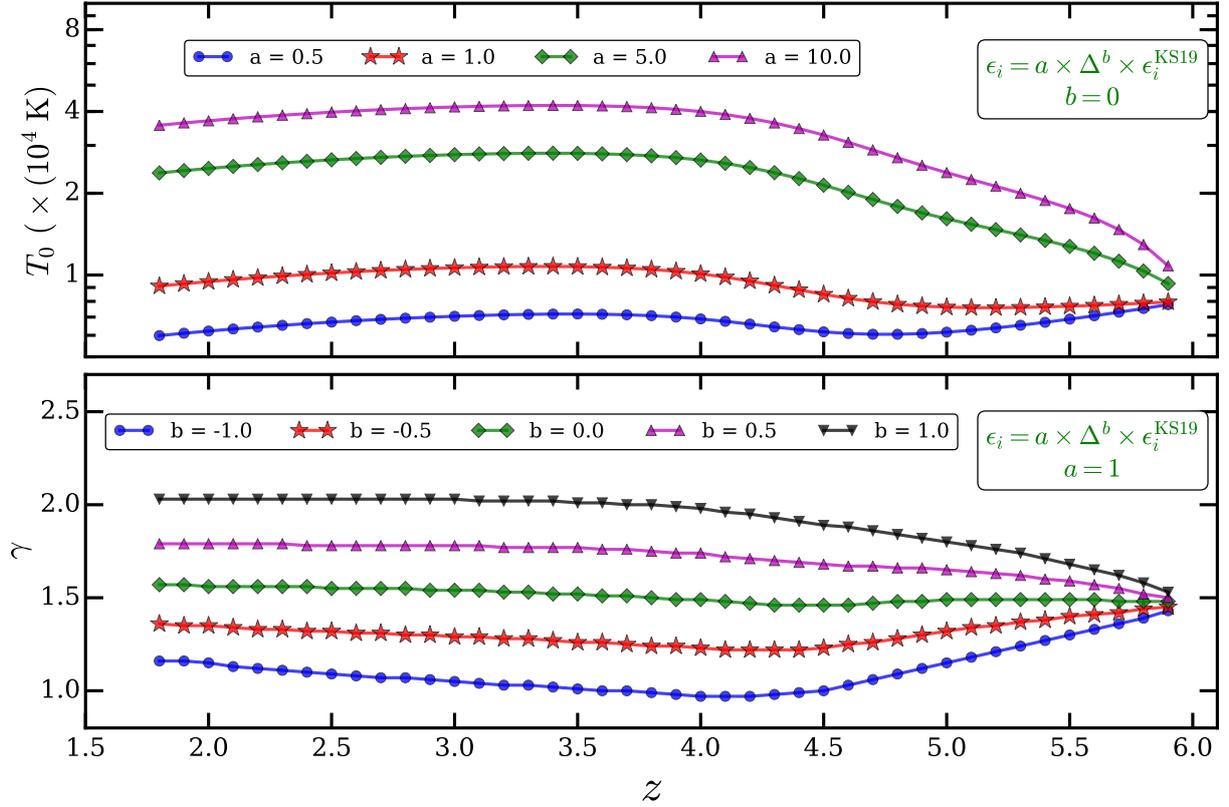
{The top and bottom panels show the variation in thermal parameters $T_0$ and $\gamma$
     obtained by rescaling the \HI, \HeI and \HeII photo-heating rates
    of the  KS19 UVB with our post-processing module \citecode.  A simple constant scaling of
    photo-heating rates (by factor $a$) leads to variation in $T_0$ while
    $\gamma$ changes relatively little. A density dependent scaling
    (by a factor $b$) leads to a variation in $\gamma$ for similar values of $T_0$.
    \citecode evolves the thermal and ionization state of the IGM by using the
    outputs of a \gthree simulation. A self-consistent \gthree simulation has
    been performed with the KS19 UVB (the curves with $a=1$ and $b=0$.). For
    generating the variation in thermal parameters, we assume ionization
    equilibrium  (but no thermal equilibrium) in  \citecode.
    We generate 999 different thermal parameter evolution histories.
}{\label{fig:thermal-history-variation}}
In this section, we explain our method of generating different thermal histories for a 
given UVB model.
We vary the thermal parameter evolution by modifying the \HI, \HeI and \HeII
photo-heating rates of the \citetalias{khaire2019a} UVB. Following the approach of \citet{becker2011}, we scale
the photo-heating rates ($\epsilon_{\rm i}$) by two free parameters $a,b$ such that $\epsilon_{\rm i} =
a \: \Delta^b \: \epsilon_{\rm i}^{\rm KS19}$ where ${\rm i} \equiv $ [\HI, \HeI,
\HeII]. A change in factor $a$ ($b$) leads to variation in $T_0$ ($\gamma$)
while keeping the evolution of $\gamma$ ($T_0$) relatively unchanged. 
It is important to note that the variation in thermal parameters obtained
by scaling the photo-heating rate is not physical. Ideally, one would like to vary
the properties of the ionizing sources such as, luminosity function, QSO SED index, redshift of
reionization etc., and generate the UVB model from a cosmological 1D radiative transfer code. 
However, such an approach cannot produce a large variation in $T_0$ and $\gamma$
\citep[but see][]{onorbe2019}. Our approach, even though not physical, is
useful in practice to probe a large range of the parameters $T_0$ and $\gamma$
on a finely sampled grid.
\figref{fig:thermal-history-variation} shows examples of the evolution in $T_0$ and
$\gamma$ obtained by varying $a$ and $b$ for several thermal histories.  The case $a=1$ and $b=0$ shows the $T_0$ and $\gamma$ evolution obtained using the default
photo-heating rates from the \citetalias{khaire2019a} UVB.  Note that the shock
heated gas ($T>10^5$ K) is mostly unaffected by the photo-heating rate
scaling,  while the low density gas responds to this scaling differently
and changes  normalization ($T_0$) and slope ($\gamma$) of the TDR.

When measuring the thermal parameters, it is important to sample the
$T_0-\gamma$ plane densely enough.  A coarse sampling of $T_0-\gamma$ plane can
lead to incorrect best fit values and excessive smoothing of the $\chi^2$ field
which, in turn, can result in underestimation of the uncertainties of the
thermal parameters. On the other hand a very dense sampling of $T_0-\gamma$
would be computationally very expensive.  We found it to be the best compromise
to choose the   $T_0-\gamma$ grids such that $T_0$ is varied from $\sim 6000$ K
to $\sim 24000$ K  in steps of 500 K while $\gamma$ is varied from $\sim 0.7$
to $\sim 2.0$ in steps of 0.05 at $z=3$\footnote{The corresponding $T_0$ and
$\gamma$ at other redshifts differ by $\sim 7$ percent}. We have thus simulated
the \lya forest for $37 \times 27 = 999$ different thermal histories.  The
computational time required to run the thermal histories and extract mock
spectra  for 999 UVB models is $\sim 2.25$ million cpu hours.


\section{The sensitivity of \lya forest statistics to $T_0$, $\gamma$}
\label{app:statistics-sensitivity}





\subsection{Kernel Density Estimation}
\label{app:kde}
\InputFigCombine{BPDF_KDE_Illustration.pdf}{170}%
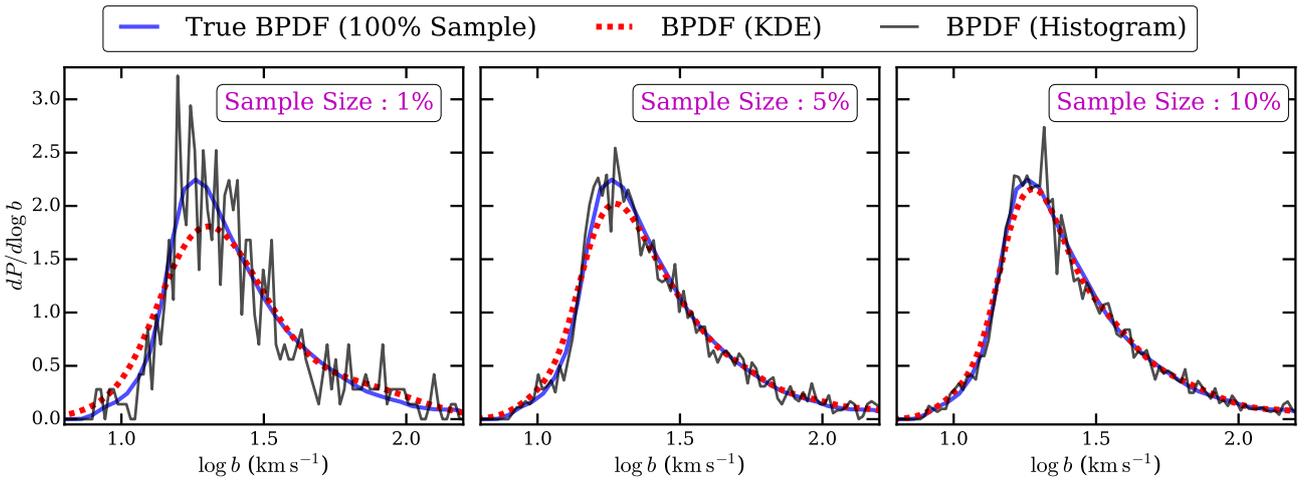
{ The figure illustrates the convergence of the BPDF using the histogram method (black
    solid curve) and kernel density estimation (KDE, red dashed line). The true
    BPDF (blue solid line) is calculated from Voigt profile fits to 500
    sight-lines at $2.9 \leq z \leq 3.1$.  The left, middle and right panel show the
    BPDF calculated using the histogram method and KDE for 1\%, 5\% and 10\% of sightlines.
    The PDF calculated using KDE converges faster with a smaller sample than
    the histogram method. For the KDE method, we use a Gaussian kernel with bandwidth 
    given by  Scott's rule. The choice of bins and bin width is the same for all
   cases.
}{\label{fig:kde-illustration}}
Kernel density estimation is a  non-parametric method to estimate the probability
distribution function. KDEs are a generalization of histograms with the advantage that
the loss of information due to binning of the data is minimal  with the  KDE method.
Unlike  histograms,  KDE is not sensitive to the choice of bins.
The PDF calculated using the KDE method converges faster than the 
simple histogram method.

We use KDE to estimate the PDF of wavelet amplitudes, curvature and $b$
parameters. There are 3 main reasons for using KDE instead of histograms
(i) the number of observed sightlines per redshift bins are usually limited
(ii) the observed spectra have finite \SNR and (iii) characterizing the shape
of the PDF is important as it is sensitive to thermal and ionization
parameters.

When calculating PDFs with  KDE, we chose a Gaussian function as a kernel. The
bandwidth ($h$) of the kernel is set by Scott's rule, $h = 3.5 \: \sigma \:
n^{-1/3}$, where $\sigma$ is the variance of the quantity whose PDF is to be
estimated and $n$ is number of samples.  \figref{fig:kde-illustration}
illustrates the advantage of using KDE over histograms to estimate the $b$
distribution (BPDF) from one of our models.  First we generate a true BPDF by
fitting Voigt profiles to 500 sightlines. We then randomly chose a fraction of
sightlines and estimate the BPDF using histograms  and KDE.  The left, middle
and right panel in \figref{fig:kde-illustration} show the comparison of the
BPDF using the histogram and KDE methods for 1\%, 5\% and 10\% of the total
sample.  \figref{fig:kde-illustration} clearly demonstrates that with KDE the
estimated BPDF reaches a good approximation of the  the true BPDF with a
smaller  sample than with the histogram method. PDFs estimated using KDE are
smoother and  converge faster than PDFs estimated with the  histogram
method\footnote{The bin center and bin width is fixed in
\figref{fig:kde-illustration}.  One can use coarse binning with the histogram
method to estimate PDFs. However, the coarse binning can  lead to loss of
information with regard to the  detailed shape of the PDF.}.


\section{Resolution and Convergence Tests}
\label{app:resolution-study}
The size of the simulation box used in this work (L10N512) is 10 Mpc/h and is
likely to be smaller than the typical size of \HeIII bubbles. As discussed in \S
\ref{sec:simulation} the choice of our simulation box is motivated by the
trade-off between dynamic range needed to resolve the \lya forest at
\zrange{2}{4} and the large thermal parameter space to probe.
However, it is important to demonstrate the sufficient convergence of our
simulation by comparing the properties of the \lya forest from this work with
that from a range of simulations in which box sizes and particle numbers are
varied.  We illustrate the convergence of our simulation in
\figref{fig:resolution-study-resolution} to \ref{fig:resolution-study-box-size-high-z}. 
We compare the FPDF, FPS, curvature PDF,
wavelet PDF, BPDF and CDDF statistics from our model with that from the
Sherwood simulation suite \citet{bolton2017} at \zrange{2.9}{3.1}.  For a fair
comparison we chose  thermal parameter $T_0,\gamma$ consistent with that from
the Sherwood simulations. For the models shown in
\figref{fig:resolution-study-resolution} and \ref{fig:resolution-study-box-size}, 
we post-process the simulated spectra to match
the observed properties of our sample at \zrange{2.9}{3.1}. Note that the
initial conditions of the density field used for the L10N512 simulation in this
work is different from those used in the \sherwood simulation suite.  When we
rescale the optical depth to match the mean flux, we find that the amount of
rescaling required is very similar in all the models.
\figref{fig:resolution-study-resolution} to \ref{fig:resolution-study-box-size-high-z} 
shows a good agreement between our model
statistics with that from the L10N512 Sherwood simulation suggesting that our
method of varying the thermal history is  consistent with the self-consistent
\gthree simulation.  
The slight mismatch between our \lya forest statistics
and that from the \sherwood simulation suite are primarily due to \citecode
and not due to differences in initial conditions or the UVB models used. 
In \citet{gaikwad2018}, we have 
shown this explictly by comparing the \lya forest statistics from spectra 
post-processed with \citecode  and those from self-consistent simulations for
the same initial condition and UVB model.
However, these differences are smaller than the observational uncertainties and are 
compensated by the ability of our method to probe the large thermal parameter 
space which  would be
computationally expensive for simulations with larger dynamic range/particle
numbers.
\figref{fig:resolution-study-box-size}, \ref{fig:resolution-study-box-size-low-z} and
\ref{fig:resolution-study-box-size-high-z} also illustrate that the statistics are
converged with regards to box sizes. We compare the statistics from the L10N512,
L20N512, L40N1024 and L80N2048 Sherwood simulations. The good agreement
among the Sherwood models L40N512, L40N1024 and L40N2048 (Sherwood) indicates
that the statistics are converged with regard to resolution/ particle mass.  

%

\InputFigCombine{Resolution_Study_Sherwood_With_G3_CITE_With_Residual_Intermediate_z_2.pdf}{170}%
{
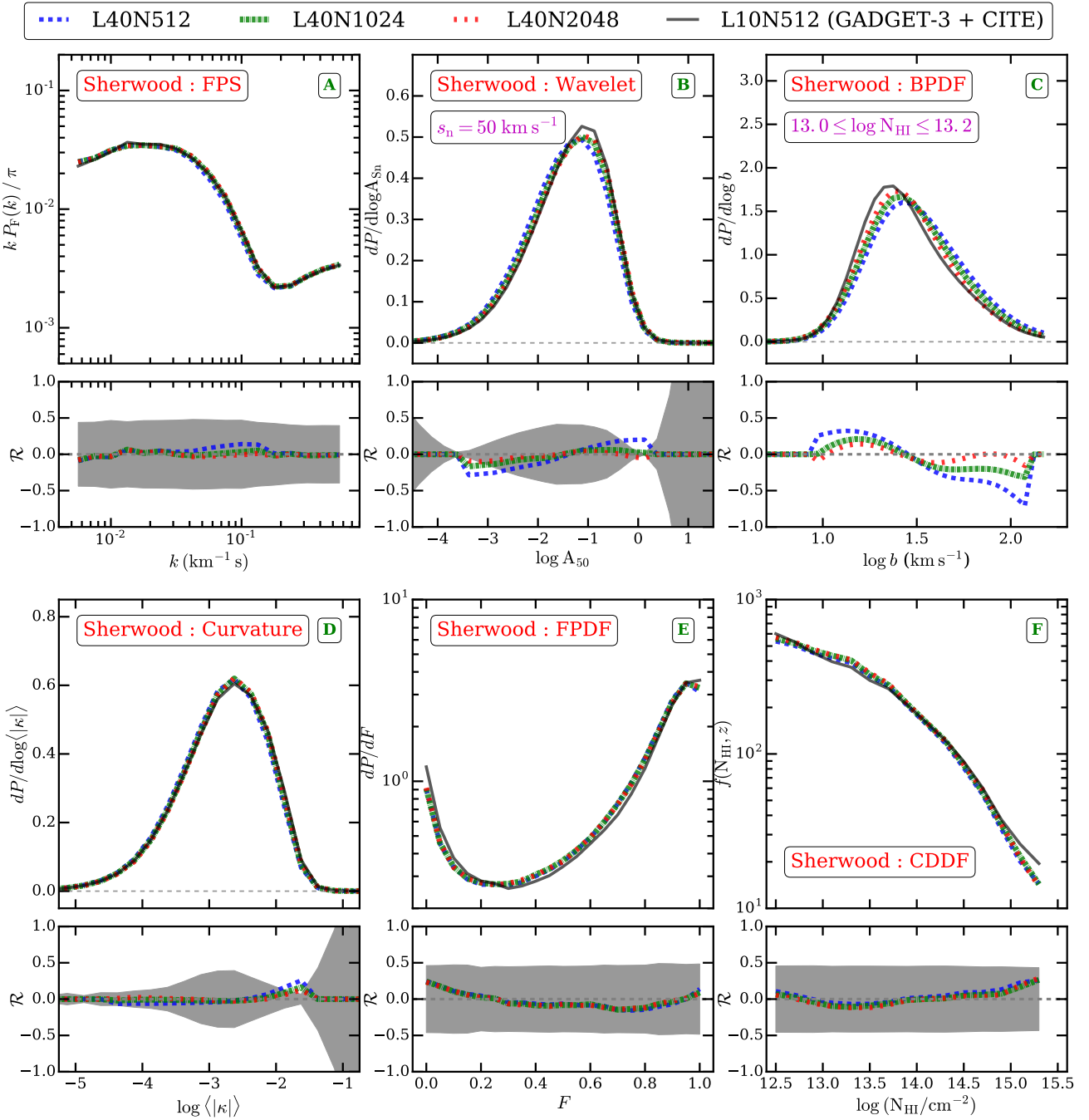
Here we show the effect of resolution on \lya statistics by comparing
these statistics for a range of simulations from the Sherwood simulation suite.
All other parameters such as $T_0$, $\gamma$, $\Gamma_{\rm HI}$ and box size
($40 \: h^{-1}$ cMpc) are the same for all the models.  The flux  statistics of
our default simulation are sufficiently  converged with regard to  resolution.
For comparison, we also show the flux statistics using our \gthree + \citecode
method. The flux statistics calculated for our post-processed simulation is
consistent with that for the corresponding self-consistent simulation from the
Sherwood simulation suite. The difference between the \gthree  + \citecode model
and the self-consistent \sherwood simulation suite are within the observational
uncertainty for each statistics (see \figref{fig:T0-gamma-best-fit-model}) All
statistics are shown at \zrange{2.9}{3.1}. The shaded region shows the 
typical observational uncertainty for each statistics.
}{\label{fig:resolution-study-resolution}}

\InputFigCombine{Resolution_Study_Sherwood_With_G3_CITE_With_Residual_Intermediate_z_4.pdf}{170}%
{
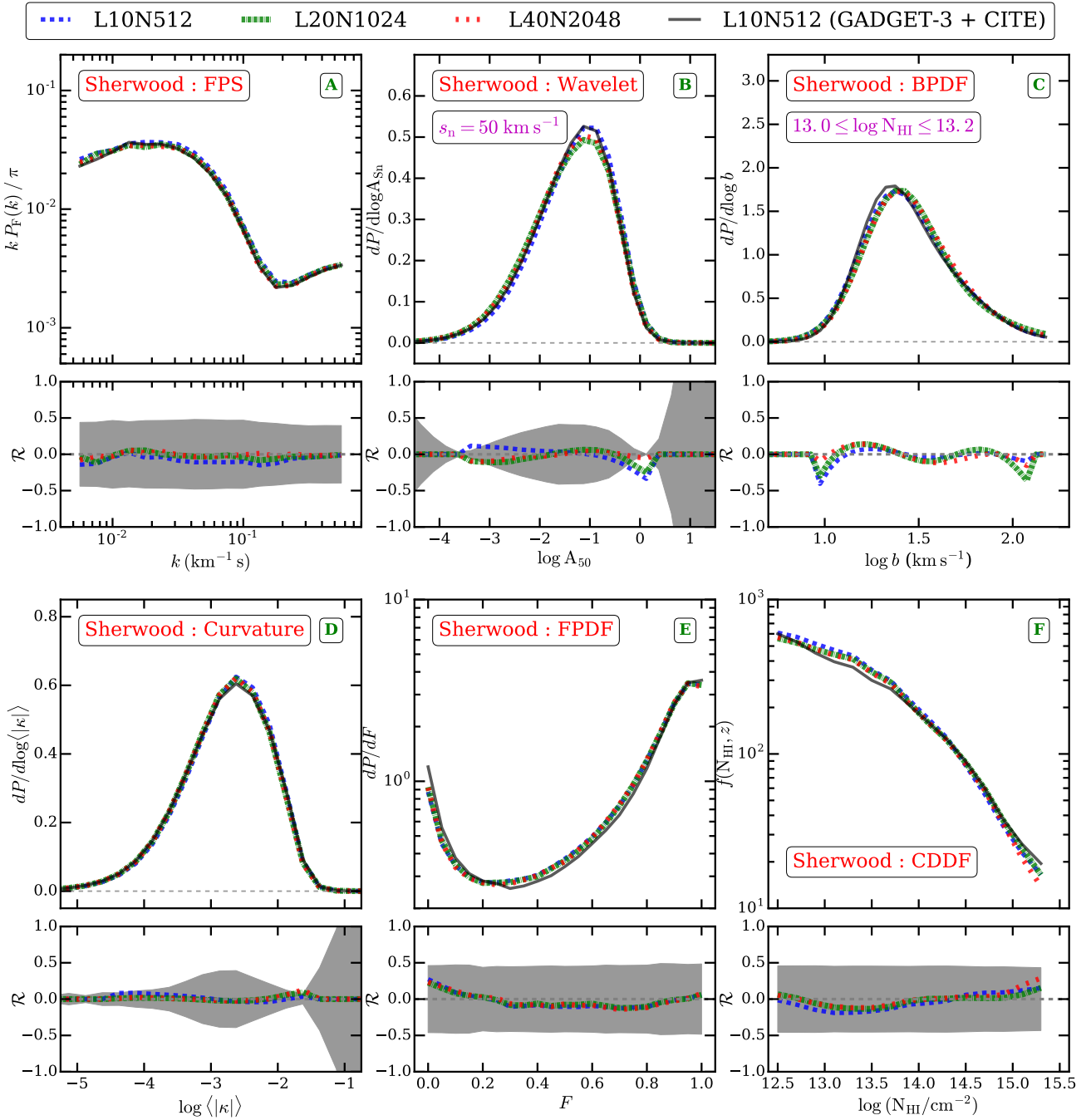
Same as \figref{fig:resolution-study-resolution} except that the effect of
box size (resolution is same) on \lya statistics is shown by comparing these
statistics for a range of simulations from the Sherwood simulation suite.  The
flux  statistics of our default simulation are sufficiently converged with
regard to box size.  For comparison, we also show the flux statistics using our
\gthree + \citecode method. In this figure the resolution of the \sherwood model
is same as that of the \gthree + \citecode model.
}{\label{fig:resolution-study-box-size}}

\InputFigCombine{Resolution_Study_Sherwood_With_G3_CITE_With_Residual_Low_z_2.pdf}{170}%
{
Same as \figref{fig:resolution-study-resolution} except  for \zrange{1.9}{2.1}.
}{\label{fig:resolution-study-resolution-low-z}}

\InputFigCombine{Resolution_Study_Sherwood_With_G3_CITE_With_Residual_Low_z_4.pdf}{170}%
{
Same as \figref{fig:resolution-study-box-size} except for \zrange{1.9}{2.1}. 
}{\label{fig:resolution-study-box-size-low-z}}

\InputFigCombine{Resolution_Study_Sherwood_With_G3_CITE_With_Residual_High_z_2.pdf}{170}%
{
Same as \figref{fig:resolution-study-resolution} except for \zrange{3.7}{3.9}.
}{\label{fig:resolution-study-resolution-high-z}}

\InputFigCombine{Resolution_Study_Sherwood_With_G3_CITE_With_Residual_High_z_4.pdf}{170}%
{
Same as \figref{fig:resolution-study-box-size} except for \zrange{3.7}{3.9}. 
}{\label{fig:resolution-study-box-size-high-z}}

\section{Comparison of best fit statistics with observations}
\label{app:best-fit-comparison}


\begin{table*}
\centering
\caption{Reduced $\chi^2$ ($\chi^2$ per degree of freedom) between data and best fit $T_0$ and $\gamma$ measurements for various statistics.}
\begin{threeparttable}
\begin{tabular}{cccccc}
\hline  \hline 
Redshift  & $b$ distribution\tnote{[a]}  & Flux power spectrum  & Wavelet PDF\tnote{[b]} & Curvature PDF & Joint Analysis \\ \hline
2.0 $\pm$ 0.1 & 1.45 & 1.15 & 1.70 & 1.51 & 1.50 \\ 
2.2 $\pm$ 0.1 & 1.51 & 1.33 & 1.80 & 1.43 & 1.57 \\
2.4 $\pm$ 0.1 & 1.85 & 1.75 & 1.17 & 1.64 & 1.34 \\
2.6 $\pm$ 0.1 & 1.02 & 0.85 & 1.94 & 1.78 & 1.40 \\
2.8 $\pm$ 0.1 & 1.77 & 1.61 & 1.01 & 0.45 & 1.21 \\
3.0 $\pm$ 0.1 & 1.35 & 0.71 & 1.29 & 0.92 & 1.10 \\
3.2 $\pm$ 0.1 & 1.53 & 0.87 & 1.23 & 1.29 & 1.24 \\
3.4 $\pm$ 0.1 & 1.45 & 0.71 & 1.45 & 0.69 & 1.12 \\
3.6 $\pm$ 0.1 & 1.61 & 0.60 & 0.54 & 0.82 & 0.94 \\
3.8 $\pm$ 0.1 & 1.72 & 0.73 & 0.72 & 1.23 & 1.14 \\
\hline \hline
\end{tabular}
\begin{tablenotes}
\item[a] The reduced $\chi^2$ is calculated by adding the $\chi^2$  between model and data $b$ distribution in different \logNHI bins.
\item[a] The reduced $\chi^2$ is calculated by adding the $\chi^2$ 
between model and data wavelet PDF for different wavelet scales ($s_{\rm n}$).
\end{tablenotes}
\end{threeparttable}
\\
\label{tab:chi-sq-dof}
\end{table*}
%

\InputFigCombine{{Observation_Simulation_Best_Fit_Model_With_Residuals_z-2.0}.pdf}{170}%
{ Same as \figref{fig:T0-gamma-best-fit-model} except for \zrange{1.9}{2.1}.
}{\label{fig:best-fit-stat-2-0}}

\InputFigCombine{{Observation_Simulation_Best_Fit_Model_With_Residuals_z-2.2}.pdf}{170}%
{ Same as \figref{fig:T0-gamma-best-fit-model} except for \zrange{2.1}{2.3}.
}{\label{fig:best-fit-stat-2-2}}

\InputFigCombine{{Observation_Simulation_Best_Fit_Model_With_Residuals_z-2.4}.pdf}{170}%
{ Same as \figref{fig:T0-gamma-best-fit-model} except for \zrange{2.3}{2.5}.
}{\label{fig:best-fit-stat-2-4}}

\InputFigCombine{{Observation_Simulation_Best_Fit_Model_With_Residuals_z-2.6}.pdf}{170}%
{ Same as \figref{fig:T0-gamma-best-fit-model} except for \zrange{2.5}{2.7}.
}{\label{fig:best-fit-stat-2-6}}

\InputFigCombine{{Observation_Simulation_Best_Fit_Model_With_Residuals_z-2.8}.pdf}{170}%
{ Same as \figref{fig:T0-gamma-best-fit-model} except for \zrange{2.7}{2.9}.
}{\label{fig:best-fit-stat-2-8}}

\InputFigCombine{{Observation_Simulation_Best_Fit_Model_With_Residuals_z-3.2}.pdf}{170}%
{ Same as \figref{fig:T0-gamma-best-fit-model} except for \zrange{3.1}{3.3}.
}{\label{fig:best-fit-stat-3-2}}

\InputFigCombine{{Observation_Simulation_Best_Fit_Model_With_Residuals_z-3.4}.pdf}{170}%
{ Same as \figref{fig:T0-gamma-best-fit-model} except for \zrange{3.3}{3.5}.
}{\label{fig:best-fit-stat-3-4}}

\InputFigCombine{{Observation_Simulation_Best_Fit_Model_With_Residuals_z-3.6}.pdf}{170}%
{ Same as \figref{fig:T0-gamma-best-fit-model} except for \zrange{3.5}{3.7}.
}{\label{fig:best-fit-stat-3-6}}

\InputFigCombine{{Observation_Simulation_Best_Fit_Model_With_Residuals_z-3.8}.pdf}{170}%
{ Same as \figref{fig:T0-gamma-best-fit-model} except for \zrange{3.7}{3.9}.
}{\label{fig:best-fit-stat-3-8}}


\InputFigCombine{b_vs_NHI_scatter.pdf}{170}
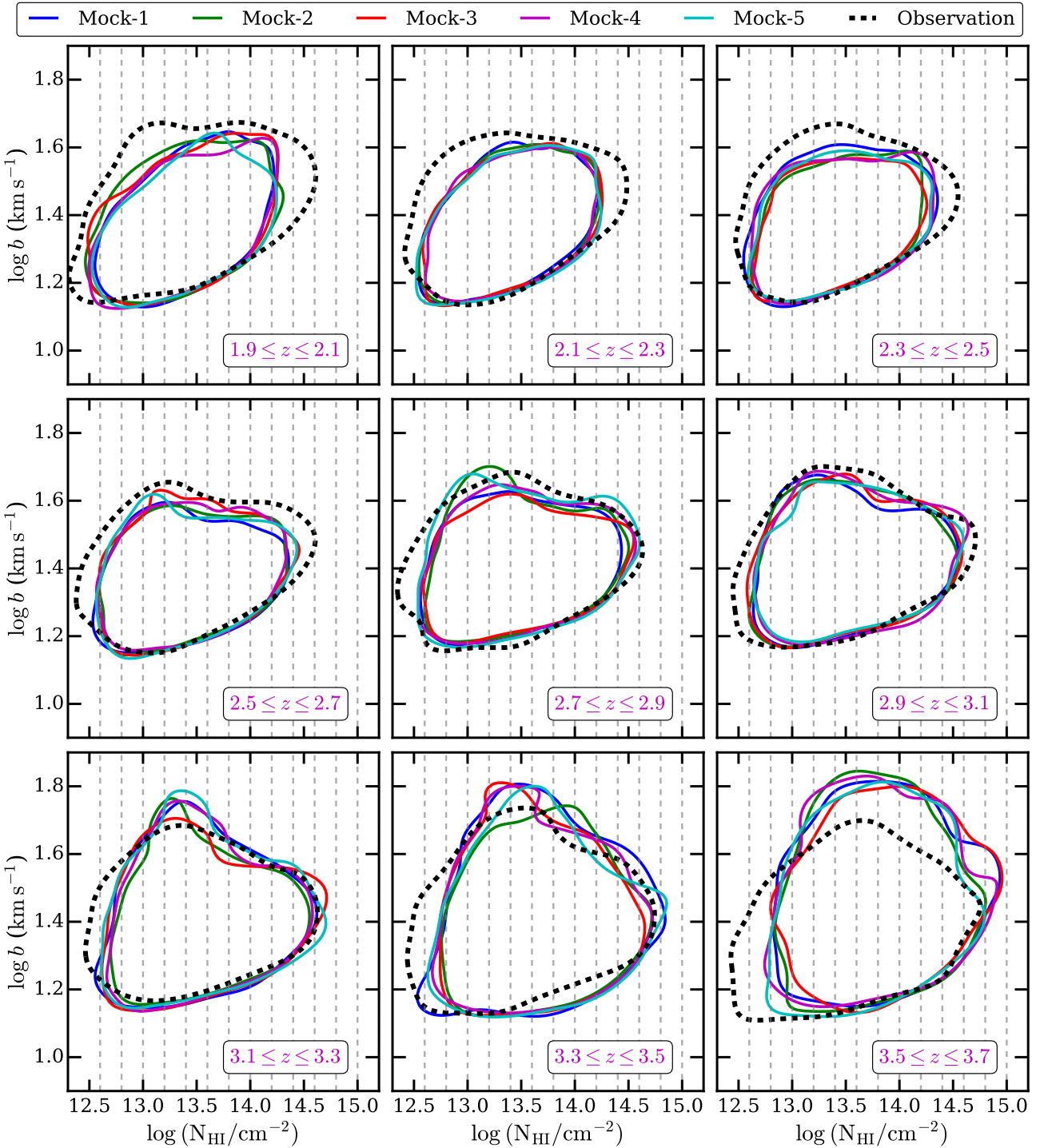
{Each panel shows the contour enclosing 90 percent of the points in
the \blogNHI plane. Black dashed contours shows the distribution of
\blogNHI points from observed samples in this work. The solid contours
(blue, green, red, magenta and cyan) show the regions of the \blogNHI distribution
for our mock samples.  Each mock sample matches the total observed redshift
path length in a given redshift bin.  For visual purposes, we show contours for 5
mock samples. However, we have computed $b$ distribution for 100 such mock
samples.  The thermal parameters of the mock samples in each redshift bin
correspond to the best fit thermal parameters obtained in this work. The \blogNHI
distributions from our mock samples is similar to the observed \blogNHI
distribution. The shape of the contours for lower $b$ values from the simulations
are in good agreement with that from observations.  Traditionally the lower
envelope of $b$ values is used to measure the thermal parameters.  The variations
in the \blogNHI distribution in different mock samples are small indicating that the
distribution is converged with respect to cosmic variance.
}{\label{fig:b-vs-NHI-scatter}}

\InputFig{Evolution_Temperature_at_Delta_c.pdf}{80}%
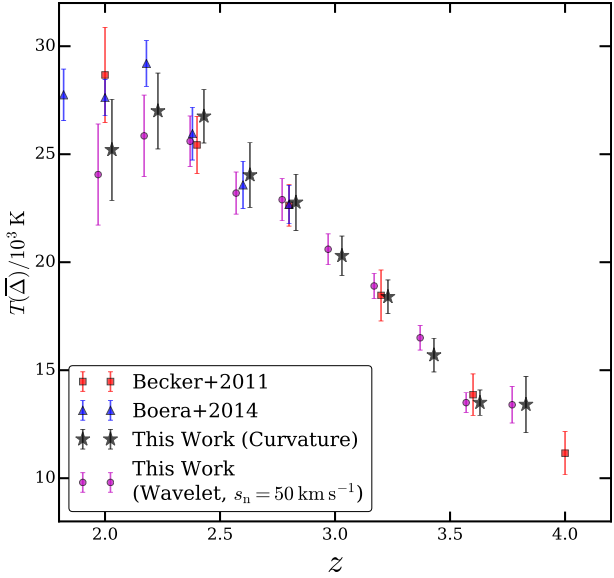
{The figure shows a comparison of temperature at the characteristic density,
$T(\overline{\Delta})$ from this work with that from
\citet{becker2011,boera2014}.  Our temperature measurements from the curvature
(black stars) and wavelet (magenta circles) statistics are in good agreement
with those from the literature at $z>2.4$. However, our measurements at $z<2.4$ are
systematically lower than those in the literature, but are consistent within 
$1.5 \sigma$.
}{\label{fig:T-Delta-c-Evolution}}


Similarly to \figref{fig:T0-gamma-best-fit-model}, we show the comparison of FPS,
wavelet, BPDF and curvature statistics of our best fit models with those obtained from
observations at different redshift bins in \figref{fig:best-fit-stat-2-0} to \ref{fig:best-fit-stat-3-8}.
In general, the predictions of the best fit models are in good agreement with 
the observed statistics. We summarize the reduced $\chi^2$ between best fit model
and observations in \tabref{tab:chi-sq-dof}. The reduced $\chi^2$ is close to 
1 in most cases. 
We also plot a comparison of the observed \blogNHI contours (that
includes 90 percent of points) with that from 5 mock samples (out of 100) in
\figref{fig:b-vs-NHI-scatter}. The mock \blogNHI distribution corresponds to the
best fit thermal parameters model.  Traditionally, the lower $b$ envelope in the
\blogNHI distribution is used to measure the thermal parameters. The limitations of
this method are described in \S \ref{app:bpdf}.
\figref{fig:b-vs-NHI-scatter} shows that the 2D observed \blogNHI distribution
is similar to that from the mock samples at all redshifts. However, in general we 
see a mismatch between observed \blogNHI and best fit \blogNHI distribution at
low and high \logNHI. The low \logNHI systems are usually contaminated 
due to the finite \SNR of the spectra. Note here that the Voigt decomposition of high \logNHI 
systems may not be unique. This effect is more prominent at higher ($z>3.3$) 
redshifts. Due to combination of these effects the mock \blogNHI
distributions further show some differences compared to observations.
It is noteworthy that the shape of the contours for lower $b$ values from the simulations
are in good agreement with that from observations.  The variations
in the \blogNHI distribution in different mock samples is also small indicating that the
distribution is converged with respect to cosmic variance.
In summary, the \blogNHI distribution from our best fit models is in general good agreement with that from observations.

In  \figref{fig:T0-gamma-evolution-literature} (panel B1), we show a comparison of $T_0$ from our method using 
the characteristic density with the results from \citet{becker2011,boera2014}. To get the $T_0$
evolution in our model we assume the best fit $\gamma$ evolution (with errors).
To perform a fair comparison with the \citet{becker2011,boera2014} method, we show 
the temperature at the characteristic density from our work and \citet{becker2011,boera2014}
in \figref{fig:T-Delta-c-Evolution}. Our temperature measurements from the curvature
and wavelet statistics are in good agreement
with that from the literature at $z>2.4$. However, our measurements at $z<2.4$ are
systematically lower than that from the literature, but are still consistent within 
$1.5 \sigma$. It is interesting to note that the characteristic density predicted
from our model (\oDelta $\sim 4.85$) is also smaller at $z \sim 2$ compared 
to that from \citet[][\oDelta $\sim 6$]{becker2011}.  

\section{Sources of uncertainty in $T_0$, $\gamma$ measurements}
\label{app:error-analysis}

\InputFig{T0_gamma_Constraints_Single_z_bin_Single_Panel_Chi_Square.pdf}{83}%
{ Each panel is same as \figref{fig:T0-gamma-constraints-single-z-bin}
    except in panel A2 and A3, we show reduced $\chi^2$ ($\chi^2_{\rm dof}$).
    The minimum $\chi^2_{\rm dof}$ is close to $\sim 1$ suggesting that the 
    errors estimated on statistics are not overestimated/underestimated.
}{\label{fig:T0-gamma-constraints-single-z-bin-chi-square}}

\InputFigCombine{T0_gamma_Constraints_Continuum_Metal_Effect_All_z.pdf}{170}
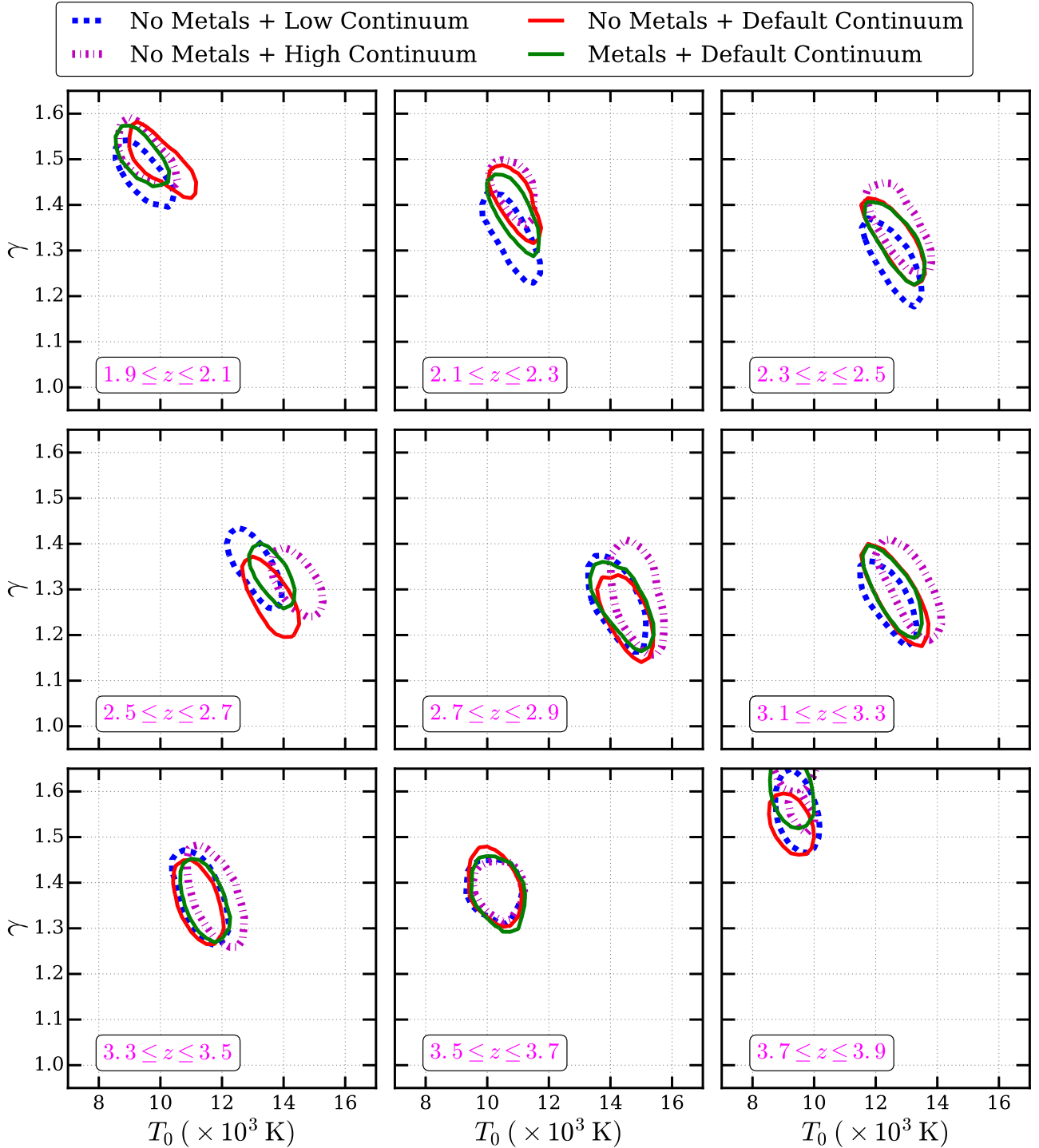
{ The figure shows the effect of continuum placement uncertainty and metal
    contamination on the $T_0$ and $\gamma$ constraints. Each panel is similar to
    panel A1 in \figref{fig:T0-gamma-constraints-continuum-metal-single-zbin}
    and shows the other 8 redshift bins spanning \zrange{1.9}{3.9}. 
}{\label{fig:T0-gamma-constraints-continuum-metal-all-zbin}}


In addition to statistical uncertainty, the $T_0$, $\gamma$ measurements are
also affected by the modeling and observational uncertainties.  We have
considered  four main sources of uncertainties contributing to the total
uncertainty on the measured  parameters. These are uncertainties due to (i)
continuum placement, (ii) contamination by unidentified metal lines, (iii)
(numerical) modeling and (iv) cosmological parameters. The first two
uncertainties are observational uncertainties while the later two uncertainties
are modeling uncertainties.  We discuss below the contribution of each to
the measurement uncertainty  in detail.


\begin{table*}
\centering
\caption{$T_0$ and $\gamma$ measurements and the associated error budget
}
\begin{threeparttable}
\begin{tabular}{ccccccc}
\hline  \hline 
Redshift  & Statistical  & Model  & Cosmological & Metal line & Continuum & Best fit $\pm$ Total \\
$z \: \pm \: dz$  & uncertainty \tnote{[a]} & uncertainty \tnote{[b]} & parameters \tnote{[c]} & contamination \tnote{[d]} & uncertainty \tnote{[e]} &  uncertainty \tnote{[f]}\\ 
& $(\sigma_{\rm stat})$ & $(\sigma_{\rm model})$ & $(\sigma_{\rm cosmo})$ & $(\sigma_{\rm metal})$ & $(\sigma_{\rm cont})$ &  $(\sigma_{\rm tot})$ \\ \hline
\multicolumn{7}{c}{Contribution of various uncertainties to the total uncertainty on $T_0$ (in K) measurements} \\ \hline
2.0 $\pm$ 0.1 & 1254 & 285 & 31 & 125 & 100 &  9500 $\pm$ 1393 \\
2.2 $\pm$ 0.1 &  831 & 330 & 36 & 125 & 125 & 11000 $\pm$ 1028 \\
2.4 $\pm$ 0.1 &  952 & 382 & 41 & 100 & 100 & 12750 $\pm$ 1132 \\
2.6 $\pm$ 0.1 &  894 & 405 & 44 & 125 & 400 & 13500 $\pm$ 1390 \\
2.8 $\pm$ 0.1 &  848 & 442 & 48 & 125 & 375 & 14750 $\pm$ 1341 \\
3.0 $\pm$ 0.1 &  827 & 442 & 48 & 125 & 375 & 14750 $\pm$ 1322 \\
3.2 $\pm$ 0.1 & 1045 & 382 & 41 & 100 & 375 & 12750 $\pm$ 1493 \\
3.4 $\pm$ 0.1 &  796 & 338 & 37 & 125 & 250 & 11250 $\pm$ 1125 \\
3.6 $\pm$ 0.1 &  885 & 308 & 33 & 125 & 125 & 10250 $\pm$ 1070 \\
3.8 $\pm$ 0.1 &  690 & 278 & 30 & 100 & 125 &  9250 $\pm$  876 \\
\hline 
\multicolumn{7}{c}{Contribution of various uncertainties to the total uncertainty on $\gamma$ measurements} \\ \hline
2.0 $\pm$ 0.1 & 0.081 & 0.018 & 0.003 & 0.010 & 0.012 & 1.500 $\pm$ 0.096 \\
2.2 $\pm$ 0.1 & 0.080 & 0.017 & 0.003 & 0.013 & 0.050 & 1.425 $\pm$ 0.133 \\
2.4 $\pm$ 0.1 & 0.095 & 0.016 & 0.003 & 0.010 & 0.025 & 1.325 $\pm$ 0.122 \\
2.6 $\pm$ 0.1 & 0.092 & 0.015 & 0.003 & 0.025 & 0.025 & 1.275 $\pm$ 0.122 \\
2.8 $\pm$ 0.1 & 0.094 & 0.015 & 0.003 & 0.010 & 0.012 & 1.250 $\pm$ 0.109 \\
3.0 $\pm$ 0.1 & 0.106 & 0.015 & 0.003 & 0.013 & 0.012 & 1.225 $\pm$ 0.120 \\
3.2 $\pm$ 0.1 & 0.115 & 0.015 & 0.003 & 0.013 & 0.013 & 1.275 $\pm$ 0.129 \\
3.4 $\pm$ 0.1 & 0.096 & 0.016 & 0.003 & 0.012 & 0.010 & 1.350 $\pm$ 0.108 \\
3.6 $\pm$ 0.1 & 0.088 & 0.017 & 0.003 & 0.012 & 0.010 & 1.400 $\pm$ 0.101 \\
3.8 $\pm$ 0.1 & 0.066 & 0.018 & 0.003 & 0.038 & 0.062 & 1.525 $\pm$ 0.140 \\
\hline \hline
\end{tabular}
\begin{tablenotes}
\item[a] Statistical uncertainty corresponds to uncertainty from joint
    constraints of respective parameters.
\item[b] Model uncertainty corresponds to uncertainty in modeling of the \lya forest in our simulation.
\item[c] Cosmological parameter uncertainty is calculated using the  Gunn-Peterson approximation. 
    Note that the GP approximation does not account for the uncertainty in $\sigma_8$.
\item[d] Contribution of unidentified metal  lines to the uncertainty in derived quantity. 
\item[e] Continuum placement uncertainty ($\sigma_{\rm cont}$) is assumed to be $\pm 5$ percent. For $T_0$ and $\gamma$ measurements, $\sigma_{\rm cont}$ includes the uncertainty due to the mean transmitted flux for a given continuum. For \logGTW  measurements, $\sigma_{\rm cont}$ include the uncertainty
due to variation of  $T_0$ and $\gamma$ parameters.
\item[f] Total uncertainty is obtained by $\sigma_{\rm tot} = \sqrt{\sigma^2_{\rm stat} + \sigma^2_{\rm model} + \sigma^2_{\rm cosmo} + \sigma^2_{\rm metal}} \: + \: \sigma_{\rm cont}$
\end{tablenotes}
\end{threeparttable}
\\
\label{tab:T0-gamma-error-budget}
\end{table*}
%


\subsection{Continuum placement uncertainty}
When comparing simulated \lya forest spectra with observations, it is
customary to use normalized observed spectra ($F_{\rm norm} = F_{\rm unnorm} /
F_{\rm cont}$).  The continuum fitting of spectra is usually based on high
order polynomial fitting connecting points in the spectrum believed to have no
absorption. The uncertainty in continuum placement $(\delta F_{\rm cont} /
F_{\rm cont})$ affects the observed flux statistics of \lya forest data and
influences  the inferred  physical parameters. In the \kodiaq DR2 sample, all
the spectra are normalized by the procedure described in
\citet{omeara2015,omeara2017}. The continuum placement uncertainty in the
\kodiaq DR2 sample is of the order of a few percent \citep[see][]{omeara2015}. We assume
a conservative value for the continuum placement uncertainty of $\pm 5$ percent.
If $F_{\rm default}$ is the flux corresponding to the default continuum and
$\delta F_{\rm cont} / F_{\rm cont}$ is the uncertainty in continuum placement
then we define a  low /high continuum as $F_{\rm default} / [1 \pm (\delta
        F_{\rm cont} / F_{\rm cont})]$.

We quantify the effect of continuum placement uncertainty by rescaling the
observed flux, deriving all the statistics from the rescaled spectra and
constraining the $T_0$, $\gamma$ parameters using our simulations.
\figref{fig:T0-gamma-constraints-continuum-metal-single-zbin} shows the effect
of continuum placement on  the (joint) $T_0$ and $\gamma$ constraints. For low
(high) continuum placement, $T_0$ and $\gamma$ are systematically
underpredicted (overpredicted) relative to measurements with the  default
continuum.  One can qualitatively understand this systematic effect of
continuum placement on flux.  Since dividing a flux by a low (high) continuum
stretches (compresses) the flux in the vertical direction, this artificially
reduces (increases) the line widths and hence one gets systematically smaller
(higher) temperature.
\figref{fig:T0-gamma-constraints-continuum-metal-all-zbin} shows a similar
effect of continuum placement uncertainty in the other redshift bins.  The
uncertainty in $T_0$ and $\gamma$ measurement due to continuum placement is
systematic in nature hence we add this uncertainty in the total error budget
(see \tabref{tab:T0-gamma-error-budget}). 

\subsection{Metal line contamination uncertainty}
The observed \lya forest is usually contaminated by absorption lines arising
from metals. For a given temperature, the metal lines are
typically narrower than \HI lines because for thermally broadened lines,
the line width $b \propto m^{-1/2}$ and
$m$ is larger for metals than hydrogen. The presence of such narrower metal lines
can potentially bias the temperature measurement to lower values.  In this
work, we do not model the intergalactic metal absorption systems as our
simulations neither follow chemical evolution of metals nor include feedback
processes. To mitigate this, we
chose QSO sightlines that do not contain DLAs or sub-DLAs. Thus our sample
selection criteria minimizes the number of metal lines contaminating the \lya
forest.  Furthermore, we identify  metal lines and replace them with continuum 
and add noise. We use manual and automated approaches
to identify metal lines.  We find that both approaches give consistent results.

In the manual approach, we first find the redshift of absorption systems using part of the spectra redward of the \HI emission line. Corresponding to this
redshift, we look for other transitions contaminating \lya forest.
Due to insufficient wavelength coverage, it is possible
that we may not identify all the metal lines. Hence we manually look at each
spectrum and look for obvious metal line systems. We then replace metal lines
with continuum  and add noise \footnote{We account for the effect of metal lines
blended with \HI absorption lines.}.

The manual method is tedious and somewhat  subjective. Hence, we have also used
an automatic method to identify the metal lines based on line widths. Since
metal lines are narrower than corresponding \HI lines, we fit all the spectra
with multi-component Voigt profiles using \viper. We treat all the lines with
$b \leq 8$ \kmps as metal lines\footnote{The narrowest line that can be
detected in spectra is limited by the line spread function (FWHM $\sim 6$
\kmps) of the spectrograph.}. We replace these metal lines with continuum and
add noise accounting for any blending effects.

\figref{fig:T0-gamma-constraints-continuum-metal-single-zbin} and
\figref{fig:T0-gamma-constraints-continuum-metal-all-zbin} shows the effect of
metal contamination on $T_0$ and $\gamma$ constraints in all the redshift bins.
As expected, when we use flux  statistics for the observed spectra without subtracting metal
contamination, we get systematically smaller $T_0$ and $\gamma$.  Even though we
identify and remove the metal lines, it is still possible that some of the
metal lines are not identified by our method.  Since the metal lines
contamination introduces an uncertainty of $\sim 2$ percent, we add this
uncertainty in the total error budget of constrained parameters (see
\tabref{tab:T0-gamma-error-budget}).

\subsection{Cosmological parameter uncertainty}
All the results presented in this work assume a flat $\Lambda$CDM cosmology
consistent with \citet{planck2014}. However the \lya optical depth depends on
cosmological parameters $\Omega_m, h$ and $\Omega_b$ as \begin{equation} \tau
    \propto  \: (\Omega_b \: h^2)^2 \: \Omega^{-0.5}_m\:  \frac{T^{-0.7}_0 \:
    \Delta^{2-0.7(\gamma-1)}}{\Gamma_{\rm HI}} \end{equation} where we assume
    that recombination rate scales with temperature as $T^{-0.7}$.  It would be
    computationally expensive to perform all the analysis with varying
    cosmological parameters. Hence we use a simplified approach of propagating
    the error using the above equation. For this we use the $1\sigma$ uncertainty
    on $\Omega_{m}, h$ and $\Omega_{b}$ from \citet{planck2014}. The
    cosmological parameters derived from the CMB are correlated with each other.
    Some of these correlation may lead to a cancellation of errors. However, while
    performing the error analysis, we always add errors due to different
    cosmological parameters.  Thus the assumed uncertainty due to cosmological
    parameters is conservative.  We do not account for 
    the uncertainty in $\sigma_8$ and $n_s$ as this would require to perform
self-consistent \gthree simulations. However it is important to note that, 
    we find that the uncertainty in $T_0$ and
    $\gamma$ measurements due to cosmological parameters uncertainty  is small,
    $\leq 0.5$ percent (see \tabref{tab:T0-gamma-error-budget}).

\subsection{Modeling uncertainty due to Jeans smoothing}
In the simulations, we vary the thermal state of the IGM by post-processing
\gthree simulations using \citecode.  Pressure smoothing effects due to the
hydrodynamical response of the gas are important during reionization
\citep{park2016,daloisio2020}. As shown in \citet{gaikwad2018}, we account for the
pressure smoothing of the gas by convolving the SPH kernel with a Gaussian
kernel that depends on the pressure smoothing scale
\citep{gnedin1998,kulkarni2015}. This approach is based on the ansatz that the
\lya absorbers are in local hydrostatistic equilibrium \citep{schaye2001}.
However, this assumption may not be valid during the process of reionization as
the photo-heating time scales are typically shorter than the dynamical response time of the gas. In addition to this, the \lya forest traces the neutral gas in
large scale filaments which are undergoing gravitational collapse onto  nearby
over-dense objects. The hydrostatic equilibrium approximation may thus not be
correct. To quantify the effect of this ansatz, we perform a self-consistent
\gthree simulation with enhanced photo-heating rates where $T_0$ is
doubled. The dynamical effect of pressure smoothing is thus captured accurately
in this self-consistent simulation. We compute the \lya forest statistics for this
self-consistent model and assume it as the fiducial model. We constrain $T_0$,
$\gamma$ and $\Gamma_{\rm HI}$ using our model, i.e. \gthree (at lower
temperature) + \citecode (with convolution of SPH kernel with Gaussian). We
find that the fiducial $T_0$, $\gamma$ and $\Gamma_{\rm HI}$ are recovered
within 2.5, 1.3 and 0.6 percent.  Such
a good level of agreement is partly expected because the smoothing of
transmitted flux at these redshifts is dominated by temperature, while the
pressure smoothing effects are sub-dominant.  We account for this modeling
uncertainty in the total error budget of derived parameters (see
\tabref{tab:T0-gamma-error-budget}). We also refer the reader to \S
\ref{sec:simulation} for a related discussion.


\section{Thermal parameter evolution in UVB models}
\label{app:T0-gamma-evolution-uvb-models}
In this section we discuss the effect of uncertainty in observed and 
modeling parameters in the UVB models on the thermal parameter evolution 
during \HeII reionization.

\subsection{Effect of uncertainty in $T_0-\gamma$ at $z=6$ on thermal parameter
evolution at \zrange{2}{4}}
\label{app:initial-T0-gamma-effect}

In this section we discuss the effect of varying the  thermal parameters $T_0$
and $\gamma$ at $z=6$ (the initial redshift chosen for \citecode in this work)
on the thermal evolution at \zrange{2}{4}. Recently \citet{gaikwad2020}
measured  $T_0$ ($\gamma$) at $z=6$ and found it to be larger  than predicted
by the \citetalias{haardt2012,onorbe2017,khaire2019a,faucher2020} UVB models.
Since we run  \citecode from $z=6$ to $z=2$, it is possible that an uncertainty
in $T_0$ and $\gamma$ at $z=6$  affects the thermal parameter evolution at
\zrange{2}{4}. \figref{fig:T0-gamma-evolution-modified-FG18} shows that the
effect of even a large uncertainty in thermal parameters at $z=6$ is subdominant
at lower redshifts. This is because the lower redshift thermal evolution is
mainly affected by the \HeII reionization prescriptions used in these models.
Thus the effect of uncertainty in $T_0,\gamma$ at $z=6$ is less than 4 percent
(maximum) at \zrange{2}{4}.  Note that even though we show this result only for
the \citetalias{faucher2020} UVB models, the results are similar for the other
UVB models.

\subsection{Effect of QSO spectral energy distribution on thermal parameter
evolution at \zrange{2}{4}}
\label{app:qso-sed-effect}
The observed QSO spectral index ($\alpha$, $f_{\nu} \propto \nu^{-\alpha}$) is
another  uncertain quantity in UVB calculations. The observations of stacked
QSO spectra in the far ultra-violet  suggests that QSO spectral indices vary
from $\alpha=1.4$ to $2.0$ \citep{lusso2014,stevans2014,shull2020}. Recently
\citet{gaikwad2019} showed that the effect of the QSO spectral index on thermal
parameter evolution is degenerate with the redshift of \HeII reionization in
\citetalias{khaire2019a} such that a flatter QSO SED corresponds to an earlier
\HeII reionization.  They further  showed that $\alpha =1.8$ is consistent with
the evolution of \taueffHeII from \citet{worseck2019}.  In
\figref{fig:T0-gamma-evolution-qso-sed-uvb-f-XI}, we show the comparison of our
thermal parameter measurements  with that predicted by the
\citetalias{khaire2019a} UVB for different QSO spectral indices. Irrespective
of the QSO spectral  index, the total emissivity is kept constant  in all the
models. The shape of the thermal parameter evolution is hence similar for all the
models. The redshift of maximum $T_0$ occurs at slightly higher redshift and is
slightly higher for flatter (small $\alpha$) QSO SED.  The earlier temperature
rise is mainly because of the \fHeII evolution  (see panel B1
\figref{fig:T0-gamma-evolution-qso-sed-uvb-f-XI}) in these models which is also
different such that \HeII reionization occurs slightly earlier for flatter QSO
SED \citep{gaikwad2019}.    However,  even after changing the  QSO SED spectral 
index in the observationally allowed range, the thermal parameter evolution in
the \citetalias{khaire2019a} UVB model is significantly different from our measurements. This suggests that additional modifications (such as $\eta$
evolution) may be needed  for the  \citetalias{khaire2019a} UVB models to become
consistent with our temperature measurements.


\subsection{Modifications to UVB models}
\label{app:uvb-modification}

\InputFigCombine{T0_gamma_Evolution_FG20_photo_heating_rate_variation.pdf}{170}
{%
    Each panel here is the same as \figref{fig:T0-gamma-evolution-all-uvb}
    except the sensitivity of thermal parameters to the photo-heating rates and
    initial $T_0$ and $\gamma$ at $z=6$ is illustrated for the
    \citetalias{faucher2020} non-equilibrium model.  Since the $T_0$ evolution
    for the default \citetalias{faucher2020} model is systematically higher
    than our measurements, we rescale the \HI, \HeI and \HeII photo-heating
    rates in panel A1 and A2.  The observed thermal parameter evolution (along
    with $1\sigma$ uncertainty) from this work is in good agreement with
    the  non-equilibrium prediction of the \citetalias{faucher2020} UVB model
    if photo-heating rates are reduced by a factor of 0.7$\pm0.1$.  A simple
    scaling of photo-heating rates does not affect the $\gamma$ evolution which
    is consistent with \figref{fig:thermal-history-variation}.  Given that the
    uncertainty in thermal parameters after \HI reionization is large, panel
    B1,B2 show the effect of initial $T_0$ and $\gamma$ at $z\sim6$ on
    the $T_0,\gamma$ evolution at late times \zrange{2}{4} \citep{gaikwad2020}.
    The uncertainty in $T_0,\gamma$ at $z \sim 6$ has only a small  (less than
    3 percent at $z\sim 3$) effect on the $T_0$ and $\gamma$ evolution at late
    times. This suggests that the thermal effect of \HI reionization  at
    \zrange{2}{4} is sub-dominant relative to that of \HeII reionization.
}{\label{fig:T0-gamma-evolution-modified-FG18}}

\InputFigCombine{T0_gamma_Evolution_all_UVB_photo_heating_rate_variation.pdf}{170}
{
    Same as panel A1 and A2 in \figref{fig:T0-gamma-evolution-modified-FG18}
    except photo-heating rates are varied here for the 
    \citetalias{haardt2012,onorbe2017,khaire2019a,puchwein2019} UVB models.
}{\label{fig:T0-gamma-evolution-modified-uvb}}

\InputFigCombine{T0_gamma_Evolution_QSO_SED_UVB_Models_f_XI_Evolution.pdf}{175}
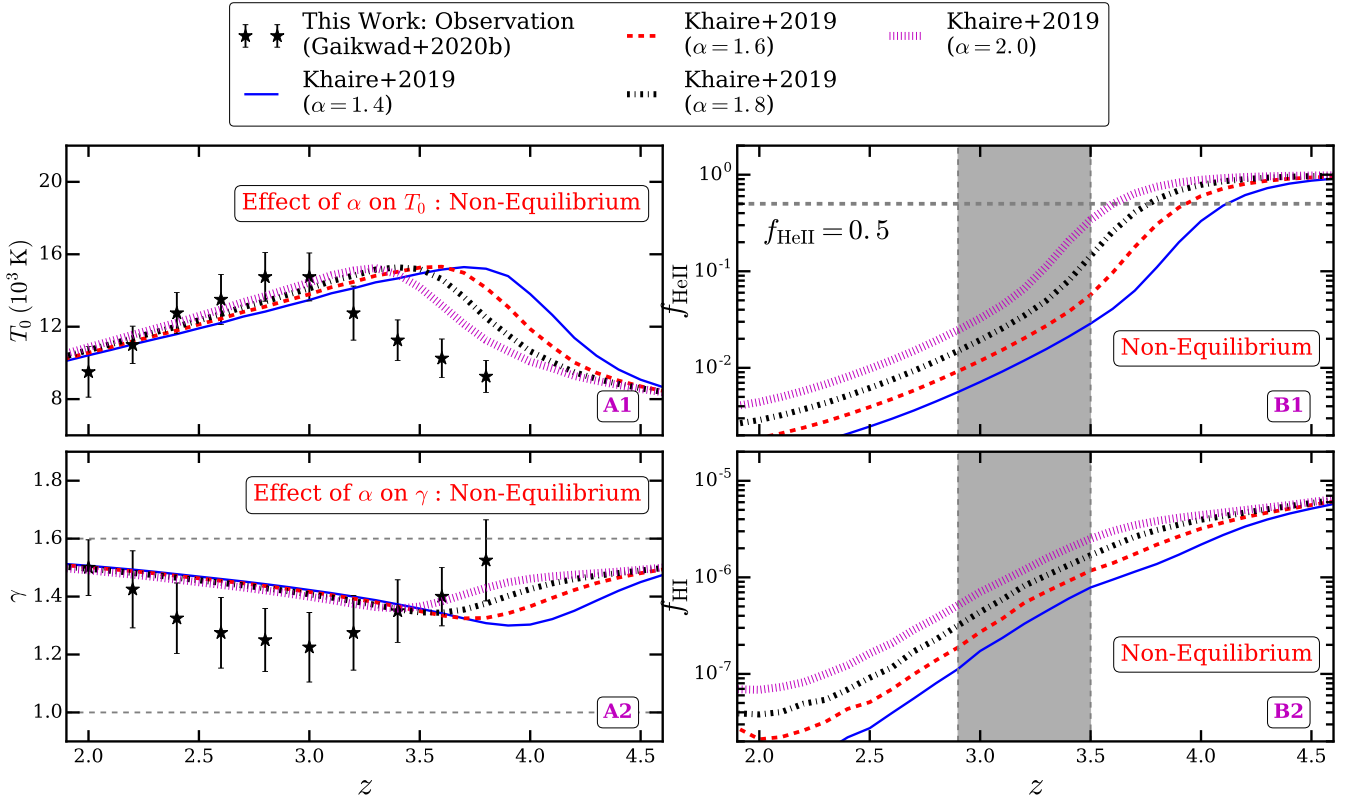
{ Each panel here is the same as
    \figref{fig:T0-gamma-evolution-modified-uvb-f-XI} except the variation of
    thermal parameters and \HI, \HeII fractions with QSO spectral index
    ($f_{\nu} \propto \nu^{-\alpha}$) in \citetalias{khaire2019a} is shown.
    Flatter slopes of the QSO spectra correspond to earlier  \HeII
    reionization  consistent with \citet{gaikwad2019}. All the models predict
    similar $T_0$ and $\gamma$ evolution at $z<3$.
}{\label{fig:T0-gamma-evolution-qso-sed-uvb-f-XI}}

Recently \citetalias{faucher2020} updated their \citet{faucher2008} UVB model with
updated constraints on galaxy and AGN luminosity functions \citep{kulkarni2019b},
intergalactic \HI and \HeII opacity models, and calibrations to measurements of \GHI and \GHeII etc. In
their UVB model, AGN (galaxy) dominate the  \HI ionizing background at $z \leq 3$
($z>3$).  Their UVB model crucially depends on parameters related to the
reionization history mainly the total temperature increments due to \HI (\DeltaTHI)
and \HeII (\DeltaTHeII) reionization heating. The \HI and \HeII photo-heating
rates are directly proportional to \DeltaTHI and \DeltaTHeII respectively.
Since the timing and amount of photo-heating during \HeII reionization is
poorly constrained at present, they parameterize this ignorance in \DeltaTHI
and \DeltaTHeII.  \citetalias{onorbe2017,faucher2020} assumed \DeltaTHI $=20000$ K
and \DeltaTHeII $=15000$ K and obtain a $T_0$ and $\gamma$ evolution similar
to previous measurements. However, both  groups assumed that photo-ionization
equilibrium is valid at all  times. But as shown by
\citet{puchwein2015,gaikwad2018,puchwein2019}, this is not a good assumption
during \HeII reionization as the photo-ionization time scales are shorter than
the recombination time scales. Thus non-equilibrium ionization evolution is more
physically motivated. \citet{puchwein2015,gaikwad2018} shows that the $T_0$ of
the IGM in non-equilibrium is significantly larger than in equilibrium calculations.
Thus when we use non-equilibrium ionization evolution for \citetalias[][which are
calibrated to match observation assuming photo-ionization equilibrium
]{onorbe2017,faucher2020}, we get significantly larger temperature predictions for  their
UVB models. In order to match the observations with non-equilibrium ionization
evolution, we need to reduce \DeltaTHI and \DeltaTHeII. The reduction of 
\DeltaTHI and \DeltaTHeII by  a factor of $f_{\rm scale}$ is equivalent to 
reducing \HI, \HeI and \HeII photo-heating rates by a factor $f_{\rm scale}$.

\figref{fig:T0-gamma-evolution-modified-FG18} shows the variation in $T_0$ and
$\gamma$ evolution with the variation in \DeltaTHI and \DeltaTHeII predicted by
the \citetalias{faucher2020} model using \gthree + \citecode.  We show results
in \figref{fig:T0-gamma-evolution-modified-FG18} for four $f_{\rm scale} =
1.0,0.8,0.7,0.6$ for \HI, \HeI and \HeII photo-heating rates.  The main effect
of changing this parameter is to reduce $T_0$ systematically at all
redshifts while  $\gamma$  changes very little.  This is expected as scaling
the parameters \DeltaTHI, \DeltaTHeII is equivalent to scaling \HI and \HeII
photo-heating rates and such scaling changes $T_0$ (keeping $\gamma$ relatively
same see \figref{fig:thermal-history-variation}).
\figref{fig:T0-gamma-evolution-modified-FG18} shows that the $T_0$ and $\gamma$
evolution is consistent with our measurements if  we use non-equilibrium
ionization evolution with $f_{\rm scale} = 0.7$ parameters in
\citetalias{faucher2020}.  For completeness, we summarize the best fit
\DeltaTHI and \DeltaTHeII parameters used in
\citetalias{onorbe2017,faucher2020} in \tabref{tab:heat-injection-parameter}
that are consistent with our temperature measurements.  In order to do a fair
comparison with other UVB models, we show the $T_0$ and $\gamma$ evolution for
varying the photo-heating rates in
\figref{fig:T0-gamma-evolution-modified-uvb}.  We find that other UVB models
match the $T_0$ evolution in some redshift bins but are systematically above
or below in other redshift bins. The evolution of $\gamma$ is similar to that
shown in \figref{fig:T0-gamma-evolution-all-uvb}.

Finally, we compare the evolution of $T_0$ and $\gamma$ over the full redshift
range  \zrange{2}{6} and all the different flux statistics in
\figref{fig:T0-gamma-evolution-literature-combine}.  We emphasize that
a considerable amount of work has also been done  to measure the $T_0$ and
$\gamma$ evolution at  $z>4$.  At these redshifts the thermal parameters have
been  measured using FPS \citet{walther2019,boera2019}, curvature statistics
\citet{becker2011}, proximity effects \citet{bolton2012} and the width distribution
of transmission spikes \citet{,gaikwad2020}. These measurements broadly agree
with the theoretical $T_0$ and $\gamma$ evolution predicted by the different
UVB models.


\InputFigCombine{T0_gamma_Evolution_Literature_Comparison.pdf}{170}
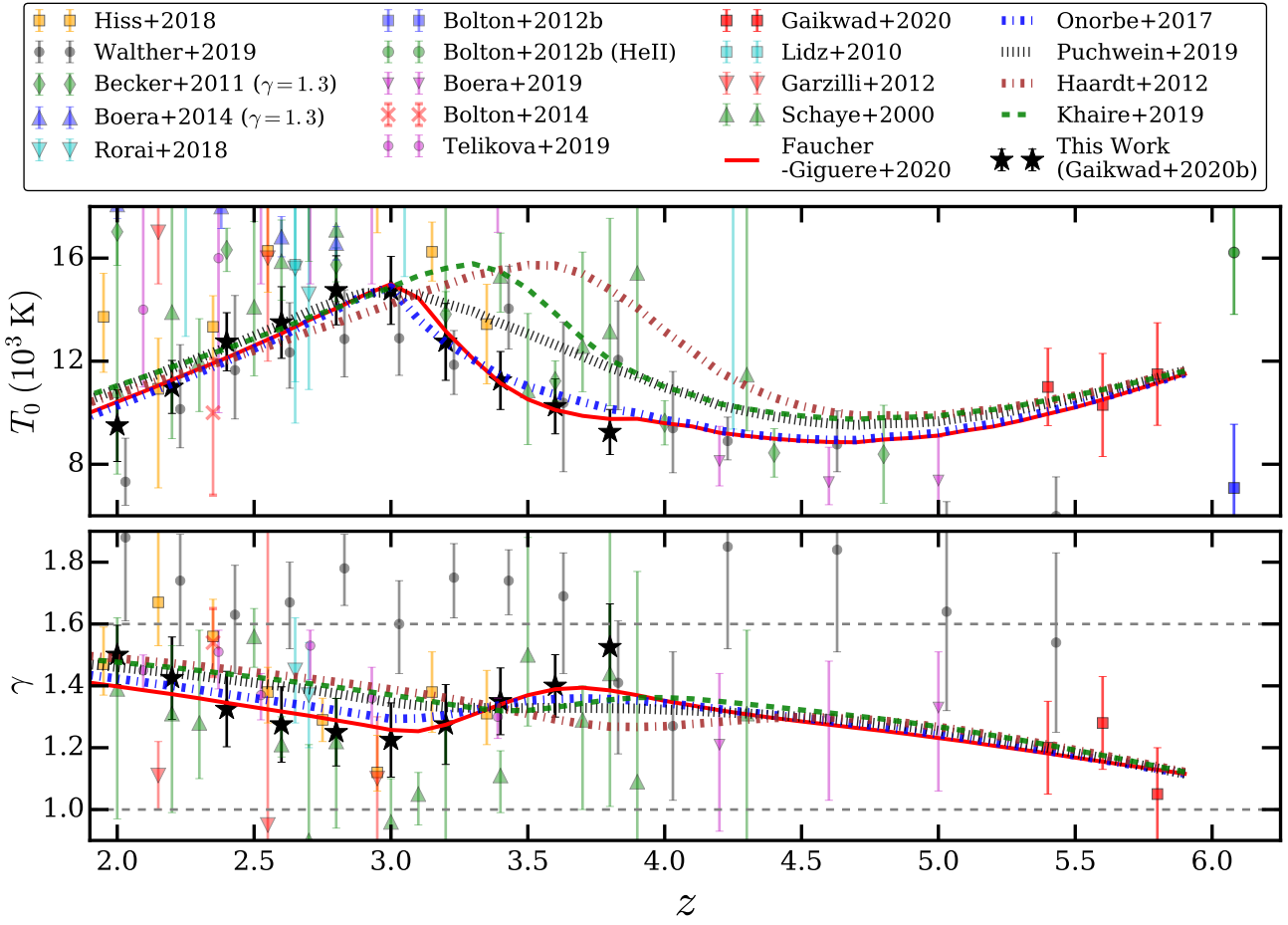
{Comparison of $T_0-\gamma$ measurements from this work with measurements in the
literature.  We also show the theoretical evolution of thermal parameters predicted by the 
 modified \citetalias{onorbe2017,puchwein2019,faucher2020} UVB models.  For the
 \citetalias{onorbe2017,puchwein2019,faucher2020} UVB
models, we have rescaled the  photo-heating rates by a factor of
0.8,0.9,0.7, respectively. 
}{\label{fig:T0-gamma-evolution-literature-combine}}

\begin{table}
\centering
\caption{Heat injection parameters (\DeltaTHI, \DeltaTHeII) in UVB models consistent with the measured $T_0$ and $\gamma$ evolution from this work.}
\begin{tabular}{ccc}
\hline  \hline 
UVB model  & \DeltaTHI (K) & \DeltaTHeII (K)  \\
\hline
\citet{onorbe2017}  & $15200 \pm 1800$ & $11400 \pm 2100$ \\
\citet{faucher2020} & $13600 \pm 2400$ & $10200 \pm 1800$ \\
\hline \hline
\end{tabular}
\\
\label{tab:heat-injection-parameter}
\end{table}

\subsection{The way forward for UVB models}
\label{app:uvb-model-forward-way}

We have rescaled the photo-heating rates in \S
\ref{subsubsec:reionization-history} to better match our temperature
measurements. Even though this is a simple modification to the UVB models, this
is not necessarily  physical because photo-ionization and photo-heating rates
are ultimately linked to the  population of ionising sources  (QSOs and /or
galaxies) and IGM properties. In this section we discuss the possible
modifications that may be needed in the various UVB models to match our
measurements. Note, however, that implementing these in the  UVB models is
beyond the scope of present work hence we discuss qualitative features of a
physically motivated UVB model that could match our measurements.

From \figref{fig:T0-gamma-evolution-all-uvb}, it is clear that the maximum in
$T_0$ occurs at systematically lower redshift than that from
\citetalias{haardt2012,khaire2019a} UVB models.  However, the maximum value of
$T_0$ from these models seem to agree well with that from observations. This
suggests that the normalization of the QSO emissivity in these models is
roughly consistent with observations. However, the \HeII reionization in these
models starts early and is more gradual than suggested by observations (see
panel B1 in \figref{fig:T0-gamma-evolution-modified-uvb-f-XI}).  Thus, in
\citetalias{haardt2012,khaire2019a} the redshift evolution of  \fHeII needs
to be modified. In \citetalias{khaire2019a}, this could be achieved by
changing the ratio of \HeII and  \HI fractions ($\eta$). Since there are
only observational limits limits on \fHeII and the fluctuations in \HeII
ionizing background are important at $z>3$, there is scope for modification
in the \fHeII evolution of the \citetalias{khaire2019a} UVB. 
The \fHeII evolution can also be modified by
varying QSO SED index \citep[$\alpha, f_{\nu} \propto
\nu^{-\alpha}$][]{gaikwad2019}.  However, we show in online supplementary 
appendix \ref{app:uvb-modification} that the observed thermal parameter evolution
can not be matched by simply varying the QSO SED index within the
observationally allowed range of $\alpha=1.4-2.0$
\citep{lusso2014,stevans2014,shull2020}.  For the \citetalias{haardt2012}
UVB model both the QSO luminosity functions and QSO SED index need to be
updated to be compatible with the most  recent compilations of the relevant
observation\citep[see e.g.][]{kulkarni2019b}.

In summary, the observed thermal parameter evolution from this work suggests a
combination of changes to existing UVB models. For
\citetalias{haardt2012,khaire2019a} UVB models, modification in \fHeII is
necessary. For the \citetalias{onorbe2017,puchwein2019,faucher2020} UVB
models modification  of the ionizing emissivity from QSOs is needed. 

\section{Cumulative energy parameter}
\label{app:cumulative-energy}
\InputFigpages{u0_Evolution_All_UVB_Models.pdf}{80}
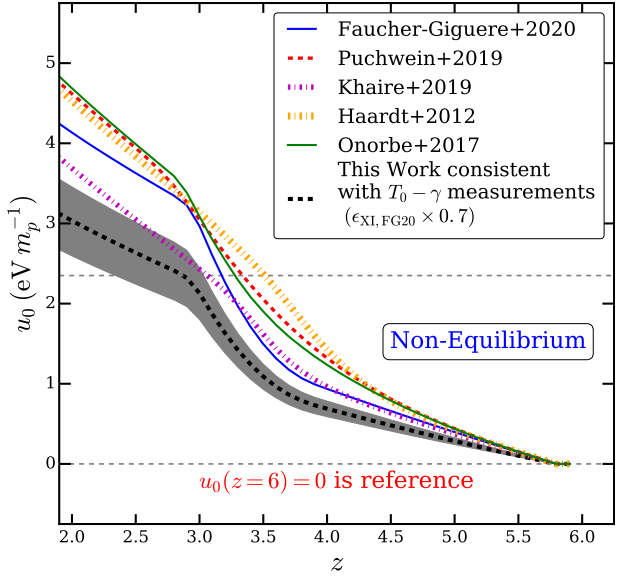
{ The evolution of the cumulative energy deposited into the IGM per unit mass ($u_0$) for
different non-equilibrium UVB models is shown. We chose $u_0(z=6) = 0$ as
reference point to minimize the effect of uncertainty in $u_0$ due to \HI
reionization.  The $u_0$ described in this work is primarily due to \HeII
reionization.  The black dashed line and the shaded region show the $u_0$
evolution consistent with the measured  $T_0-\gamma$ evolution from this
work. Without re-scaling the photo-heating rate of all the UVB models  predict
systematically larger $u_0$ than consistent with our temperature measurements.
Not that  the $u_0$ evolution accounts for  the heating but not the cooling of the IGM. 
Hence to match the $T_0-\gamma$ evolution, the whole of the
$u_0$ evolution is important and not  just matching $u_0$ at some redshift. A
sharp increase in the  $u_0$ inferred from our temperature measurements 
is evident  at \zrange{3.0}{3.6}.}{\label{fig:u0-evolution}}{1}
The default photo-heating rates of all  the UVB models predict systematically
higher $T_0$ than we measure at $z>3$ (see
\figref{fig:T0-gamma-evolution-all-uvb}). Since we have changed the photo-heating
rates to better match the observed $T_0$ evolution, it is also interesting to
have a look at  the evolution of the cumulative energy ($u_0$) deposited into
IGM by  \HeII reionization.  We calculate the cumulative energy per unit mass
($u_0$) deposited into IGM as 
\begin{equation} u_0 = \int
    \limits_{z=6}^{z_{re}} \; \frac{\mathscr{H}}{\overline{\rho}} \:\;,
    \frac{dz}{H(z) \: (1+z)}
\end{equation} 
where $\mathscr{H} = \sum n_{\rm XI} \: \epsilon_{\rm XI}$ (with
XI$=$\HI,\HeI,\HeII), $\overline{\rho},H(z),n_{\rm XI},\epsilon_{\rm XI}$ are
mean baryon density, Hubble constant, number density of species XI and
photo-heating rate of specie XI \citep{nasir2016}.
To calculate $u_0$ for a simulation box with
given UVB and at a given redshift, we first calculate the volume average
neutral fraction $f_{\rm XI}$. We then use above expression to compute $u_0$.
Given that there could be uncertainty in $u_0$ after \HI reionization, we chose
$u_0(z=6) = 0$ as a reference point. 
The $u_0$ defined in this work is mainly sensitive to heat deposited during
\HeII reionization.  

In \figref{fig:u0-evolution} we show the evolution of $u_0$ for all the UVB
models with non-equilibrium ionization evolution.  The observed $u_0$ evolution
is difficult to measure directly from observations. To nevertheless facilitate
a comparison, we show the $u_0$ evolution in \figref{fig:u0-evolution} for the
\citetalias{faucher2020} non-equilibrium UVB with scaled photo-heating rates
that matches the observed $T_0-\gamma$ evolution from this work.  We treat the
$u_0$ evolution corresponding to this model as that inferred from our
temperature measurements (black dashed line with shaded region).

\figref{fig:u0-evolution} also shows the $u_0$ evolution in all the UVB models
assuming non-equilibrium ionization evolution. The photo-heating rates here are
not scaled for the UVB models. At $z=3$ the $u_0$ predicted by all the UVB
models is systematically higher than the $u_0$ evolution inferred from our
temperature measurements (\tabref{tab:early-late-reionization}). Thus to match
the observed $T_0-\gamma$ evolution smaller values of energy deposited per unit
mass $2.13 \: \pm 0.32$ ${\rm eV} \: m^{-1}_p$ at $z \sim 3$ are required.  It
is interesting to note that $u_0$ at $z=3$ from the \citetalias{khaire2019a}
model is close to that from a model which is consistent with our measurements.
However, the $T_0$ and $\gamma$ evolution from the \citetalias{khaire2019a}
model is significantly different at $z>3$. This is because $u_0$ accounts for
the heating of the IGM, but does not account for the cooling of the IGM due to
recombination, Hubble expansion or inverse Compton cooling etc. Hence to match
the $T_0-\gamma$ evolution, the whole of the $u_0$ evolution is important and
not only  matching $u_0$ at some redshift. The comparison of $u_0$ evolution
in \figref{fig:u0-evolution} again suggests that rather rapid \HeII
reionization is needed to match our measured $T_0-\gamma$ evolution.  The $u_0$
evolution in \citetalias{haardt2012,khaire2019a,puchwein2019} is more gradual
compared to the other two models.








\bsp	
\label{lastpage}
\end{document}